\documentclass[a4paper, 11pt]{article}
\usepackage{jheppub} 
\usepackage{graphicx}
\usepackage{amsmath}
\usepackage{xspace}

\newcommand{\eV}{\text{e\kern-0.15ex V}\xspace}
\newcommand{\GeV}{\text{G\eV}\xspace}

\def \dd#1{\mathop{\textnormal{d}#1}}
\def \kstar{{K^{\!*}}}
\def \thl {{\theta_\ell}}
\def \thK {{\theta_{K}}}

\def \azeL{{{\cal A}_0^L}}
\def \azeR{{{\cal A}_0^R}}
\def \apaL{{{\cal A}_\parallel^L}}
\def \apaR{{{\cal A}_\parallel^R}}
\def \apeL{{{\cal A}_\perp^L}}
\def \apeR{{{\cal A}_\perp^R}}
\def \abs#1{\left| #1 \right|}
\def \eff{{\text{eff}}}
\def \nn {\nonumber}
\def \evtgen{\texttt{EvtGen}\xspace}

\title{A new Monte Carlo Generator for BSM physics in $B\to K^* \ell^+
  \ell^-$ decays with an application to lepton non-universality in
  angular distributions}

\author[a]{Alexei~Sibidanov,}
\author[a]{Thomas~E.~Browder,}
\author[a]{Shawn~Dubey,}
\author[a]{Shahab~Kohani,}
\author[b]{Rusa~Mandal,}
\author[c]{Saurabh~Sandilya,}
\author[a]{Rahul~Sinha}
\author[a]{and Sven~E.~Vahsen}

\affiliation[a]{University of Hawai`i at Manoa, Honolulu, HI 96822, USA}
\affiliation[b]{Department of Physics, Indian Institute of Technology Gandhinagar, Palaj, Gujarat 382355, India}
\affiliation[c]{Indian Institute of Technology Hyderabad (IITH), Telangana 502285, India }

\date{\today} \abstract{ Within the widely used \evtgen framework, we
  have added a new event generator model for $B\to K^* \ell^+ \ell^-$
  with improved standard model (SM) decay amplitudes and possible BSM
  physics contributions, which are implemented in the operator product
  expansion in terms of Wilson coefficients. This event generator can
  then be used to estimate the statistical sensitivity of a simulated
  experiment to the most general BSM signal resulting from
  dimension-six operators.  We describe the advantages and potential
  of the newly developed `Sibidanov Physics Generator' in improving
  the experimental sensitivity of searches for lepton non-universal
  BSM physics and clarifying signatures. The new generator can
  properly simulate BSM scenarios, interference between SM and BSM
  amplitudes, and correlations between different BSM observables as
  well as acceptance bias. We show that exploiting such correlations
  substantially improves experimental sensitivity. As a demonstration
  of the utility of the MC generator, we examine the prospects for
  improved measurements of lepton non-universality in angular
  distributions for $B \to K^* \ell^+ \ell^-$ decays from the expected
  50~ab$^{-1}$ data set of the Belle~II experiment, using a
  four-dimensional unbinned maximum likelihood fit. We describe
  promising experimental signatures and correlations between
  observables. The use of lepton-universality violating
  $\Delta$-observables significantly reduces uncertainties in the SM
  expectations due to QCD and resonance effects and is ideally suited
  for Belle II with the large data sets expected in the next
  decade. Thanks to the clean experimental environment of an $e^+ e^-$
  machine, Belle II should be able to probe BSM physics in the Wilson
  coefficients $C_7$ and $C_7'$, which appear at low $q^2$ in the
  di-electron channel.}
  
\begin{document}
\maketitle
\section{Introduction}
Flavor-changing neutral current (FCNC) processes in the weak
interaction do not occur at tree-level in the standard model (SM),
which makes them sensitive probes of physics beyond the standard model
(BSM). Intriguing BSM hints have been reported in $b \to s$
transitions by multiple flavor physics experiments, LHCb, Belle and
BaBar. Here we investigate the prospects for the decay $B\to K^{*}
\ell^+ \ell^-$ with $\ell=e,\mu$.  We have implemented a new Monte
Carlo (MC) signal generator model inside the widely used event
generator framework \evtgen~\cite{Lange:2001uf}, with improved SM
decay amplitudes implemented in the operator product expansion in
terms of the Wilson Coefficients $C_7, C_9, C_{10}$ and their
right-handed counterparts $C_7^{\prime}, C_9^{\prime},
C_{10}^{\prime}$. We allow for imaginary contributions to $\delta
C_i$. However, in this paper, we do not consider CP violating
observables. Amplitudes for additional BSM physics contributions to
$B\to K^{*} \ell^+ \ell^-$, which we call $\delta C_i=C_i^{\rm
  eff}-C_i^{\rm SM}$, and which correspond to a general BSM signal
resulting from dimension-six operators, can be chosen by the
user. Until recently, non-zero $\delta C_9$ and $\delta_{C_{10}}$ were
preferred by global theoretical fits to experimental
data~\cite{Altmannshofer:2021qrr,Alguero:2021anc,Hurth:2021nsi,Ciuchini:2021smi,Chrzaszcz:2018yza}. While
the significance of lepton-flavor universality violating contributions
(LFUV) is reduced and large signals as indicated in [2-6] are not
expected, LFUV contributions are still possible and might become
significant once again with larger data sets. Our study considers such
a case. Non-zero $\delta C_9$ and $\delta C_{10}$ for the di-muon
final state are preferred by theoretical fits to current
data~\cite{Ciuchini:2022wbq,Alguero:2023jeh,Hurth:2023jwr}.

We describe the advantages and the potential of the newly developed
`Sibidanov Physics Generator' in improving statistical sensitivity of BSM physics
searches and clarifying experimental signatures.  We have examined
potential experimental correlations in the four variables that
characterize $B\to K^{*} \ell^+ \ell^-$ decays
(Fig.~\ref{fig:kinematics}): the invariant mass of the lepton pair,
$q^2$, the lepton helicity angle from the virtual $Z/\gamma$-decay,
$\cos \theta_\ell$, the $K^*$ helicity angle, $\cos\theta_K$, and the
angle between the $K^*$ and di-lepton decay planes, $\chi$. In
addition, there are angular asymmetries that are a function of $q^2$.

Inclusion of BSM amplitudes in the MC generator allows one to properly
simulate acceptance bias in BSM scenarios, interference between SM and
BSM amplitudes, and correlations between different BSM
observables. For example, $\delta C_9$ will lead to correlated
signatures in the angular asymmetries $A_{FB}$ versus $q^2$ and $S_5$
versus $q^2$.  Another striking signature of $C_7^\prime$ is found in
the modulation of the angle $\chi$ at low $q^2$ in $B \to K^* e^+e^-$.
Belle II should have excellent statistical sensitivity to such BSM physics in $C_7$ and $C_7^\prime$.

We find a number of signatures that were previously not fully
appreciated.  We show that exploiting correlations between angular
observables substantially improves experimental statistical sensitivity. A
four-dimensional likelihood function is used to fit for the $\delta
C_i$ coefficients. We describe the potential for improved measurements
from the expected 50~ab$^{-1}$ data set of the Belle~II experiment
with fits to the angles (see Fig.~\ref{fig:kinematics}) and $q^2$ in
$B \to K^{*} \ell^{+} \ell^{-}$ decays.

In addition to BSM contributions, the observed angular asymmetries
could originate from SM QCD or $c \bar{c}$ resonance effects, which
appear in the experimental final state $K^*\ell^+\ell^-$. Given the
difficulties in reliably computing all of the possible hadronic
effects, the use of $\Delta$-observables appears to be the only
possible approach to unambiguously distinguish between these SM
effects and lepton-universality violating BSM physics in $b\to s\mu^+
\mu^-$. Using the BSM Monte Carlo generator, we will demonstrate the
feasibility of this experimental approach. We have used the overall
detection efficiency from Belle analyses \cite{Belle:2016fev} to
estimate statistical uncertainties. However, it should be noted that
the MC simulation results described here do not yet include
backgrounds, final state radiation, detector resolution and efficiency
functions.

The three most important $\Delta$-observables are:
\begin{align}
  \Delta A_{\rm FB} (B\to K^* \ell^+ \ell^-) &\equiv A_{\rm FB} (B\to K^* \mu^+ \mu^-)  - A_{\rm FB} (B \to K^* e^+ e^-),\\
  \Delta S_5 (B\to K^* \ell^+ \ell^-) &\equiv S_5 (B\to K^* \mu^+ \mu^-) -
  S_5(B \to K^* e^+ e^-),\ {\rm and}\\
  \Delta C_9^{\rm eff} (B\to K^* \ell^+ \ell^-) &\equiv C_9^{\rm eff} (B\to K^* \mu^+ \mu^-)  - C_9^{\rm eff} (B
  \to K^* e^+ e^-).
\end{align}
The first two of these observables can be derived from angular
asymmetries, while the third is obtained from the four-dimensional,
unbinned maximum likelihood fits described above.

An example of $\Delta$-like variables was introduced in
Ref.~\cite{Capdevila:2016ivx}. Belle reported a first attempt
to measure $\Delta$-type observables in $B\to K^* \ell^+ \ell^-$ using
a 0.7 ab$^{-1}$ data sample~\cite{Belle:2016fev}. The
$\Delta$-observables appear ideally suited for Belle~II (which has
comparable sensitivities for di-electrons and di-muons) with the large
data sets expected in the next decade~\cite{B2TIPbook}.

Here we introduce a new Monte Carlo generator to enable further
searches for BSM contributions in $B\to K^* \ell^+ \ell^-$. As an
example of the utility of this MC tool, we apply it to generate a
number of BSM scenarios and then carry out four-dimensional unbinned
maximum likelihood fits to future $B\to K^* e^+ e^-$ and $B\to K^*
\mu^+\mu^-$ Belle II datasets and estimate the statistical sensitivity
to lepton flavor violating contributions to the coefficient $C_9$.  A
proposal to perform unbinned fits to angular distributions of dimuon
and dielectron $B\to K^* \ell^+ \ell^-$ modes in LHCb datasets was
given in \cite{MauriLFV:2018vbg}. However, in contrast to
\cite{MauriLFV:2018vbg}, we have developed a MC simulation generator
to be used by experimenters to directly explore the effects of BSM
Wilson coefficients when large $B\to K^* \ell^- \ell^+$ data samples
are available. This tool allows for straightforward determinations of
acceptance and efficiency corrections in BSM scenarios. In particular,
using our BSM MC generator, the variations of selection efficiencies
and acceptance effects with respect to the underlying physics model
can be directly accessed.

\section{Overview}
\subsection{Current experimental anomalies in $b \to s$ neutral current (FCNC) processes}
Tensions with the SM in the lepton-flavor-universality (LFU) violating
ratios $R_K$ and $R_K^*$ were indicated by initial results from the
LHCb collaboration~\cite{LHCb:2021trn,LHCb:2017avl}, in the low
di-lepton mass-squared region. However, recent updated results
\cite{LHCb:2023qnv} no longer indicate a significant deviation for the
$R_K$ and $R_K^*$ ratios.

There remain significant deviations from SM expectations observed in
angular asymmetries of $B\to K^* \mu^+ \mu^-$~\cite{LHCb:2020lmf} and
$B_s \to \phi \mu^+\mu^-$~\cite{LHCb:2021zwz} decays, which seem to be
experimentally robust.

Several theoretical groups fit all $b\to s \ell^+ \ell^-$ observables
($R_K$ type results and angular asymmetries) for BSM Wilson
coefficients, $\delta C_i$, in $b \to s \mu^+ \mu^-$ and suggest a
coherent picture of BSM physics in $C_9$ and $C_{10}$.  BSM physics in
$\delta C_9$ and $\delta C_{10}$ is found with a range of
significances, though this varies between various global analyses,
based on different assumptions of hadronic
corrections~\cite{Altmannshofer:2021qrr,Alguero:2021anc,Hurth:2021nsi,Ciuchini:2021smi,Ciuchini:2022wbq,Alguero:2023jeh,Hurth:2023jwr}.

In order to explain these deviations, the global fits initially
assumed that the BSM contribution is restricted to the di-muon channel
and does not affect the di-electron channel.  This is plausible if the
BSM physics is lepton-flavor dependent and will be the basis for the
scenarios that are simulated in this paper.  Other scenarios with LFU
BSM couplings will be simulated in future work.

\subsection{Definition of observables and angular asymmetries}
\begin{figure}
  \centering\includegraphics[width=0.5\columnwidth, viewport=35 90 800 510, clip=true]{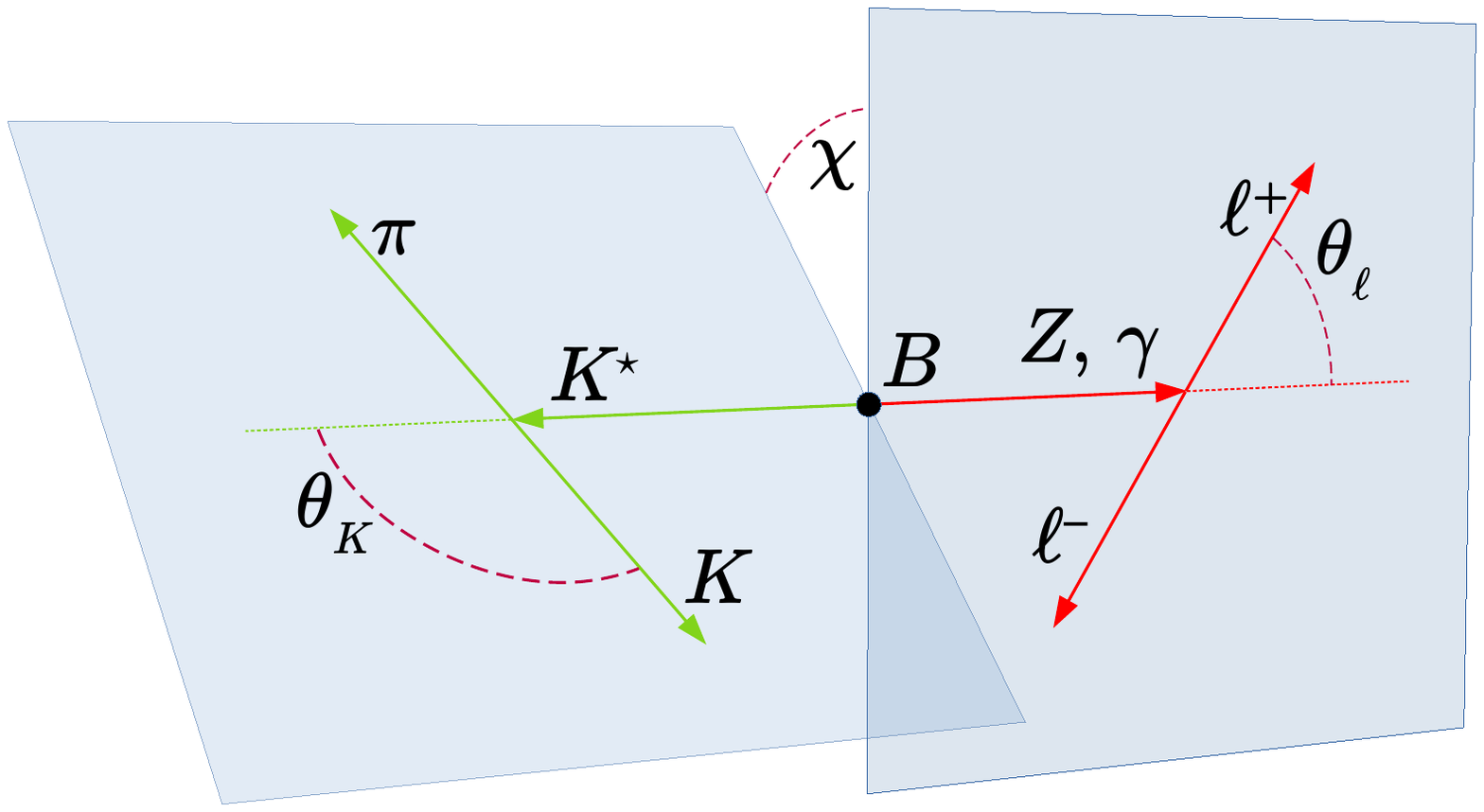}
  \caption{\label{fig:kinematics} The $B\to K^* \ell^+ \ell^-$ decay
    and the subsequent $K^*\to K\pi$ decay kinematic parameters.}
\end{figure}

For the $B\to K^* \ell^+ \ell^-$ decay with $K^*\to K\pi$ the
differential decay rate can be described in terms of the $K\pi$
invariant mass $m_{K\pi}$, the di-lepton mass squared $q^2$, and three
angles $\theta_\ell$, $\theta_K$, and $\chi$. The angle $\theta_\ell$
is defined as the angle between the direction of the positively
charged lepton and the direction of $B$ meson in the di-lepton rest
frame.  The angle $\theta_K$ is defined as the angle between the
direction of the kaon and the direction of the $B$ meson in the
$K^*$-meson rest frame. The angle $\chi$ is the angle between the
plane formed by the di-lepton pair and the plane formed by the $K^*$
decay products in the $B$-meson rest frame. A graphical representation
of the angle definitions is shown in Fig.~\ref{fig:kinematics}.

In the SM angular asymmetries such as $A_{\rm FB}$ arise due to the
interference between different decay amplitudes. In the case of BSM
physics there will be additional interference terms that are linear in
the BSM contribution, which appear in several observables: the well
known forward-backward asymmetry $A_\text{FB}(q^2)$ is defined as
\begin{align}
  A_\text{FB}(q^2) = \dfrac{\left[\left(\displaystyle\int_{0}^{1}
      - \displaystyle\int_{-1}^{0}\right)\dd{\cos{\theta_\ell}}\right]\dd{(\Gamma
      - \bar{\Gamma})}}{\displaystyle\int_{-1}^{1}\dd{\cos{\theta_\ell}}\dd{(\Gamma+\bar{\Gamma})}
  },
\end{align}
where $\Gamma$ and $\bar{\Gamma}$ denote the decay rate of $\bar{B}^0
\to \bar{K}^{0*} \ell^+ \ell^-$ and the $CP$-conjugate channel $B^0 \to
K^{0*} \ell^+ \ell^-$, respectively (see Eqs.~\ref{eq:helicity} and
\ref{eq:helicitybar} for full angular distributions).  Other important
angular asymmetries involve the angle $\chi$ between the $K^*$ and
di-lepton decay planes~\cite{Mandal:2014kma}:
\begin{align}
  S_{4}(q^2) =
  -\frac{\pi}{2}\frac{\Big[\displaystyle\int_{-\pi/2}^{\pi/2}-\displaystyle\int_{\pi/2}^{3\pi/2} \Big]\dd{\chi} \Big[\displaystyle\int_0^1
      -\displaystyle\int_{-1}^0 \Big] \dd\cos\theta_K \Big[\int_0^1
      -\int_{-1}^0 \Big] \dd\cos\theta_\ell \dd{(\Gamma+\bar{\Gamma})}}
  {\displaystyle\int_0^{2\pi}\dd{\chi}\int_{-1}^1\dd{\cos\theta_K} \int_{-1}^1\dd{\cos\theta_\ell}\, \dd{(\Gamma+\bar{\Gamma})}}\,,
\end{align}
and
\begin{align}
  S_{5}(q^2) = \frac{4}{3}\frac{\Big[\displaystyle\int_{-\pi/2}^{\pi/2}
      -\displaystyle\int_{\pi/2}^{3\pi/2} \Big]\dd{\chi} \Big[\int_0^1
      -\int_{-1}^0 \Big] \dd{\cos\theta_K} \displaystyle\int_{-1}^1 \dd{\cos\theta_\ell}~ \dd{(\Gamma-\bar{\Gamma})}}
  {\displaystyle\int_0^{2\pi}\dd{\chi}\int_{-1}^1\dd{\cos\theta_K} \int_{-1}^1\dd{\cos\theta_\ell}\, \dd{(\Gamma+\bar{\Gamma})}}.
\end{align}
Note that the angular observable $P_5^\prime$, widely used in the
literature, is related to $S_5$ via $P_5^\prime \equiv S_5/\sqrt{F_L
  (1-F_L)}$, where $F_L$ is the longitudinal polarization fraction of
the $K^*$ meson.
\subsection{SM and BSM Lorentz structures}
 
The  matrix element for the decay $B\to K^*
\ell^+\ell^-$, where $\ell$ is $e$, $\mu$ or the $\tau$ lepton, is the
following:
\begin{equation}\label{eq:matrixelement}
\begin{split}
    {\mathcal M}\ = \frac{G_F\alpha}{\sqrt{2}\pi}V_{tb}^{}V_{ts}^*\bigg\{
    \bigg[ \langle K^* | {\bar{s}\gamma^{\mu}({C_9^\text{eff}P_L+
	  C_9^{\prime} P_R})b} | {\bar B} \rangle\\
      -\frac{2m_b}{q^2}
      \langle{K^*} | {\bar{s}i\sigma^{\mu\nu}q_{\nu}(C_7^\text{eff} P_R+
	C_7^{\prime}  P_L)b}| {\bar B} \rangle \bigg]
    (\bar{\ell}\gamma_{\mu}\ell)\\
    + \langle{K^*} | {\bar{s}\gamma^{\mu}({C_{10} P_L+
	C_{10}^{\prime} P_R})b} | {\bar B} \rangle (\bar{\ell}\gamma_{\mu}\gamma_5 \ell)  \bigg \},
\end{split}
\end{equation}
where the SM Wilson coefficients are known at NNLL
accuracy~\cite{Altmannshofer:2008dz}:
\begin{eqnarray}
  \label{eq:WC}
  C_7^{\rm eff} & = & -0.304,
  \nonumber\\
  C_9^{\rm eff} & = & C_9 + Y(q^2)= 4.211 + Y(q^2)\,,\\
  C_{10} &=& -4.103\,, \nonumber
\end{eqnarray}
and where the function $Y(q^2)$ is described in
Section~\ref{App:theory}. Note that we include the lepton masses ($m_\mu$ and $m_e$) in all results. We assume BSM states are heavy, far above the
 $\mu = m_b = 4.8$ GeV/$c^2$ scale. We therefore treat
BSM contributions to the Wilson coefficients as $q^2$-independent
offsets.

\begin{figure}[tbh]
  \includegraphics[width=0.495\columnwidth]{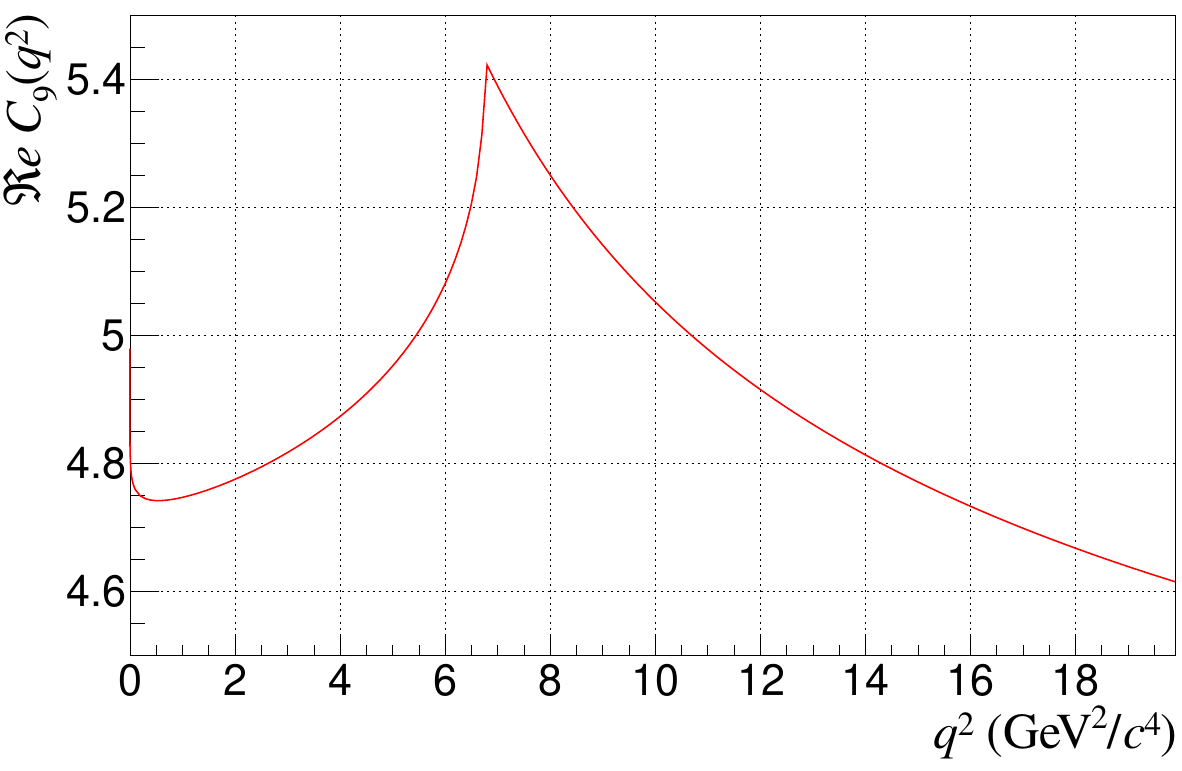}
  \includegraphics[width=0.495\columnwidth]{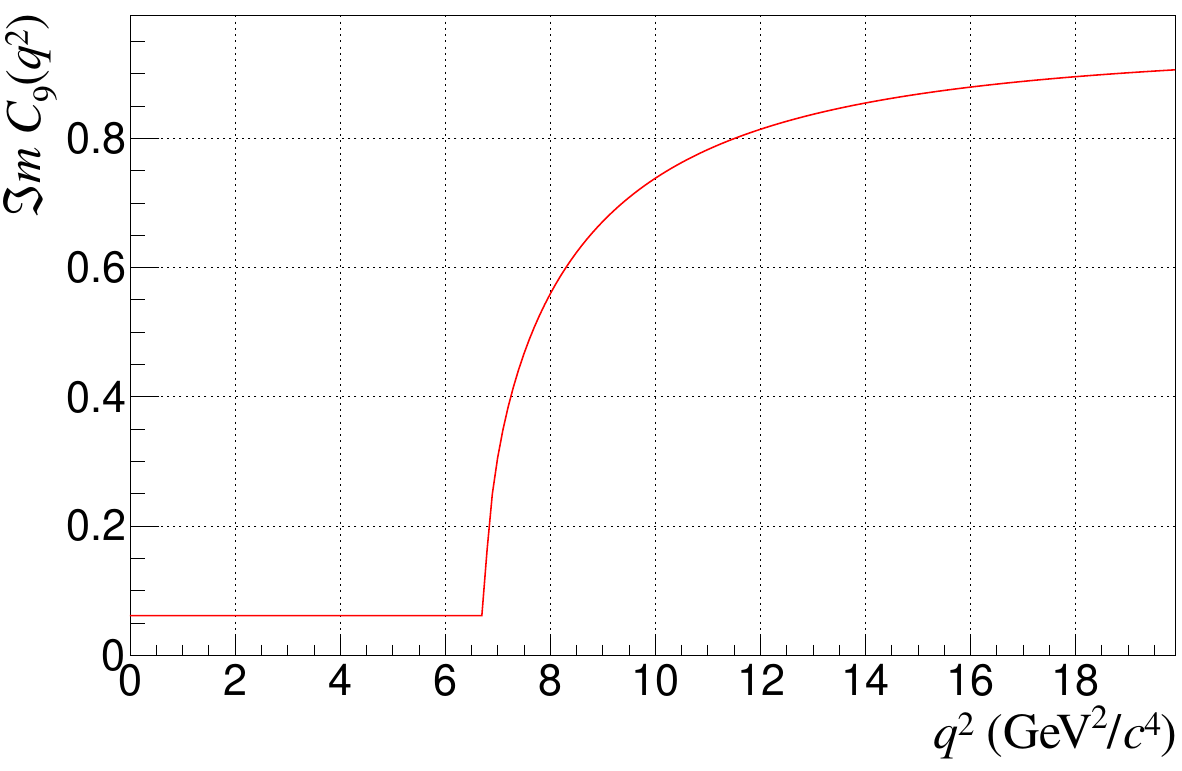}
  \caption{\label{fig:C9} The SM Wilson coefficient $C_9$ versus $q^2$
    now used in \evtgen (see Eq.~\ref{eq:WC}) without
    $c\bar{c}$ resonances.  }
\end{figure}

The chirality-flipped operator
$C_{7}^\prime$ is highly suppressed while the $C_{9}^\prime$
$C_{10}^\prime$ operators
are absent in the SM but can arise in the presence of BSM
physics. Note that the SM Wilson coefficient $C_9^{\rm eff}$ varies
with $q^2$; as shown in Fig.~\ref{fig:C9} separately for real and
imaginary parts. 
The $B\to K^*$ matrix elements in Eq.~\ref{eq:matrixelement} can be
expressed in terms of seven form factors that depend on the momentum
transfer $q^2$ between the $B$ and the $K^*$ ($q^\mu = p^\mu -
k^\mu$):
\begin{equation}
  \begin{split}
    \langle \bar K^*(k) | \bar s\gamma_\mu(1 \mp \gamma_5) b | \bar B(p)\rangle  =  
    \mp i \epsilon^*_\mu (m_B+m_{K^*})
    A_1(q^2)\\
    \pm i (2p-q)_\mu (\epsilon^* \cdot q)\,
    \frac{A_2(q^2)}{m_B+m_{K^*}} \\
    \pm i q_\mu (\epsilon^* \cdot q) \,\frac{2m_{K^*}}{q^2}\,
    \left[A_3(q^2)-A_0(q^2)\right] \\
    + \epsilon_{\mu\nu\rho\sigma}\epsilon^{*\nu} p^\rho k^\sigma\,
    \frac{2V(q^2)}{m_B+m_{K^*}},\label{eq:SLFF}
  \end{split}
\end{equation}
\begin{equation}
  {\rm with\ }A_3(q^2)  =  \frac{m_B+m_{K^*}}{2m_{K^*}}\, A_1(q^2) -
  \frac{m_B-m_{K^*}}{2m_{K^*}}\, A_2(q^2)\mbox{~~and~~} 
  A_0(0) =  A_3(0);\label{eq:A30}
\end{equation}
\begin{eqnarray}
  \langle \bar K^*(k) | \bar s \sigma_{\mu\nu} q^\nu (1\pm \gamma_5) b | \bar B(p)\rangle &= & i\epsilon_{\mu\nu\rho\sigma} \epsilon^{*\nu}
    p^\rho k^\sigma \, 2 T_1(q^2) \nonumber\\
  & & \pm T_2(q^2) \left[ \epsilon^*_\mu
    (m_B^2-m_{K^*}^2) - (\epsilon^* \cdot q) \,(2p-q)_\mu \right]\nonumber\\
  & & \pm T_3(q^2)(\epsilon^* \cdot q) \left[ q_\mu \!- \displaystyle\frac{q^2}{m_B^2-m_{K^*}^2}\, (2p-q)_\mu \right], \label{eq:pengFF}
\end{eqnarray}
with $T_1(0) = T_2(0)$.  Here $\epsilon_\mu$ is the polarization
vector of the $K^*$. We used the following convention for the
Levi-Civita tensor $\epsilon_{0123}=+1$.

We adopt the formalism developed in Ref.~\cite{Bharucha:2015bzk} where
the extrapolation of form factors from the calculated LCSR input
points ($q^2 \lesssim 0\, $GeV) to larger $q^2$ values is performed by
a simple pole form with a $z$-expansion
\begin{equation}
  \label{eq:fitform}
  F_i^{B\to K^*}(q^2) \equiv \frac{1}{1 - q^2 / m_{R,i}^2}\, \sum_{n=0}^{2} \alpha_n^{i} \left[z(q^2) - z(0)\right]^n\,,
\end{equation}
where $z(t) \equiv \displaystyle\frac{\sqrt{t_+ - t} - \sqrt{t_+ -
    t_0}}{\sqrt{t_+ - t} + \sqrt{t_+ - t_0}}$, ~$t_\pm = (m_B \pm
m_{K^*})^2$ and $t_0 \equiv t_+ \left(1 - \sqrt{1 - t_- /
  t_+}\right)$.

The values for the fit parameters $\alpha_{0,1,2}^{i}$ including the
correlation among them, and the masses of resonances $m_{R,i}$
associated with the quantum numbers of the respective form factor
$F_i$ are obtained in Ref.~\cite{Bharucha:2015bzk}.

\subsection{Form factor uncertainties}
In the old \evtgen model BTOSLLBALL, used by Belle and Belle
II up to now, the form factors were taken from
Ref.~\cite{Ball:2004rg}. In our new \evtgen model BTOSLLNP
(see Section~\ref{sec:mods}), we have now implemented the most recent
hadronic form factors from Ref.~\cite{Bharucha:2015bzk}, also known as
the ABSZ form factor parameterization. ABSZ updates the previous
computations by including current hadronic inputs and several
theoretical improvements such as inclusion of higher twist
distribution amplitudes of mesons, as well as a combined fit to
Light-Cone Sum Rule form factors for the low $q^2$ region and Lattice
QCD estimates valid in the high $q^2$ range. The form factor fit
results are expressed as the coefficients of the expansion shown in
Eq.~\ref{eq:fitform}.

\begin{figure*}[t]
  \includegraphics[width=0.495\columnwidth]{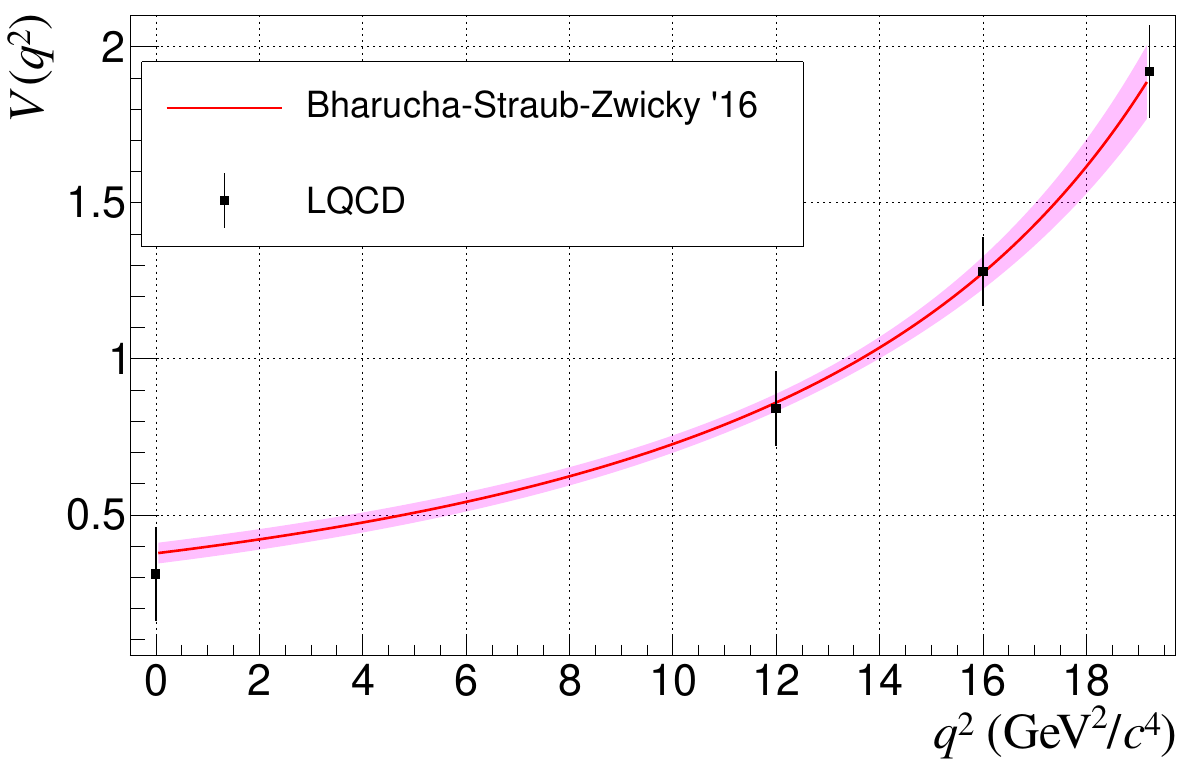}
  \includegraphics[width=0.495\columnwidth]{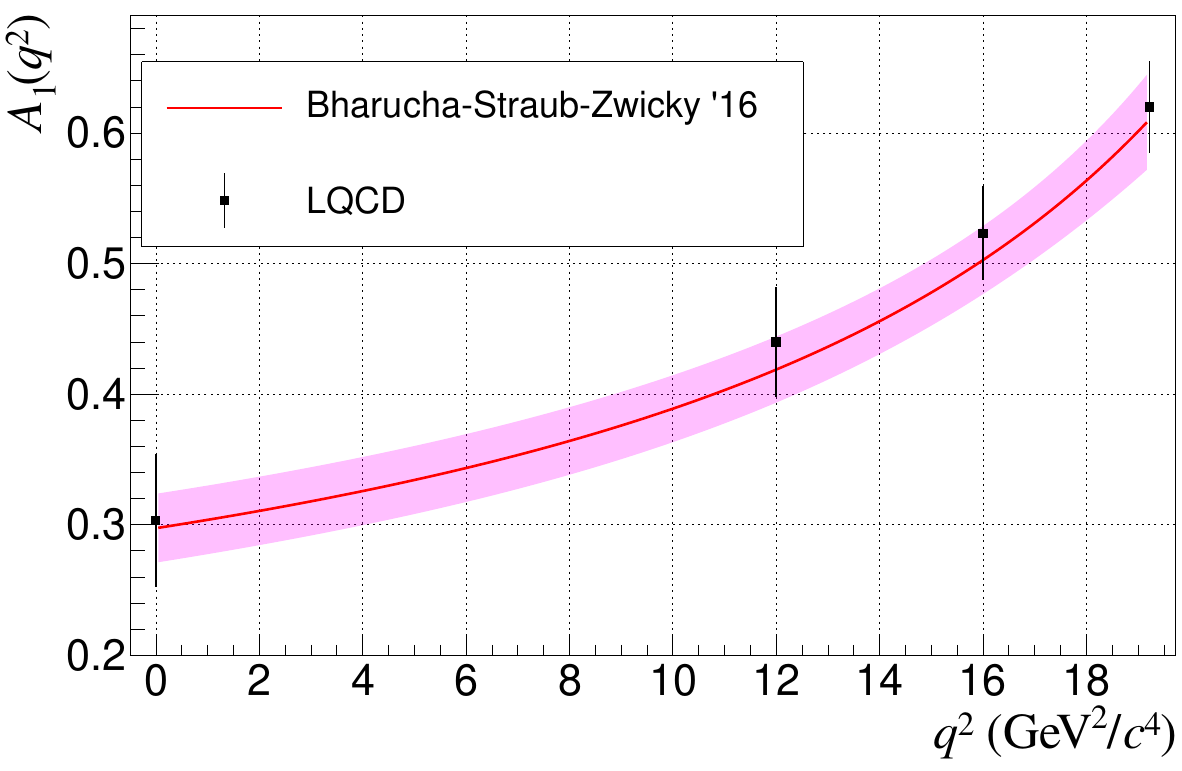}\\
  \includegraphics[width=0.495\columnwidth]{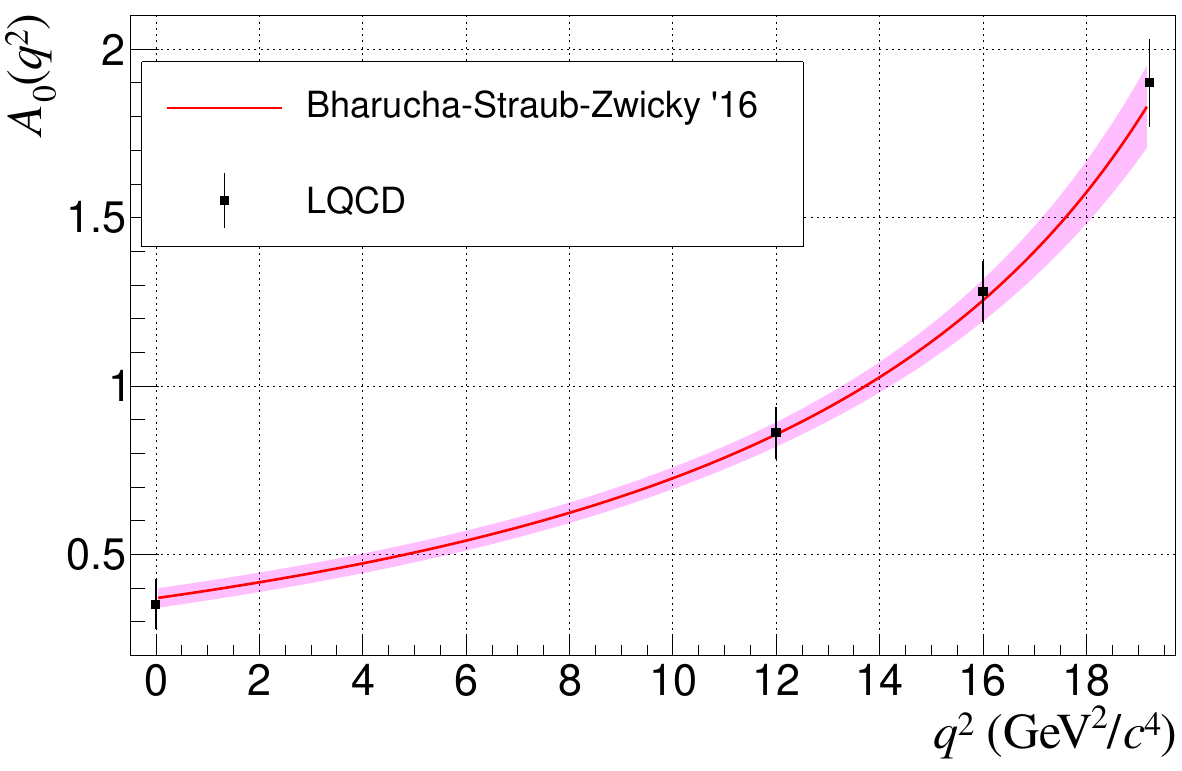}
  \includegraphics[width=0.495\columnwidth]{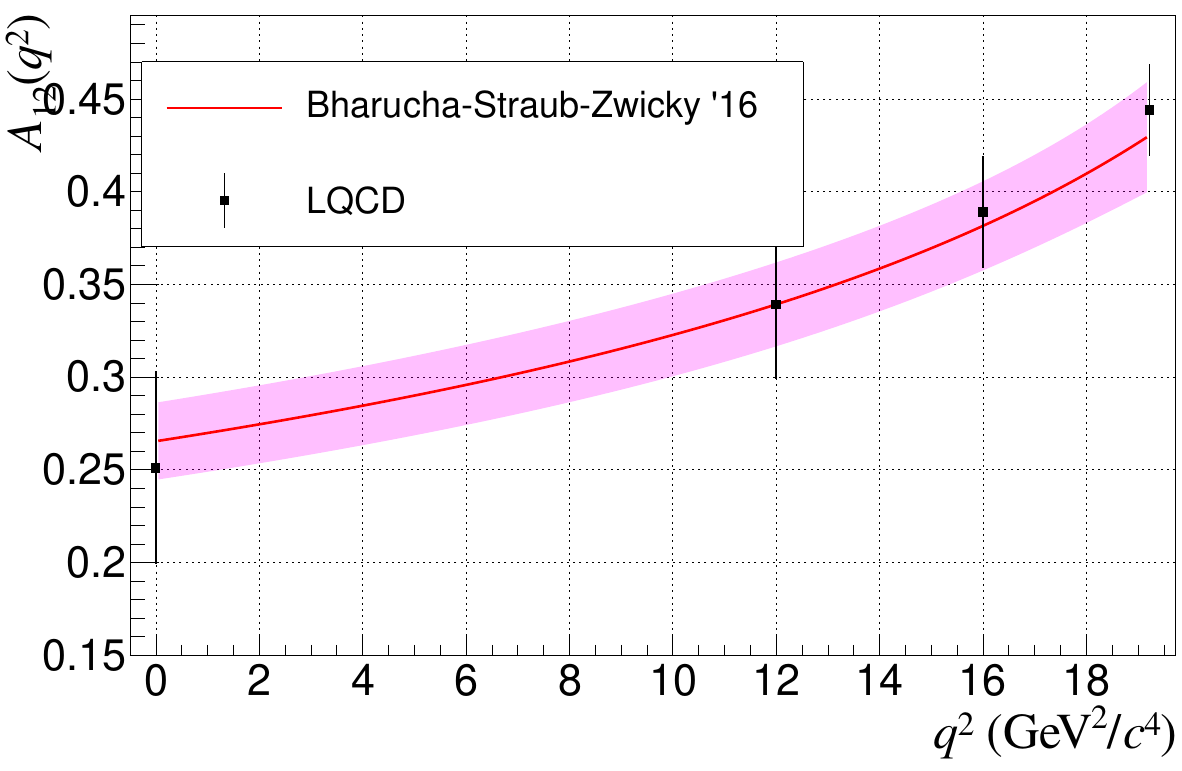}\\
  \includegraphics[width=0.495\columnwidth]{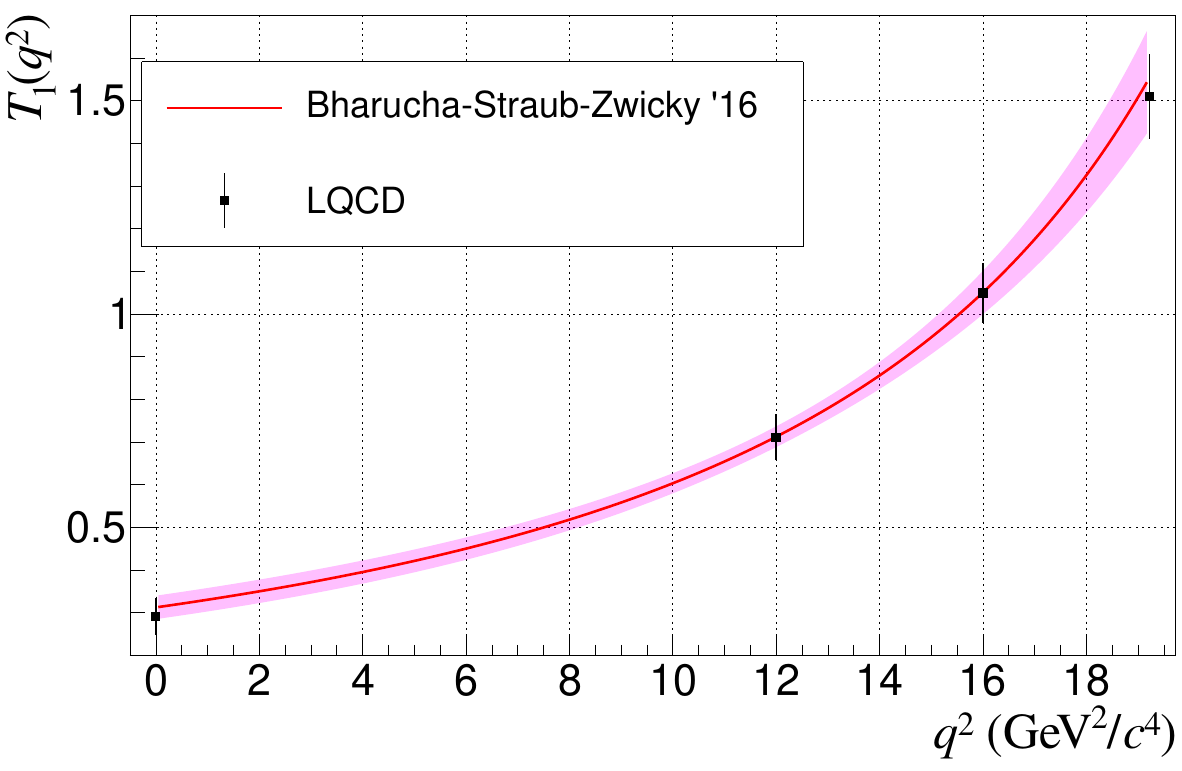}
  \includegraphics[width=0.495\columnwidth]{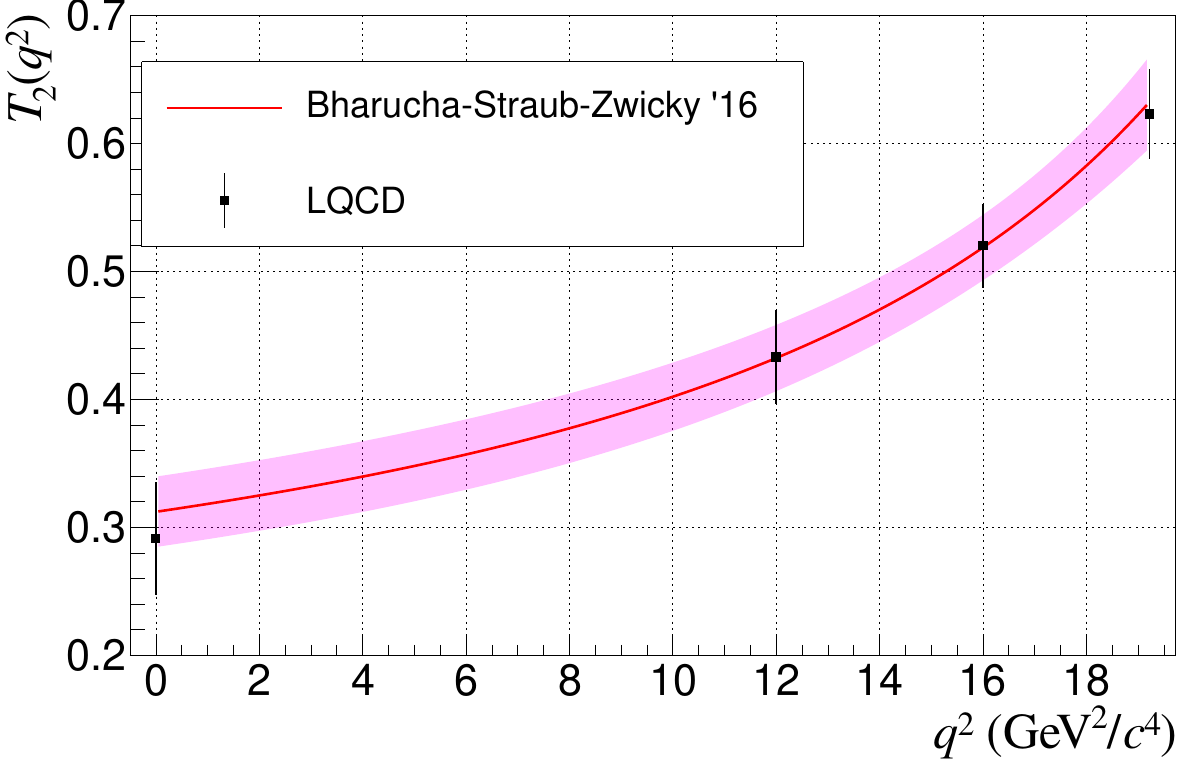}\\
  \centering\includegraphics[width=0.495\columnwidth]{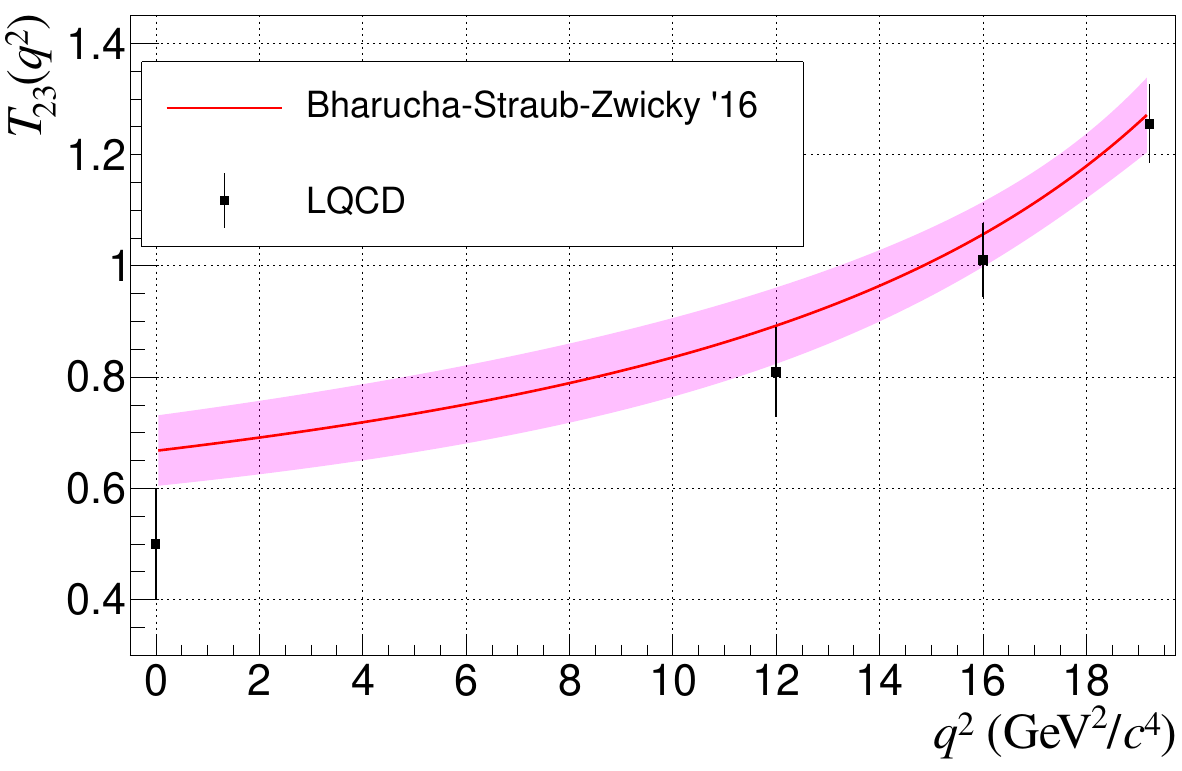}
  \caption{\label{fig:hffs}Result of the combined fit to LCSR and LQCD
    calculations from Ref.~\cite{Bharucha:2015bzk} for $B\to K^*$
    hadronic form factors. The shaded bands show the combined
    one-standard-deviation uncertainty of the fit according the
    provided covariance matrix. The points with error bars show LQCD
    results from Table 11 in Ref.~\cite{Horgan:2015vla}. These
    theoretical uncertainties in the form factors can limit extraction
    of new physics.}
\end{figure*}
\begin{figure*}
  \includegraphics[width=0.495\columnwidth]{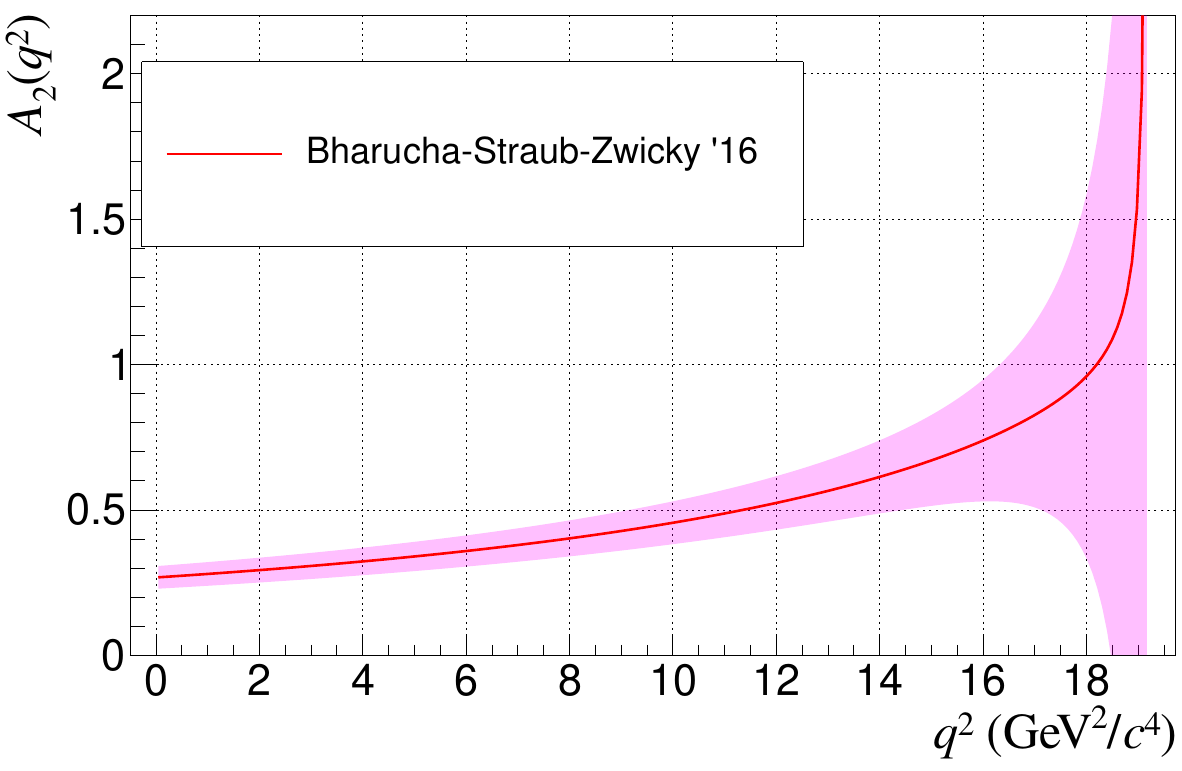}
  \includegraphics[width=0.495\columnwidth]{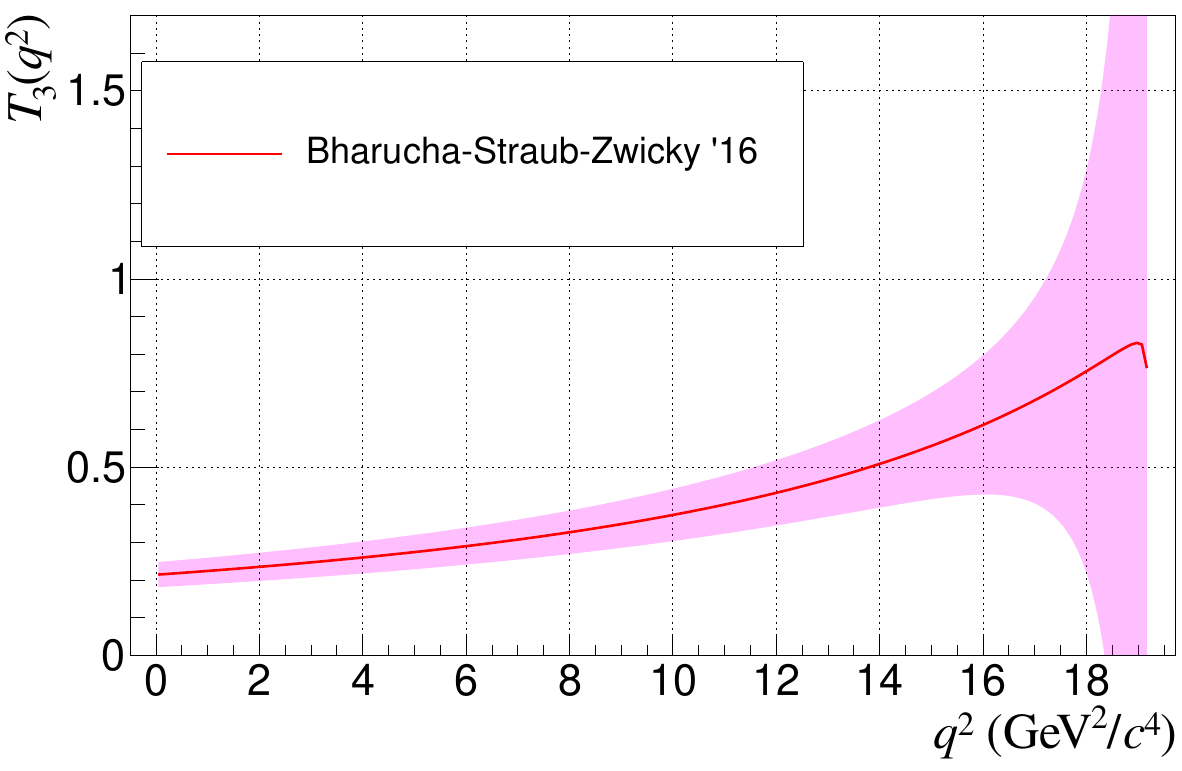}
  \caption{\label{fig:hffs2}The $B\to K^*$ hadronic form factors $A_2$
    and $T_3$ obtained using Eq.~\ref{eq:a12} and \ref{eq:t23}. The
    shaded bands show the propagated one-standard-deviation
    uncertainty. These theoretical uncertainties in the form factors
    can limit extraction of new physics.}
\end{figure*}
There are seven form factors, which are functions of $q^2$: $V(q^2)$,
$A_0(q^2)$, $A_1(q^2)$, $A_2(q^2)$, $T_1(q^2)$, $T_2(q^2)$, and
$T_3(q^2)$.  The form factors $A_{12}$ and $T_{23}$ are defined in Ref.~\cite{Bharucha:2015bzk}, where the
form factors $A_2$ and $T_3$ are extracted using the following
expression:
\begin{equation}
  A_{12}(q^2) = \frac{(m_B + m_{K\pi})^2(m_B^2 - m_{K\pi}^2-q^2)A_{1}(q^2) - \lambda(q^2)A_2(q^2)}{16 m_B m_{K\pi}^2(m_B + m_{K\pi})},
  \label{eq:a12}
\end{equation}
and
\begin{equation}
  T_{23}(q^2) = \frac{(m_B^2 - m_{K\pi}^2)^2(m_B^2 + 3m_{K\pi}^2-q^2)T_{2}(q^2) - \lambda(q^2)T_3(q^2)}{8 m_B m_{K\pi}^2(m_B - m_{K\pi})}.
  \label{eq:t23}
\end{equation}
Here $\lambda(q^2)$ is the Källén-function defined in
Eq.~\ref{eq:kallen} and $m_{K\pi}$ is the invariant mass of the kaon
and pion to take into account the finite width of $K^*$. If the $K^*$
width is omitted, a singularity appears in the physical region.

The resulting form factors we now use in the updated \evtgen
generator are shown in Figs.~\ref{fig:hffs} and ~\ref{fig:hffs2}. Note
that the theoretical uncertainties in the form factors shown in the
figure limit extraction of new physics. Below, we discuss strategies
for overcoming this limitation.
\clearpage
\subsection{BSM Modifications to the \evtgen Monte Carlo Generator\label{sec:mods}}
We implement the $B\to K^*\ell^+\ell^-$ generator with BSM physics
contributions in the \evtgen Monte-Carlo simulation framework. We use
the BTOSLLBALL model as a reference to create the BTOSLLNP model,
which incorporates the desired matrix element with BSM contributions
included. Here we use the generator, BTOSLLNP, only in the standalone
mode but it has been also successfully integrated into the Belle II
Analysis Software Framework (BASF2~\cite{Kuhr:2018lps}). In the
BTOSLLNP model, the user can specify the following BSM input
parameters: $\delta C_7$, $C_7'$, $\delta C_9$, $C_9'$, $\delta
C_{10}$, $C_{10}'$, $C_S - C_S'$, and $C_P - C_P'$. In the last two
cases only such combinations are relevant. Each of these model
parameters is complex-valued. The parameters can be specified in any
order and in any combination in the user input file. The default value
for each parameter is zero and if no parameters are specified in the
file the generator gives the SM result. 

To generate BSM physics the user inputs several arguments in the user
input file. The first argument specifies
the coordinate system---Cartesian(0) or polar(1)---to be used for the
complex-valued input arguments. The user can then enter each desired
BSM parameter as three consecutive numbers. These triplets can be
entered in any order.

Below we present several examples of
user input files to illustrate the usage of the BSM $B\to K^*
\ell^+\ell^-$ MC generator:
\begin{verbatim}
## the first argument is the Cartesian(0) or polar(1) representation of
## complex BSM coefficients, which are three consecutive numbers
##   {id, Re(C), Im(C)} or {coeff id, |C|, Arg(C)}

## id==0 delta C_7  -- BSM addition to NNLO SM value
## id==1 delta C_9  -- BSM addition to NNLO SM value
## id==2 delta C_10 -- BSM addition to NNLO SM value
## id==3 C'_7  -- BSM right-handed coefficient
## id==4 C'_9  -- BSM right-handed coefficient
## id==5 C'_10 -- BSM right-handed coefficient
## id==6 (C_S - C'_S) -- BSM scalar left- and right-handed coefficient
## id==7 (C_P - C'_P) -- BSM pseudo-scalar left- and right-handed
##                       coefficient


Decay anti-B0
## The SM Case 
1.000 anti-K*0 e+ e- BTOSLLNP;
Enddecay

Decay anti-B0
## delta C_9eff = (-0.87, 0.0) all other coefficients correspond to the 
## SM values
1.000 anti-K*0 e+ e- BTOSLLNP 0 1 -0.87 0.0 ;
Enddecay

Decay anti-B0
## delta C_9eff = (-0.87, 0.0)  and delta C_10eff = (+0.87, 0.0)
## all other coefficients correspond to the 
## SM values
1.000 anti-K*0 e+ e- BTOSLLNP 0 1 -0.87 0.0 2 0.87 0.0 ;
Enddecay

Decay anti-B0
## delta C_7prime eff = ( 0.04, 0.0) all other coefficients correspond to the 
## SM values
1.000 anti-K*0 e+ e- BTOSLLNP 0 3 0.04 0.0 ;
Enddecay


\end{verbatim}


\subsection{Signal generator performance improvements\label{sec:sig_gen_dev}}
\begin{figure}[tbh]
  \includegraphics[width=\columnwidth, viewport=0 4 565 115, clip=true]{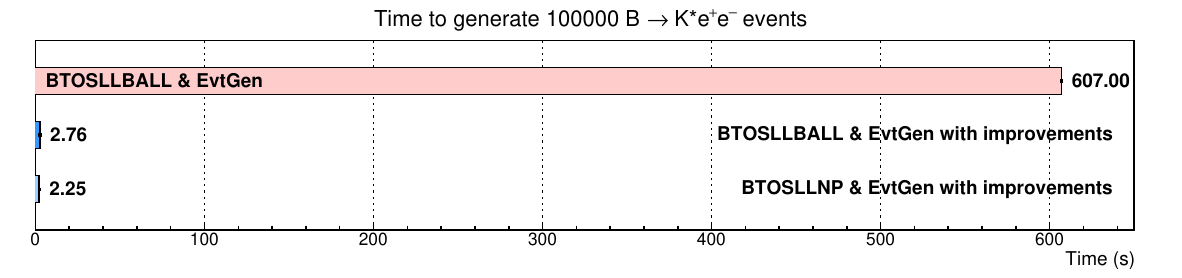}
  \caption{\label{fig:perf} The measured time to generate 100000 $B\to
    K^* e^+ e^-$ events with (bottom bars) and without (top bar)
    \evtgen modifications.}
\end{figure}

In addition to implementing BSM amplitudes and the ABSZ form factors
in \evtgen, we also significantly improved the performance of
\evtgen.  The details of these improvements are briefly
described below:
\begin{itemize}
\item The phase space generation procedure has been reviewed and
  significantly optimized.
\item Tensor contraction operations have been reviewed and optimized.     
\item The importance sampling for three body decays with a pole has
  been reviewed and an incorrect treatment of the pole has been
  identified and fixed.
\item The maximal amplitude search has been reviewed and several
  mistakes have been fixed.
\item A special importance sampling procedure has been developed
  to treat narrow resonances for optimal performance.
\end{itemize}
Combining together all the \evtgen modifications more than two
orders of magnitude performance improvement has been attained as shown
in Fig.~\ref{fig:perf}.

\section{\label{sec:bsmsignals} Signatures of BSM Physics in $B \to K^* \ell^+ \ell^- $ and improved sensitivity via multi-dimensional likelihood fits }

\begin{figure}[b!]
  \includegraphics[width=0.495\columnwidth]{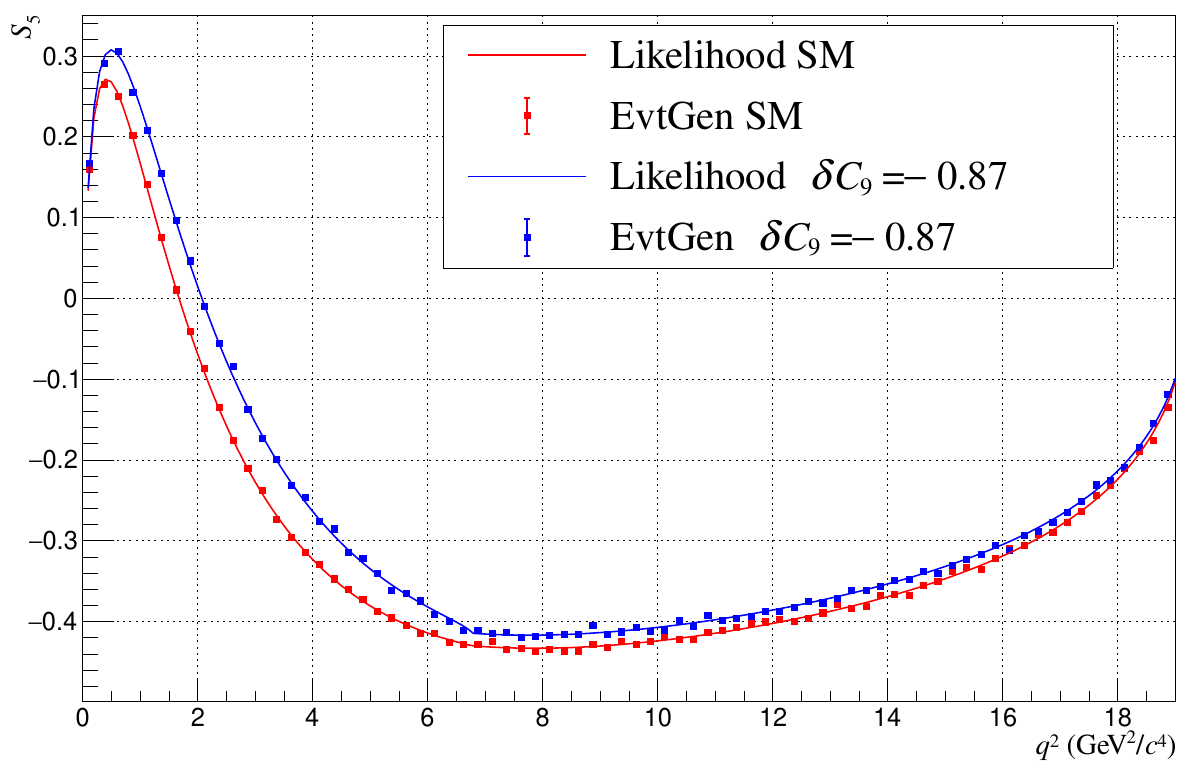}\hfill
  \includegraphics[width=0.495\columnwidth]{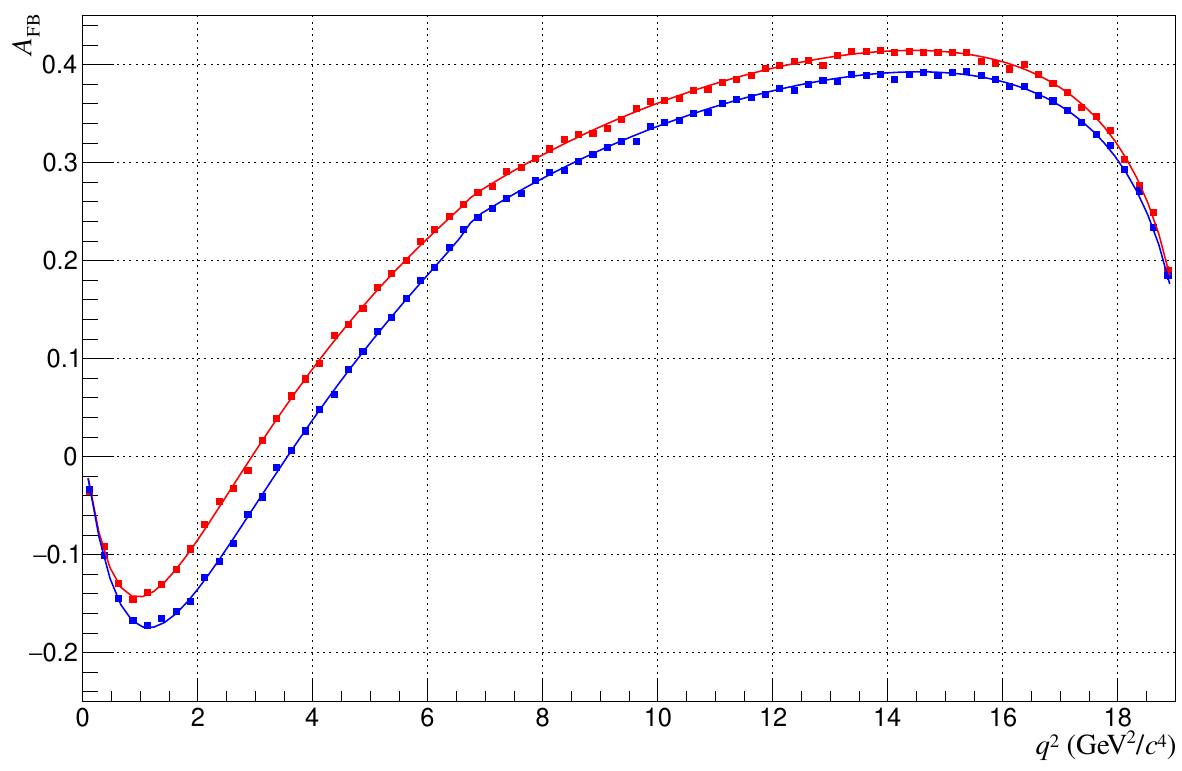}
  \caption{\label{fig:c9_obs} Comparison of $S_5$ (left plot) and
    $A_\text{FB}$ (right plot) observables with BSM $\delta C_9=-0.87$
    and SM $\delta C_9=0$ in the di-muon mode. The points are
    generated with the BSM \evtgen simulation while the curves
    are the results of integrating the four-dimensional likelihood
    function. Note that no $c \bar{c}$ resonance effects are included.}
\end{figure}

We show the expected statistical sensitivity of the unbinned likelihood fit under
ideal conditions using only the primary generator output without
taking into account detector effects or hadronic uncertainties as well
as resonance effects. Hadronic uncertainties are considered in
Section~\ref{sec:qcd}.

A likelihood function has been implemented according to
Eq.~\ref{eq:helicity}, where the matrix element defined in
Eq.~\ref{eq:matrixelement} is already squared, and which explicitly
depends only on the four variables $q^2$, $\cos\thl$, $\cos\thK$,
$\chi$, and on the Wilson coefficients as fit parameters. The
likelihood function has been cross-checked against \evtgen-generated
distribution and asymmetries. The quark masses $m_b = 4.8$ GeV/$c^2$,
$m_c = 1.3$ GeV/$c^2$ were used. Good agreement between the \evtgen
predictions and integrals of the likelihood function can be seen in
Fig.~\ref{fig:c9_obs}\footnote{Since this is a simplified sensitivity
study we do not take into account resonances and thus results in
resonance region are not physical.}.

The expected experimental data sample depends on integrated luminosity
and on the selection criteria required to sufficiently suppress
backgrounds. In Belle~II, we currently expect about 25\%
selection efficiency for the $B^0\to K^{*0}(K^+\pi^-) \ell^+\ell^-$
mode and 40 to 50 ab$^{-1}$ total integrated luminosity. Here we study
only the statistical sensitivity of the fit so all internal parameters
in the \evtgen decay generator and the likelihood function are
the same. In addition, for simplicity we do not include resonance
effects in the generator and the likelihood function and use the
entire $q^2$ range without veto windows. The total decay rate is not
fixed in the test.  Depending on assumptions we expect from 5000 to
10000 events in the di-muon mode and about 25\% more events at very
low $q^2$ for the di-electron mode. In the sensitivity tests below,
the number of events lies within this range.

In each fit we extract the deviation of a single Wilson coefficient
from the SM value:
$$ \delta C_i = C_i^\text{BSM} - C_i^\text{SM}.$$ The SM values are
taken from NNLO calculations~\cite{Altmannshofer:2008dz}.  Since the
Wilson coefficients are complex numbers, both the real and imaginary
parts are extracted in these fits, and shown in the figures that
follow.

\subsection{$\delta C_9$ correlated signatures in $A_{\rm FB}$ and $S_5$ (and $\dd{\mathcal{B}(B\to
    K^*\ell^+\ell^-)}/\dd{q^2}$)}

\begin{figure*}[h!]
  \includegraphics[width=0.495\columnwidth]{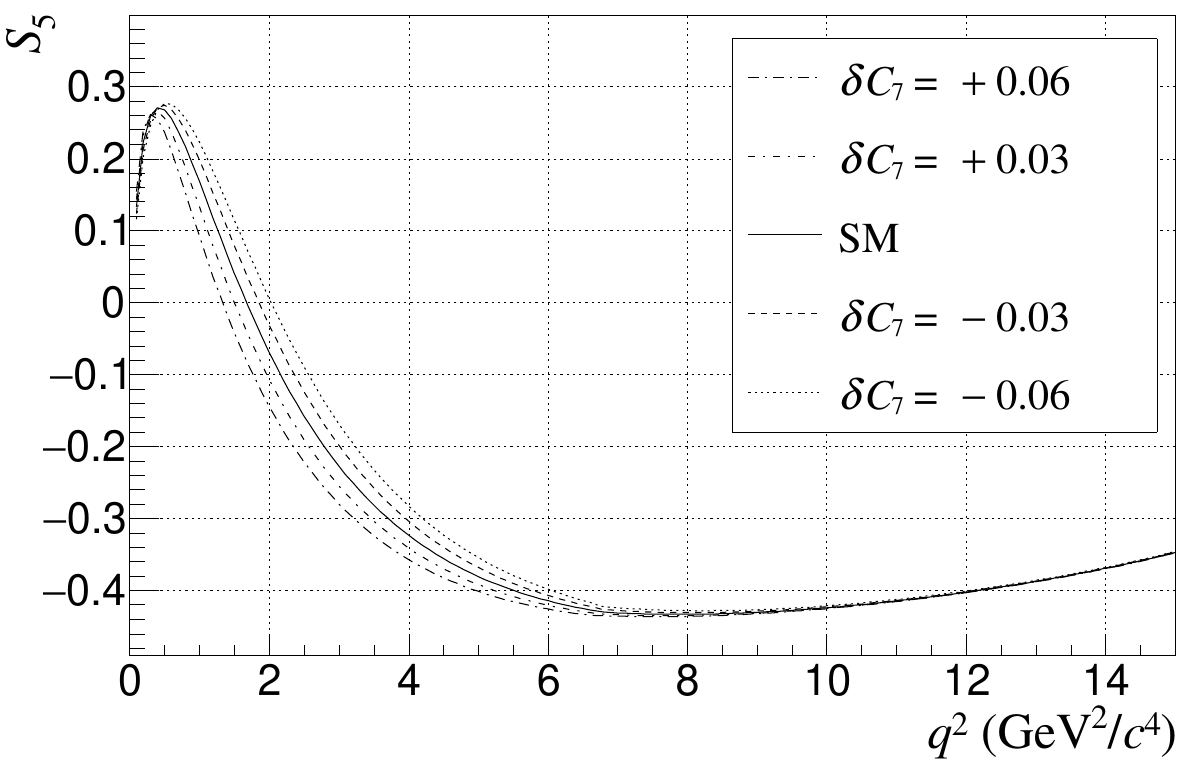}\hfill
  \includegraphics[width=0.495\columnwidth]{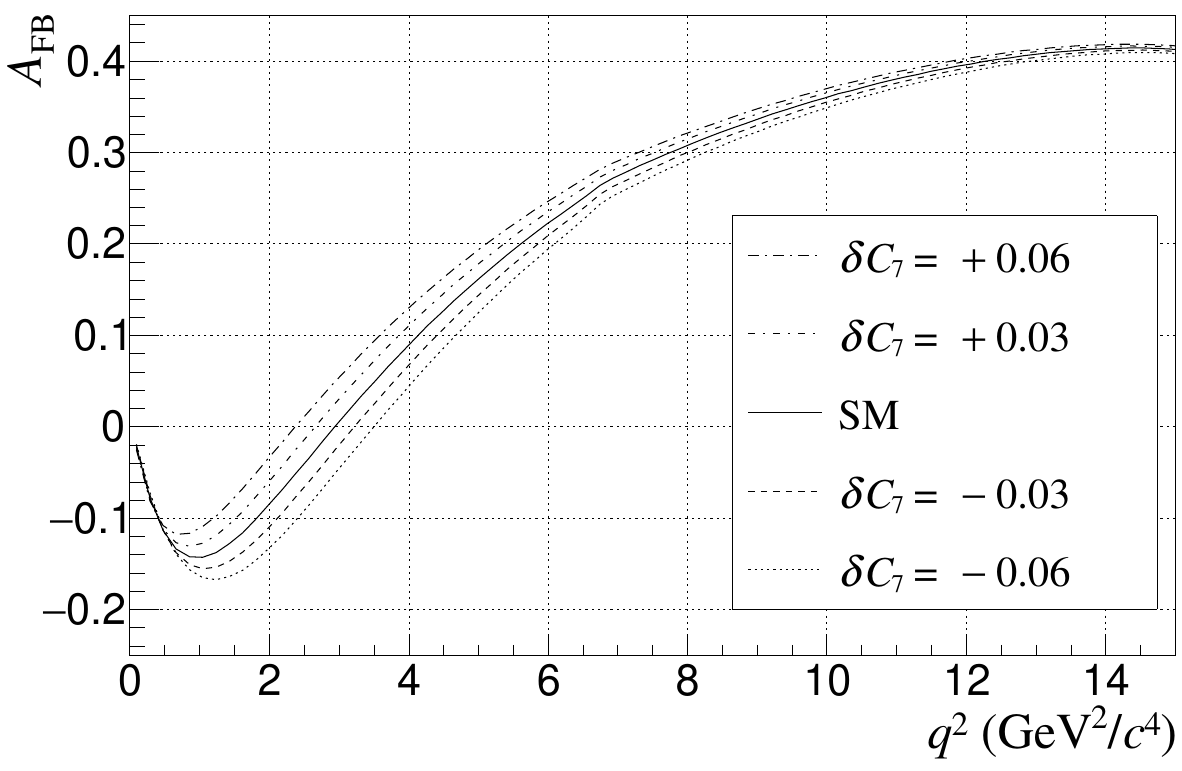}
  \caption{\label{fig:c7_sev} Comparison of $S_5$ (left plot) and
    $A_\text{FB}$ (right plot) observables with several values of
    $\delta C_7$ in the di-muon mode obtained by integrating the
    four-dimensional likelihood function. Note that no $c \bar{c}$ resonance effects are included.}
\end{figure*}

\begin{figure*}[h!]
  \includegraphics[width=0.495\columnwidth]{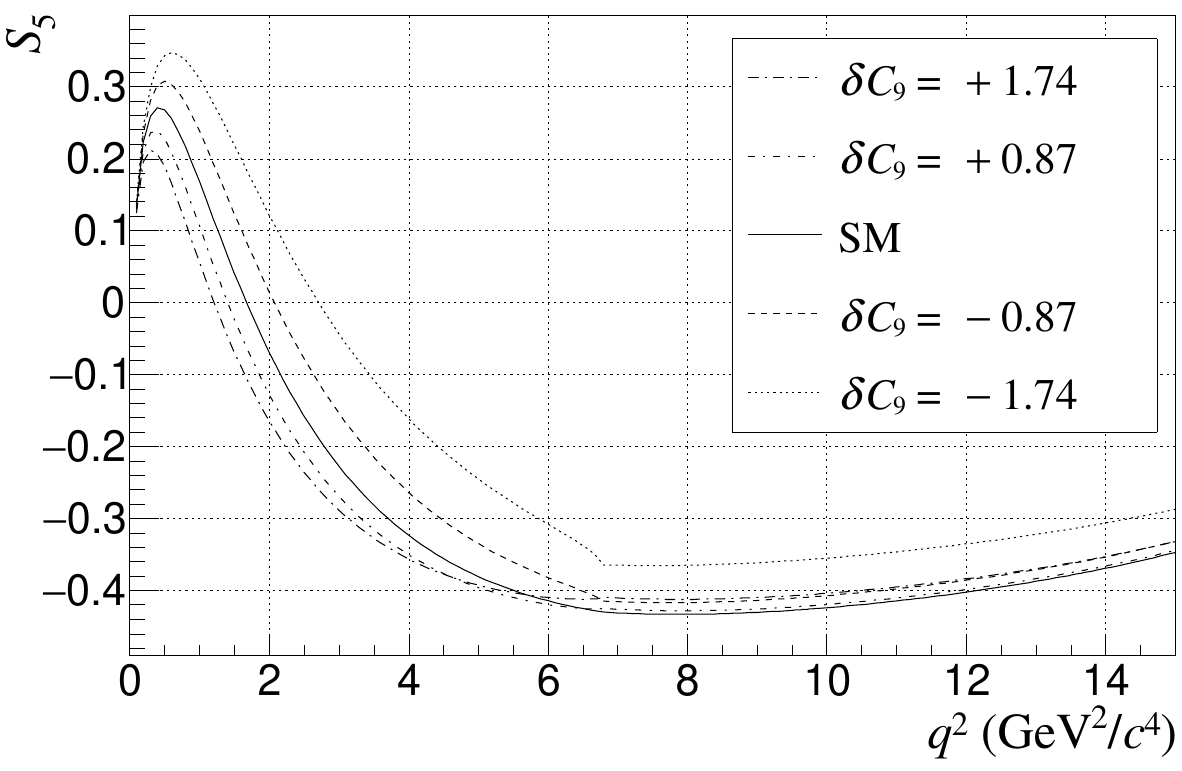}\hfill
  \includegraphics[width=0.495\columnwidth]{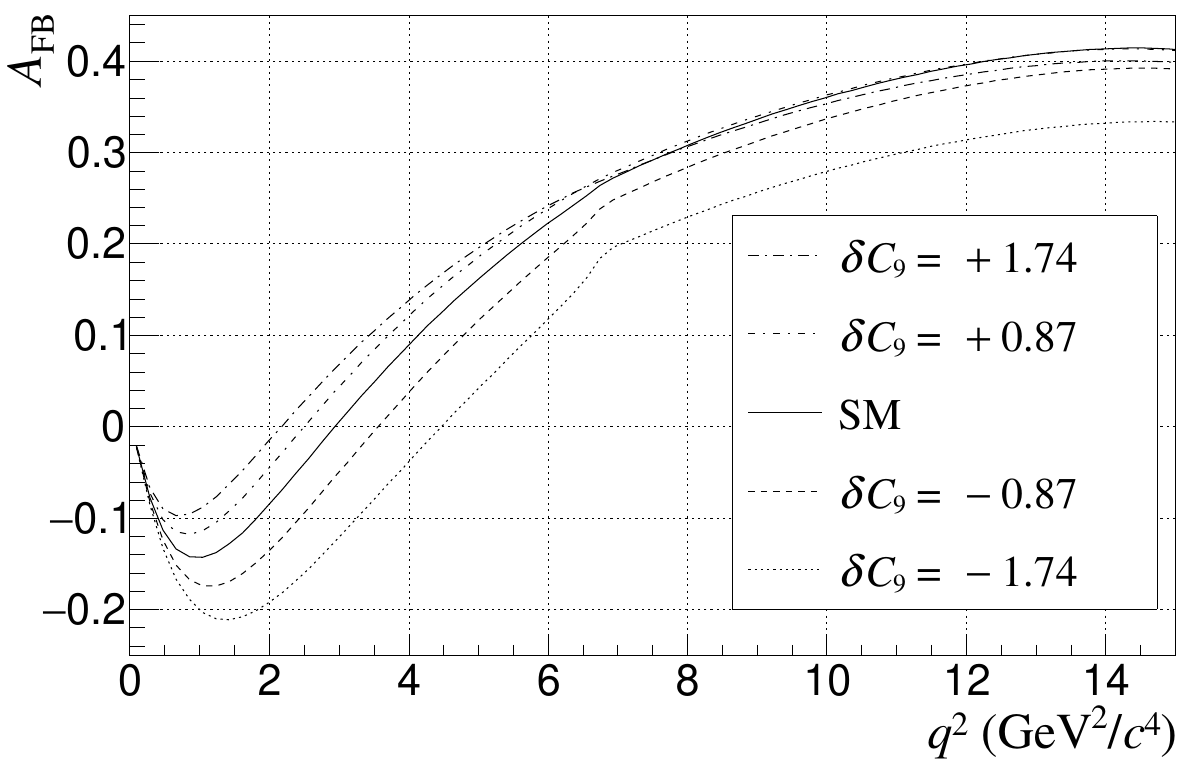}
  \caption{\label{fig:c9_sev} Comparison of $S_5$ (left plot) and
    $A_\text{FB}$ (right plot) observables with several values of
    $\delta C_9$ in the di-muon mode obtained by integrating the
    four-dimensional likelihood function. Note that no $c \bar{c}$
    resonance effects are included. The expectation that $\delta C_9$
    dependence at high $q^2$ should cancel out arises as a consequence
    of Eq.~(3.1). The dependence at high $q^2$ only occurs for a very
    large value of $\delta C_9=-1.74$. }
\end{figure*}

\begin{figure*}[h!]
  \includegraphics[width=0.495\columnwidth]{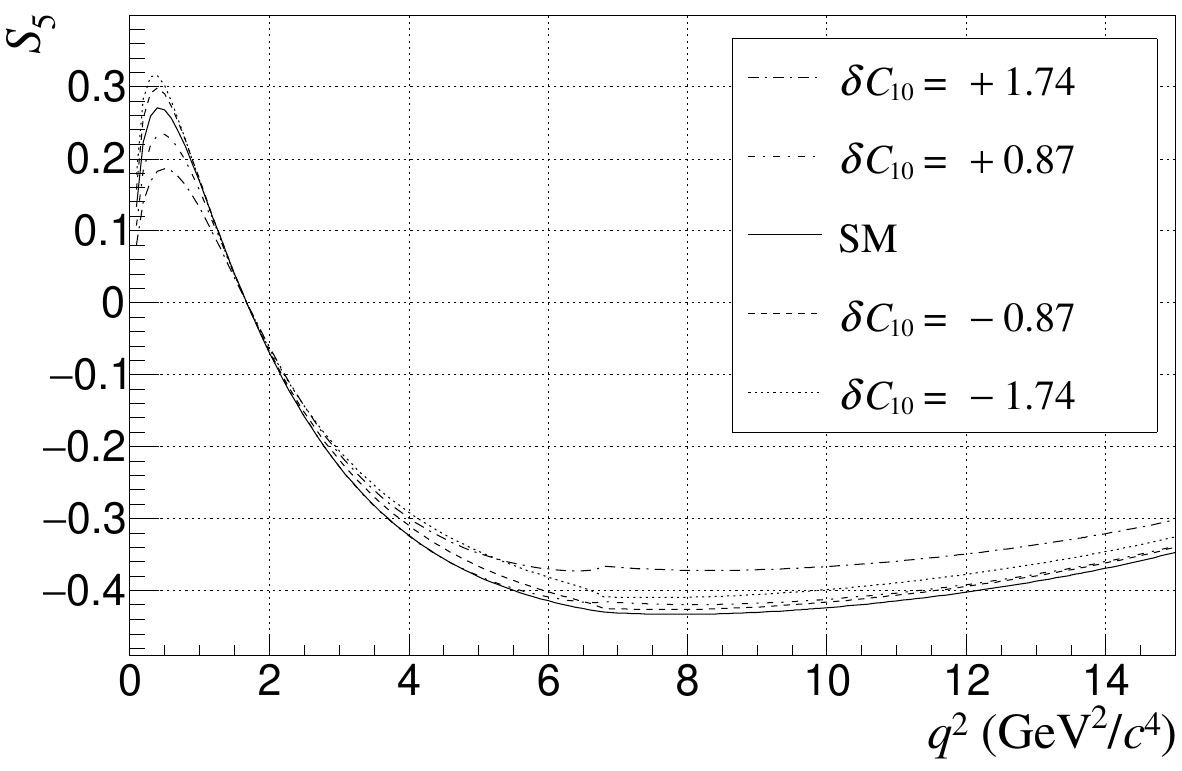}\hfill
  \includegraphics[width=0.495\columnwidth]{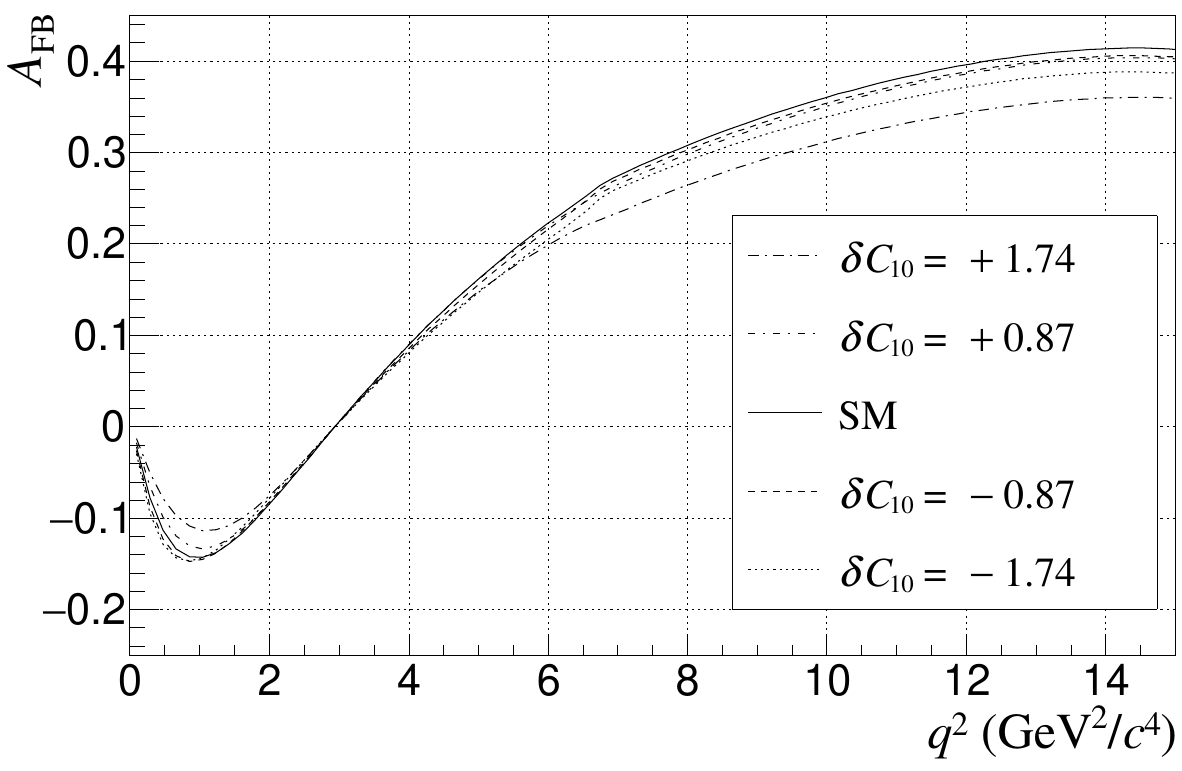}
  \caption{\label{fig:c10_sev} Comparison of $S_5$ (left plot) and
    $A_\text{FB}$ (right plot) observables with several values of
    $\delta C_{10}$ in the di-muon mode obtained by integrating the
    four-dimensional likelihood function. Note that no $c \bar{c}$ resonance effects are included.}
\end{figure*}

Figure \ref{fig:c9_obs} shows how the observables $S_5$ and
$A_\text{FB}$ are modified when a BSM contribution ($\delta C_9$) to
$C_9$ is present. We assume that $\delta C_9=-0.87$, which corresponds
to a representative value from global theory fits to $b\to s$
experimental data in the $\mu \mu$ channel
\cite{Altmannshofer:2021qrr}. There are clear differences between the
SM and BSM distributions of $S_5$ and $A_\text{FB}$, and these
signatures are correlated. However, we do not find any significant
effect of $\delta C_9$ on the differential distribution
$\dd{\mathcal{B}(B\to K^*\ell^+\ell^-)}/\dd{q^2}$.

The observables $S_5$ and $A_\text{FB}$ with the Wilson coefficients
varied in a more wide range are shown in Fig.~\ref{fig:c7_sev},
\ref{fig:c9_sev}, and \ref{fig:c10_sev}. These plots suggest that BSM
physics contributions to the different Wilson coefficients affect
different $q^2$ regions in the observables and the correlations
between them have to be selected in the proper $q^2$ regions to
maximize statistical sensitivity to BSM.

 In the absence of contributions from
  right-handed and scalar currents, 
  (i.e. assuming $C_{7}^{\prime}$, $C_{9}^{\prime}$, $C_{10}^{\prime}$ and $C_S$ are zero) 
  the correlations observed in the results from the Monte Carlo generator between $S_5$ and $A_\text{FB}$ in
Figs.~\ref{fig:c7_sev}, \ref{fig:c9_sev}, and \ref{fig:c10_sev}, can be understood from
the following analytical expression relating the two observables:
\begin{equation}
\label {eq:ratio}
\frac{S_5}{A_\text{FB} }=-\displaystyle\frac{16\sqrt{2}}{3\sqrt{q^2}}
\frac{A_{12}}{A_1}\frac{m_Bm_{K^*}}{m_B + m_{K^*}}
\displaystyle\frac{{\cal R}\text{e}(C_9)+\displaystyle\frac{m_b}{q^2}\text{Re}(C_7)
Z_1}{\text{Re}(C_9)+\displaystyle\frac{m_b}{q^2}
\text{Re}(C_7)  Z_2}
\end{equation}
where $Z_1$ and $Z_2$ are purely functions of form factors and masses and are
given by
\begin{equation}
  Z_1=\displaystyle\frac{2T_1A_{12}(m_B+m_{K^*})^2 + T_{23}Vq^2}{2VA_{12}(m_B+m_{K^*})},
\end{equation}
\begin{equation}
  Z_2=\displaystyle (m_B+m_{K^*})\frac{T_1}{V} + (m_B-m_{K^*})\frac{T_2}{A_1}.
\end{equation}

It is interesting to note that Eq.~\eqref{eq:ratio} is independent of
$\text{Im}(C_7)$, $\text{Im}(C_9)$ and $C_{10}$. Eq.~\eqref{eq:ratio}
implies that the sensitivity to both $\text{Re}(C_7)$, and
$\text{Re}(C_9)$ should disappear at large $q^2$.  Since all
asymmetries must vanish at $q^2=0$, no sensitivity is observed in
Figs.~\ref{fig:c7_sev}, \ref{fig:c9_sev}, and \ref{fig:c10_sev}, at
$q^2\approx 0$. The region $q^2=(1-8)~\GeV^2$ is therefore highly
sensitive to $C_7$ and $C_9$. It may be noted that only ratios of form
factors contribute to $S_5/A_\text{FB}$.

\subsection{Unbinned four-dimensional maximum likelihood fitting for BSM}
\begin{figure*}[b!]
  \includegraphics[width=0.495\columnwidth]{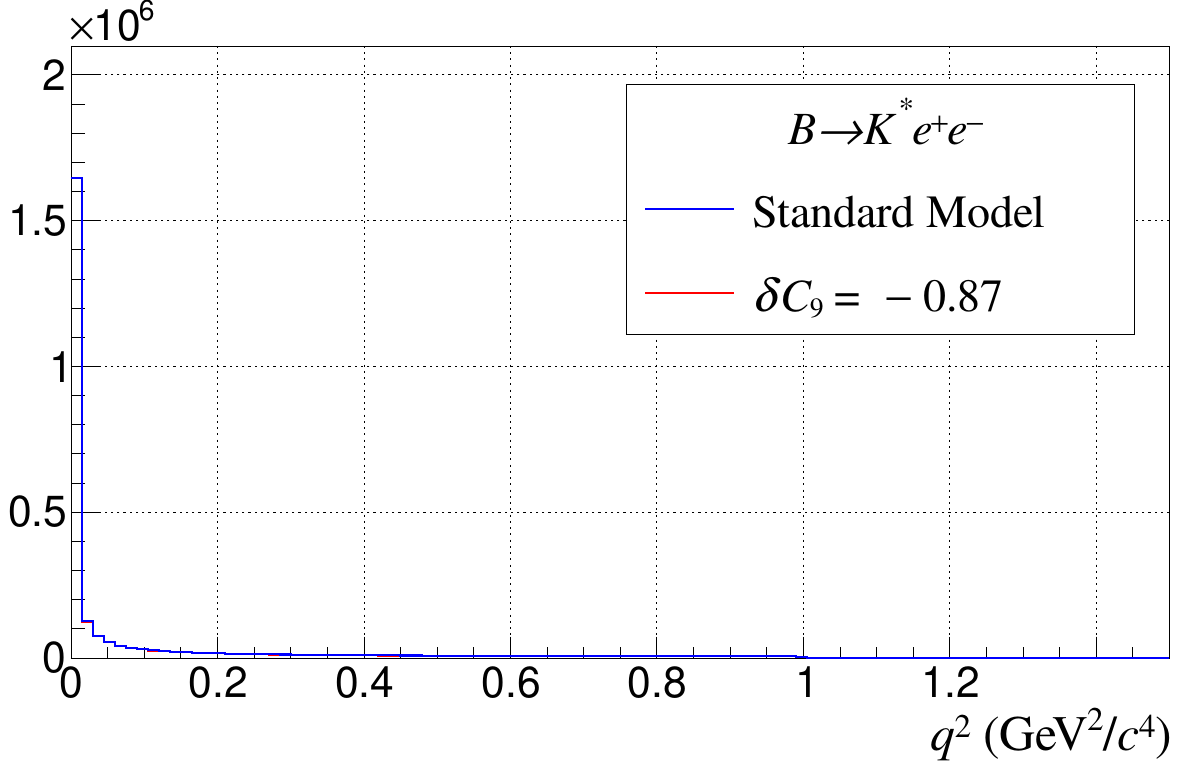}
  \includegraphics[width=0.495\columnwidth]{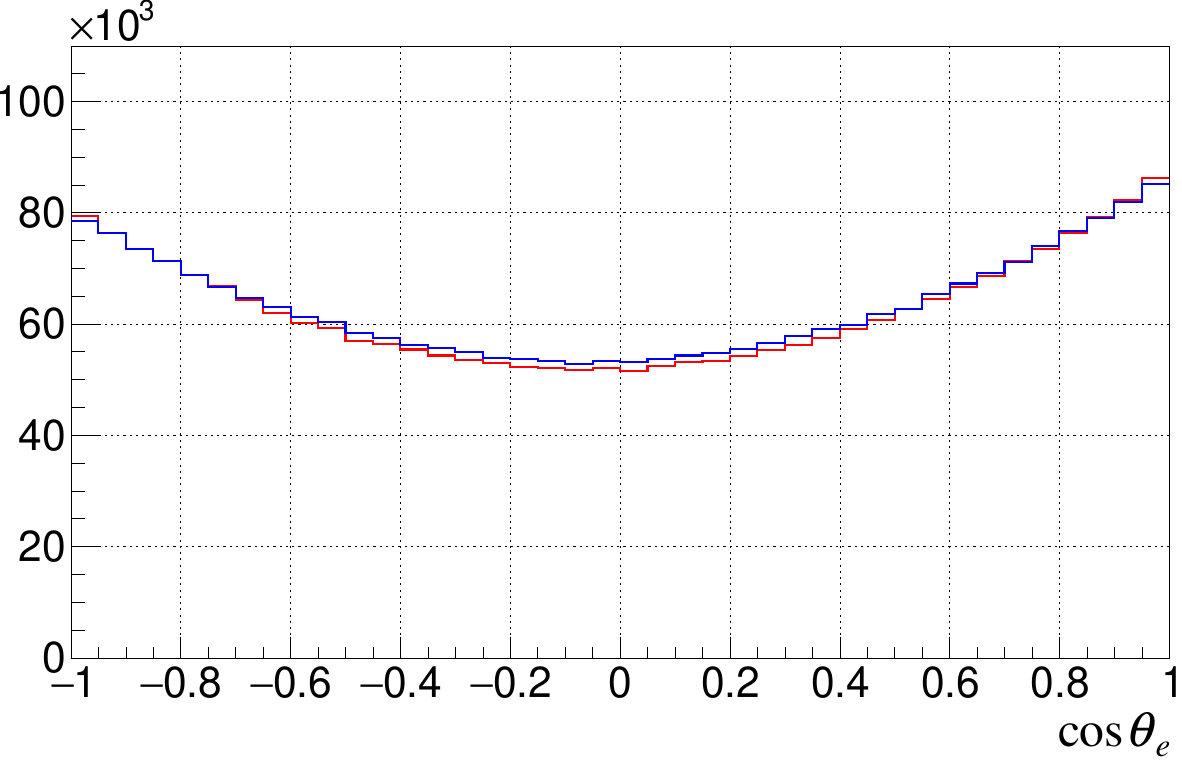}\\ 
  \includegraphics[width=0.495\columnwidth]{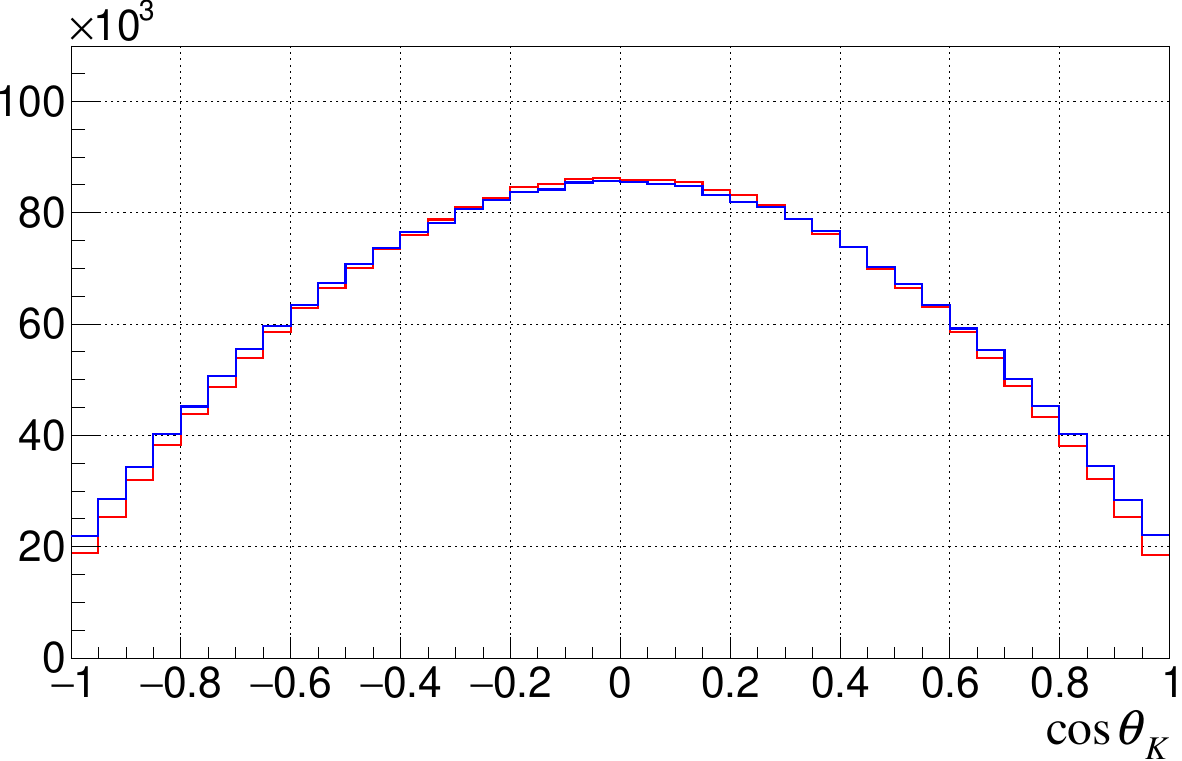}
  \includegraphics[width=0.495\columnwidth]{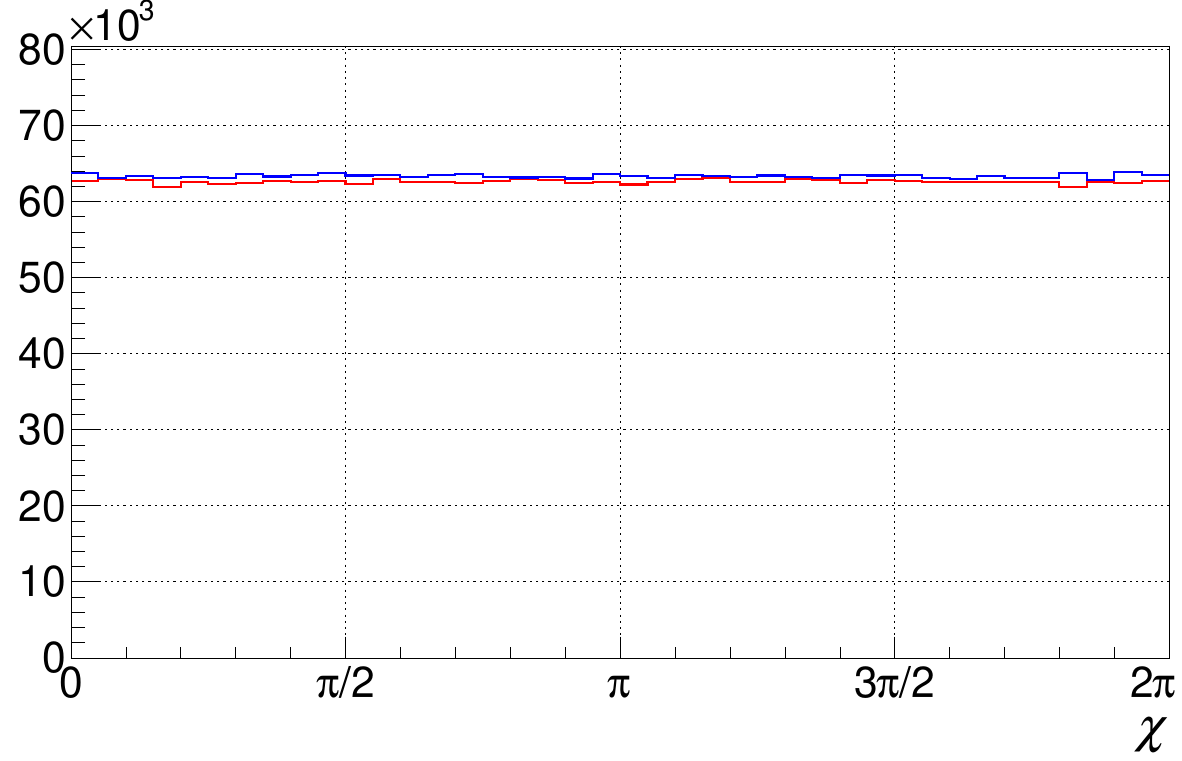}\\
  \includegraphics[width=0.495\columnwidth]{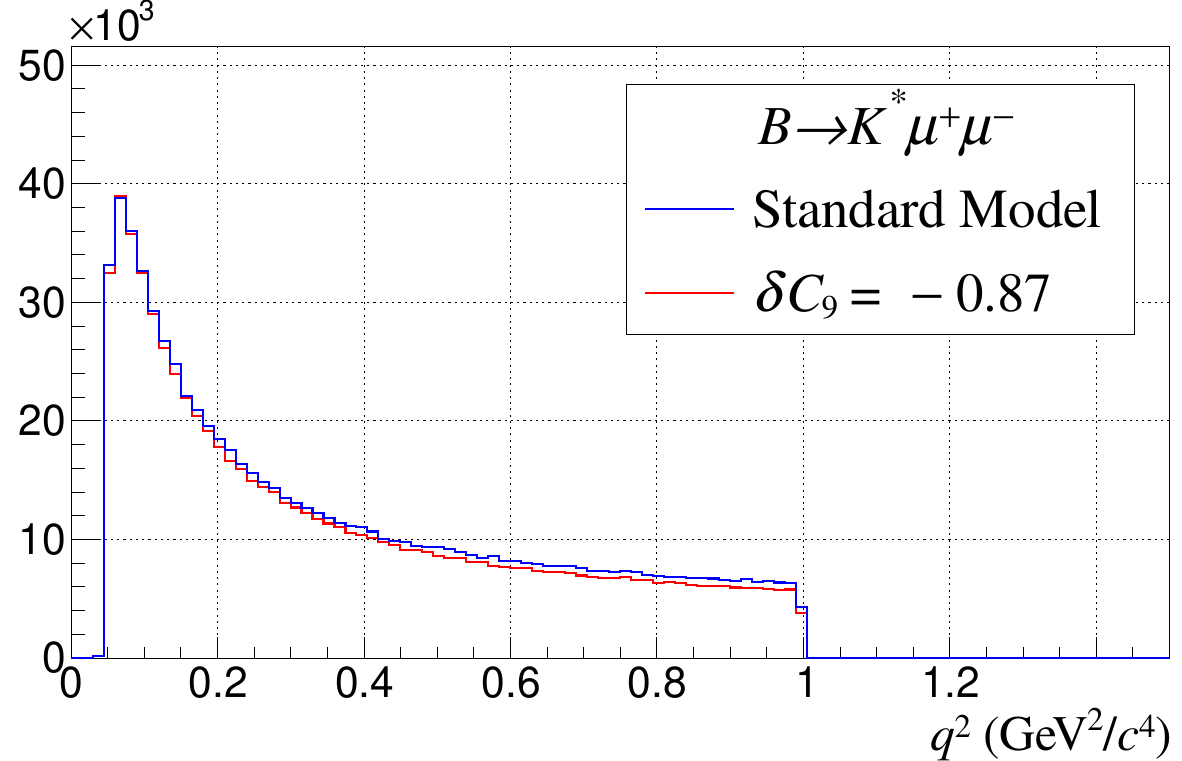}
  \includegraphics[width=0.495\columnwidth]{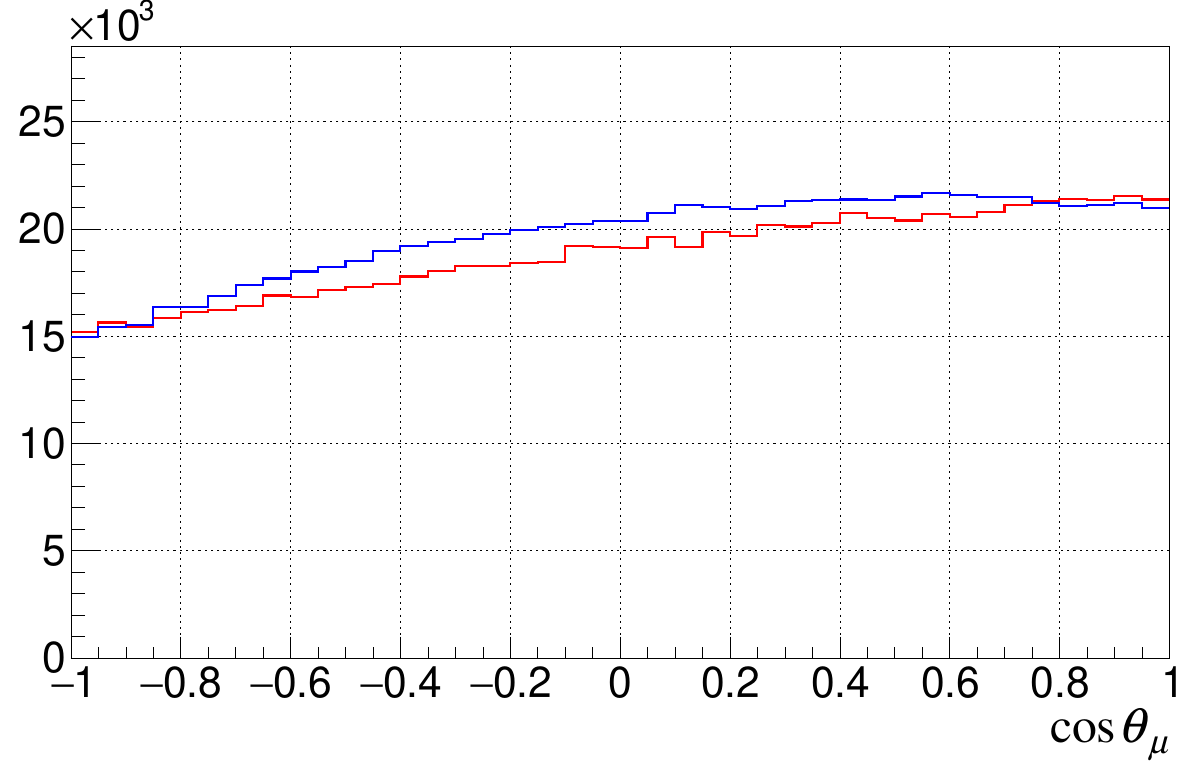}\\ 
  \includegraphics[width=0.495\columnwidth]{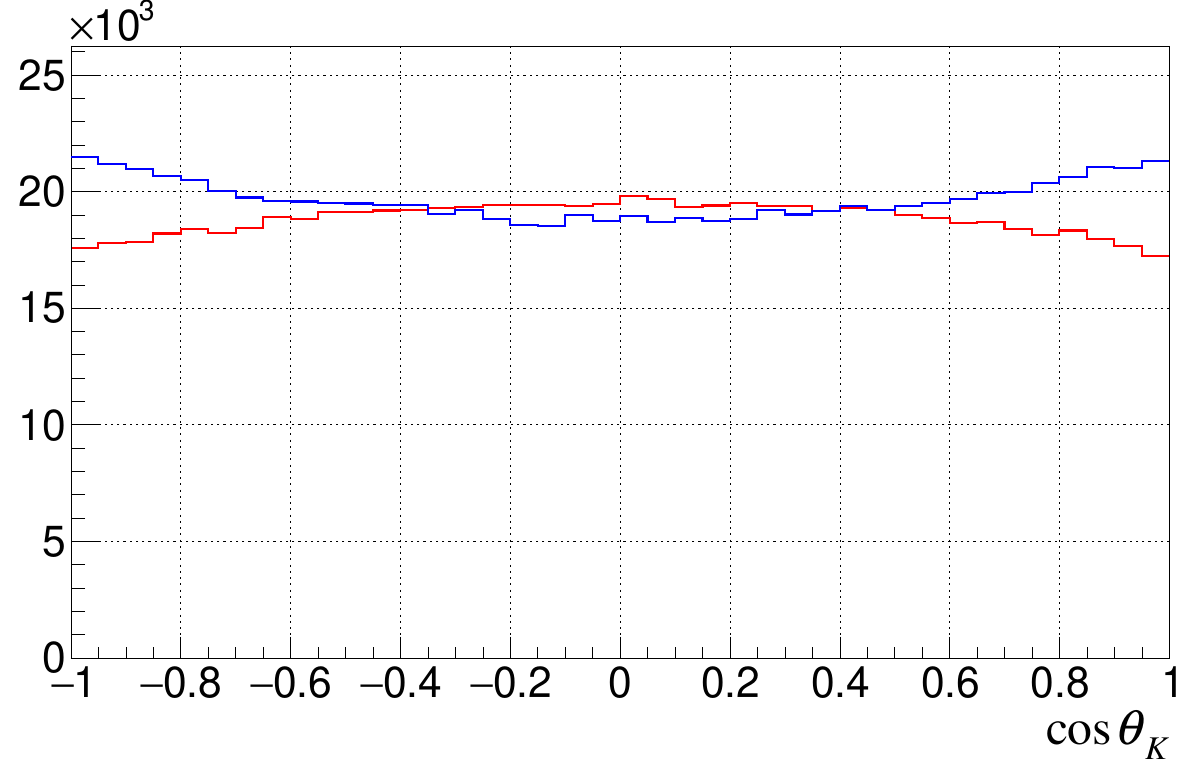}
  \includegraphics[width=0.495\columnwidth]{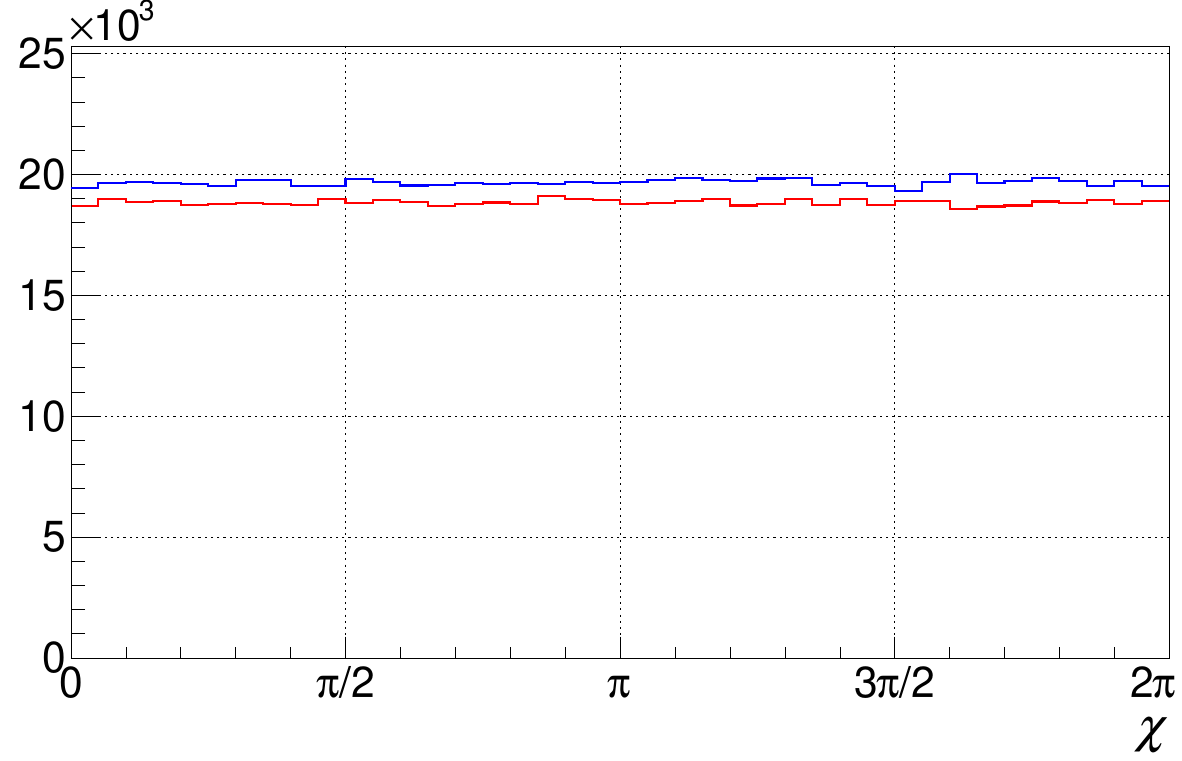}\\
  \caption{\label{fig:ab0ll_q200_01} A comparison of SM and BSM ($\delta
    C_9 = -0.87$) angular distributions in $q^2<1$ $\GeV^2/c^4$ region for
    the di-electron~(4 upper plots) and di-muon~(4 lower plots) decay modes.}
\end{figure*}

\begin{figure*}[h!]
  \includegraphics[width=0.495\columnwidth]{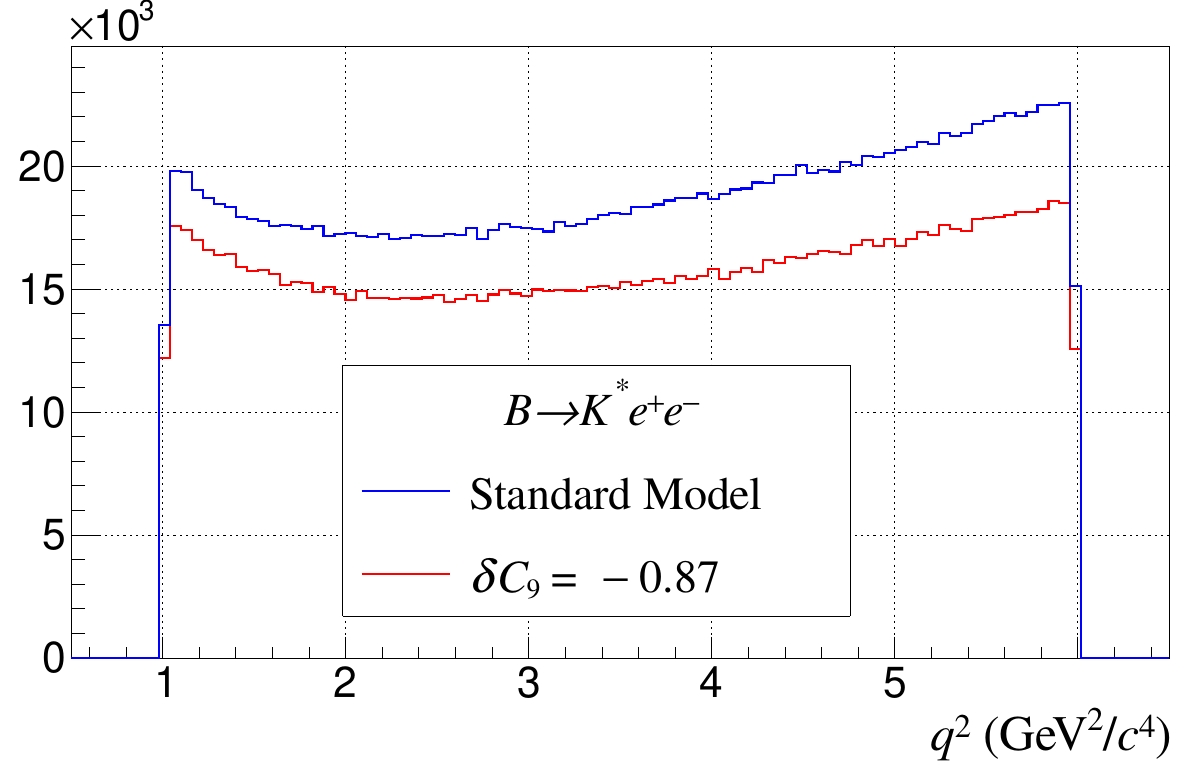}
  \includegraphics[width=0.495\columnwidth]{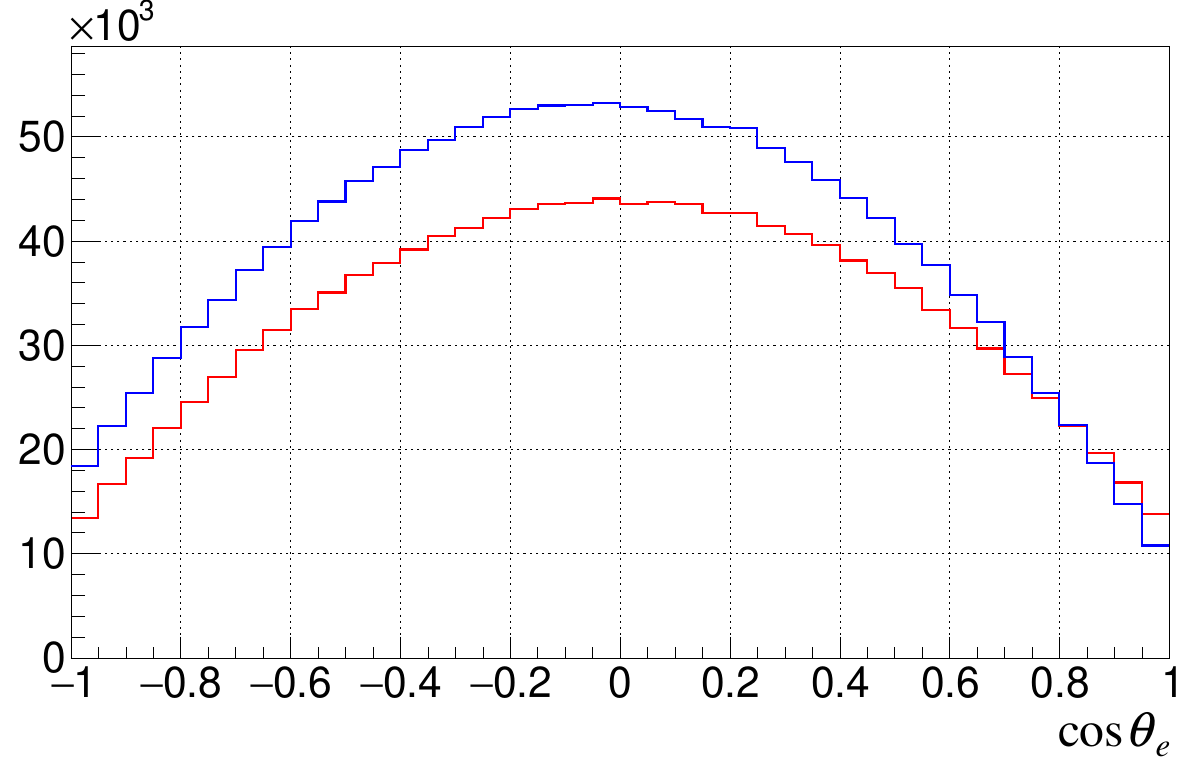}\\ 
  \includegraphics[width=0.495\columnwidth]{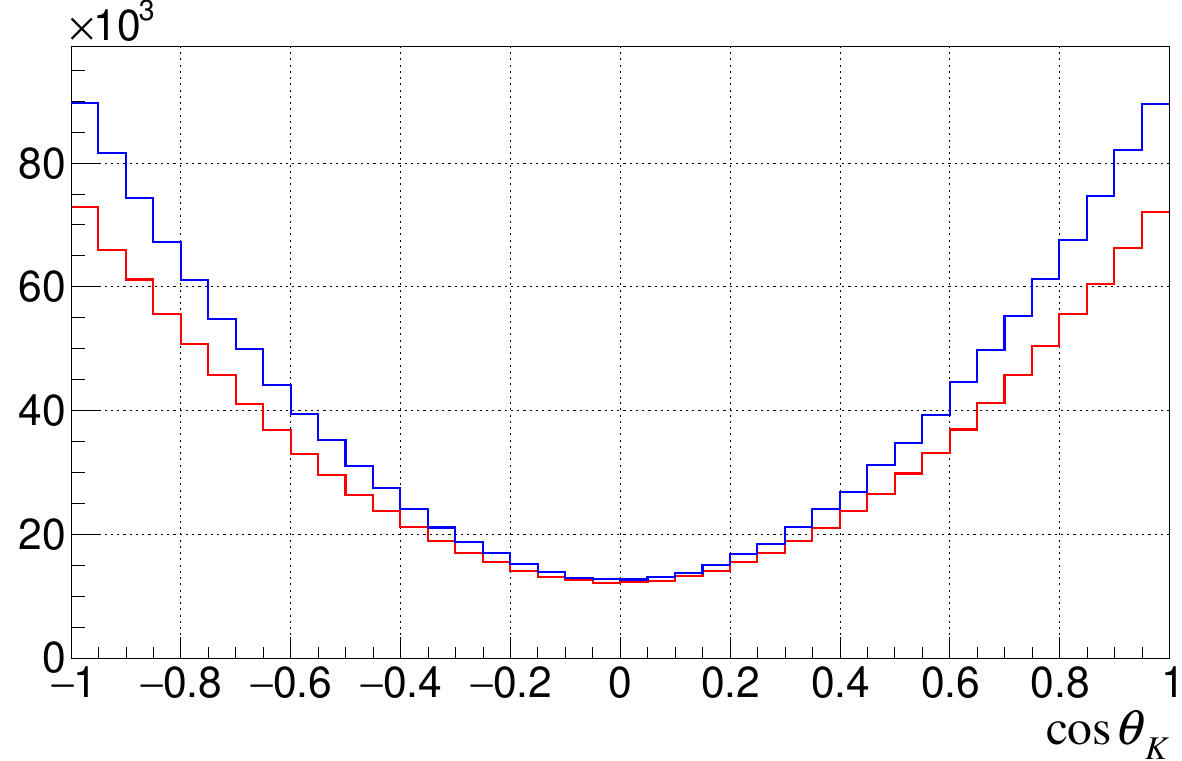}
  \includegraphics[width=0.495\columnwidth]{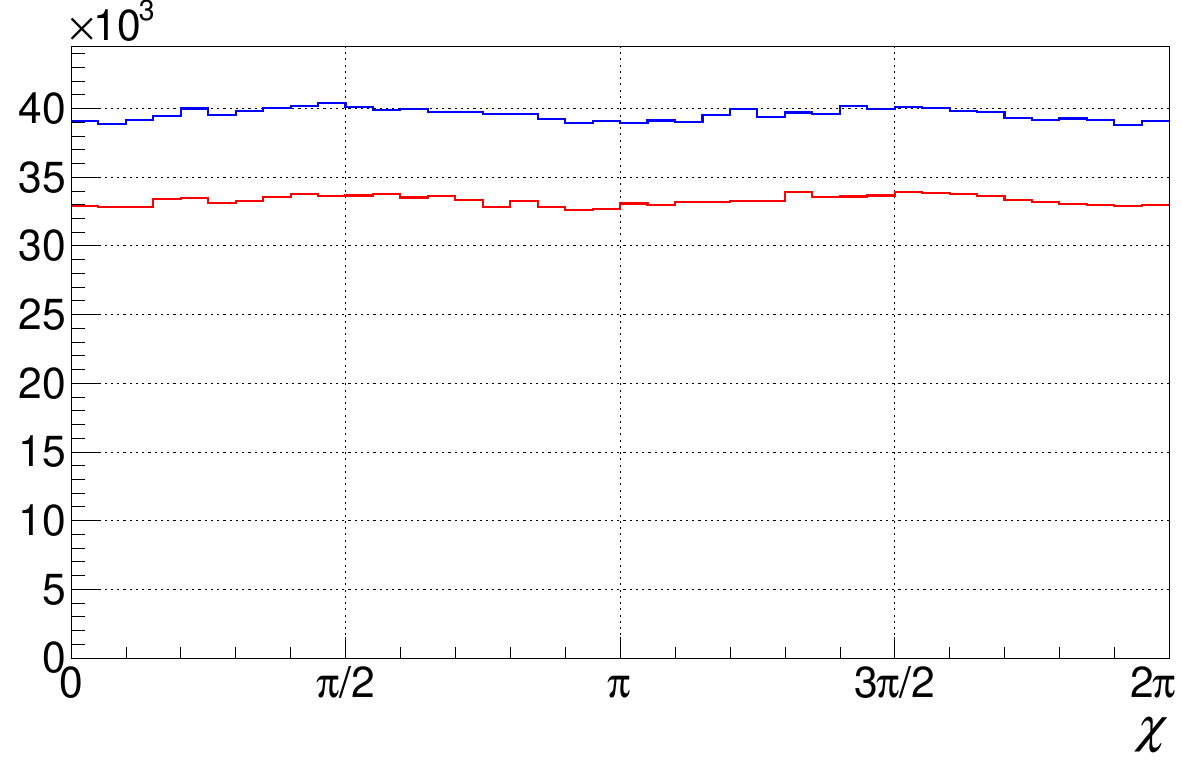}\\
  \includegraphics[width=0.495\columnwidth]{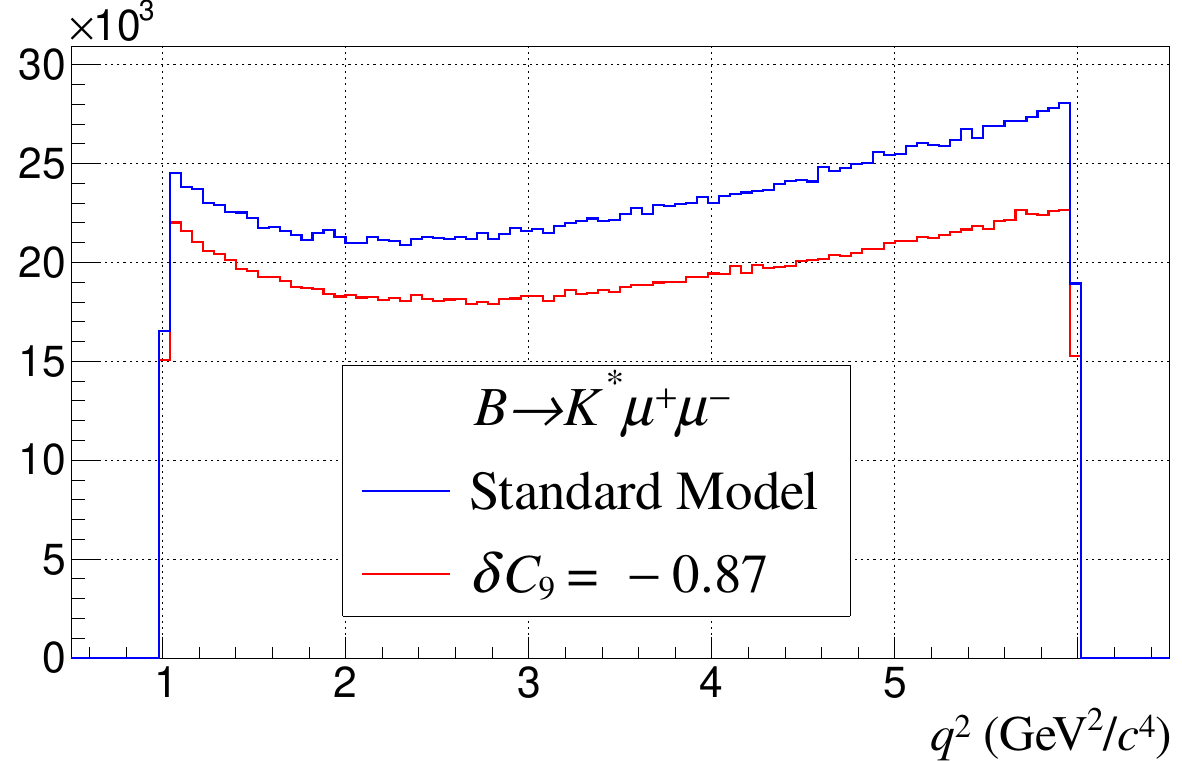}
  \includegraphics[width=0.495\columnwidth]{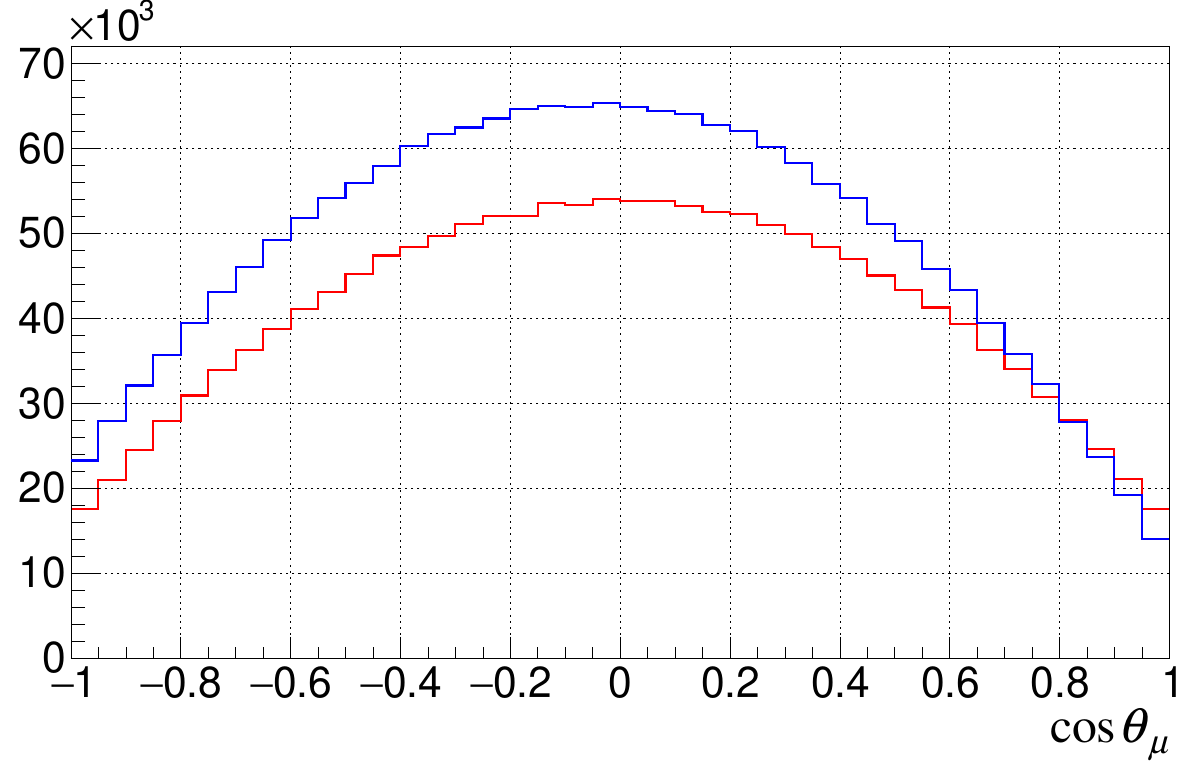}\\ 
  \includegraphics[width=0.495\columnwidth]{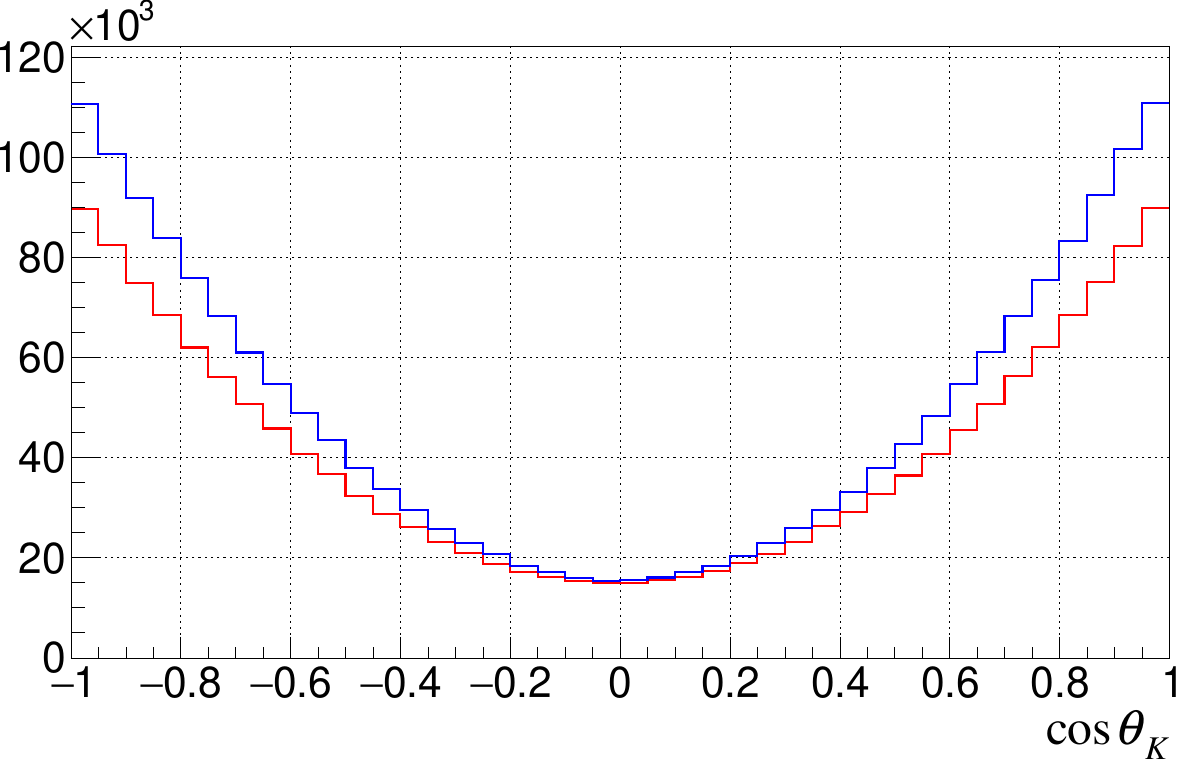}
  \includegraphics[width=0.495\columnwidth]{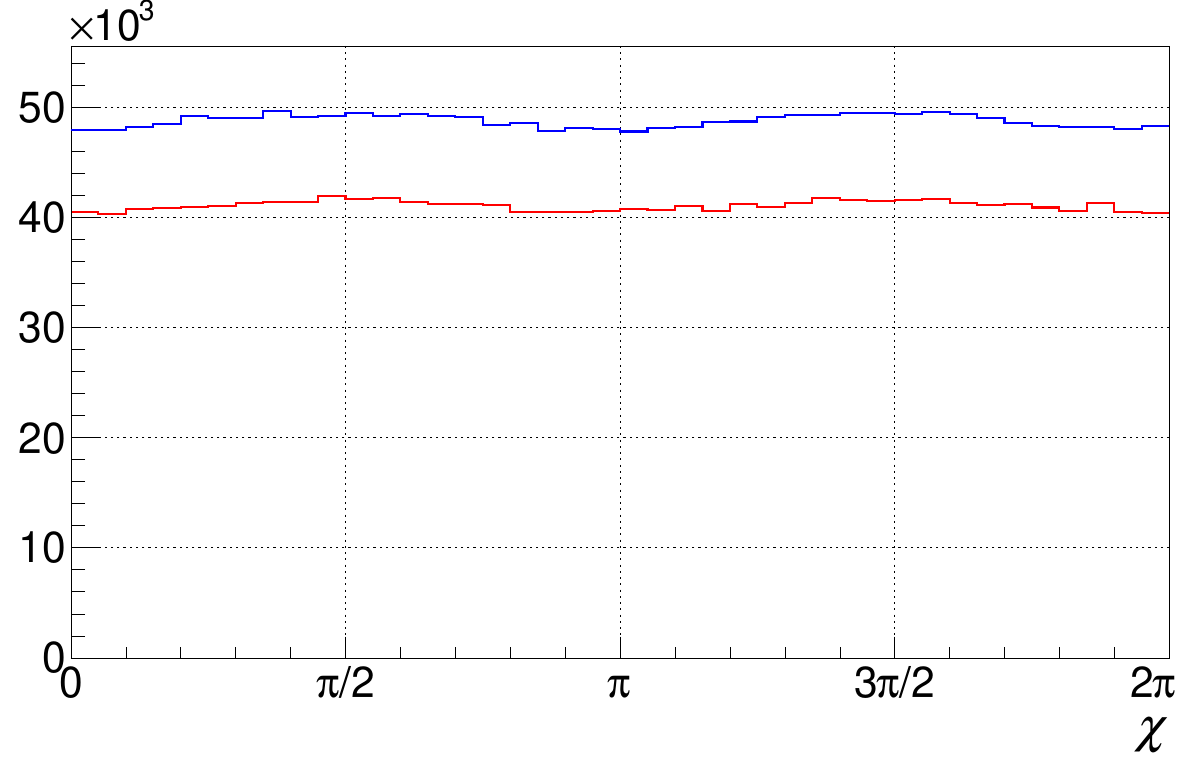}\\
  \caption{\label{fig:ab0ll_q201_06} A comparison of SM and BSM ($\delta
    C_9 = -0.87$) angular distributions in the $1<q^2<6$ $\GeV^2/c^4$ region for
    the di-electron~(4 upper plots) and di-muon~(4 lower plots) decay modes.}
\end{figure*}

\begin{figure*}[h!]
  \includegraphics[width=0.495\columnwidth]{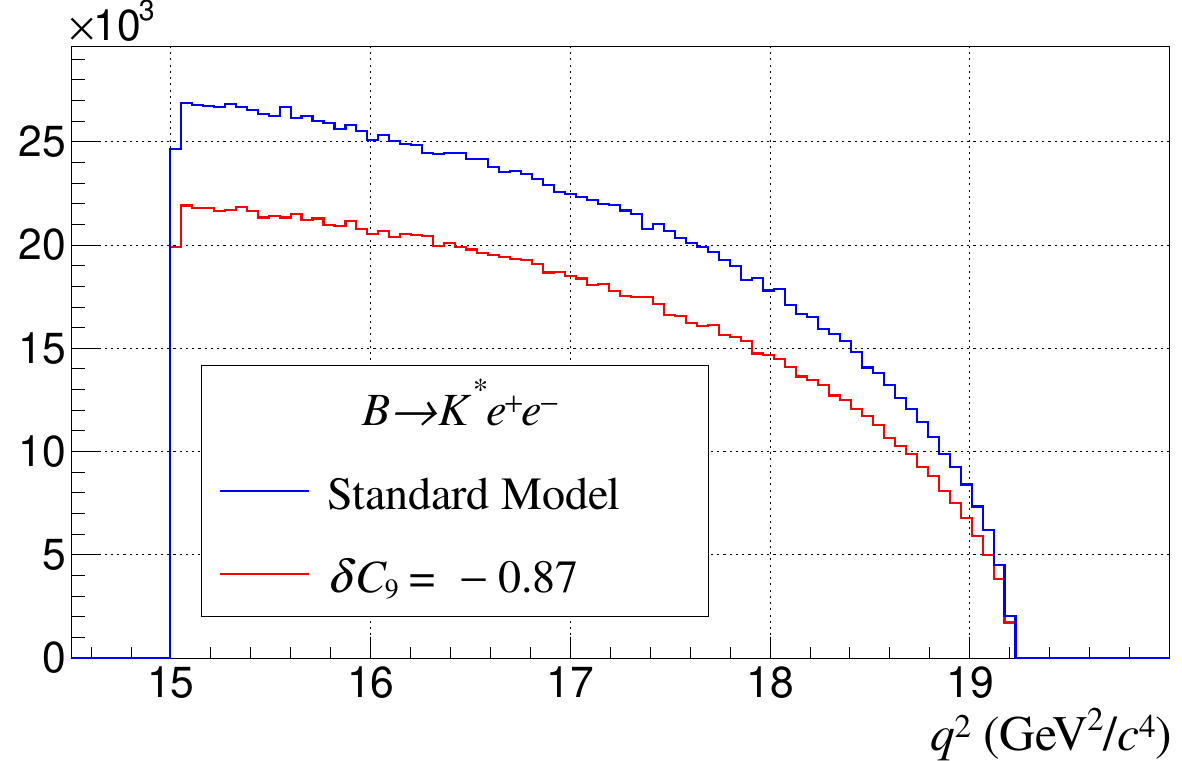}
  \includegraphics[width=0.495\columnwidth]{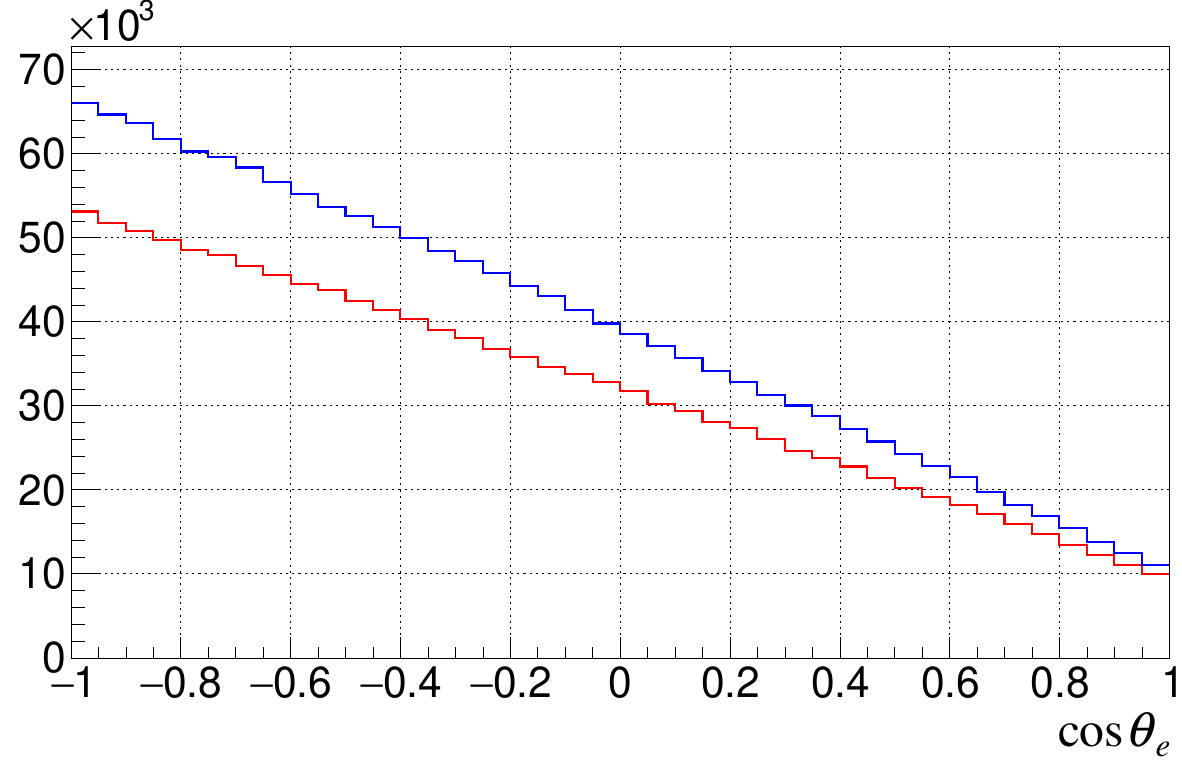}\\ 
  \includegraphics[width=0.495\columnwidth]{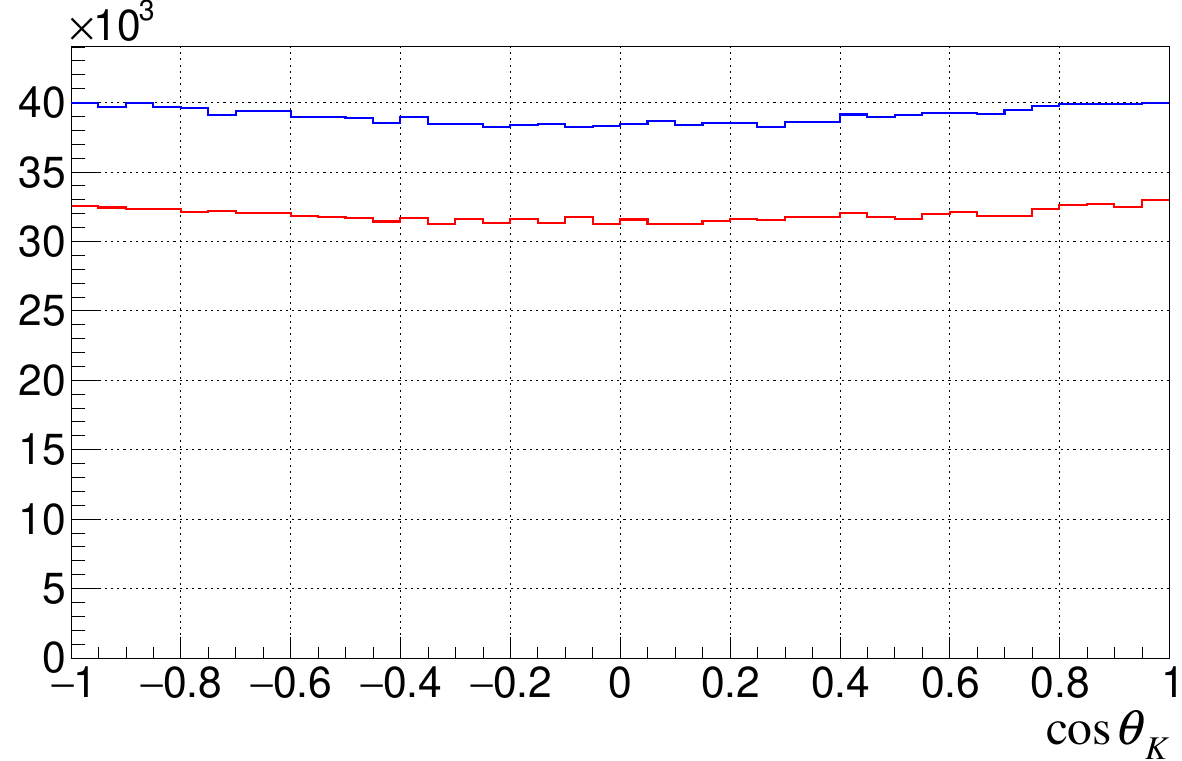}
  \includegraphics[width=0.495\columnwidth]{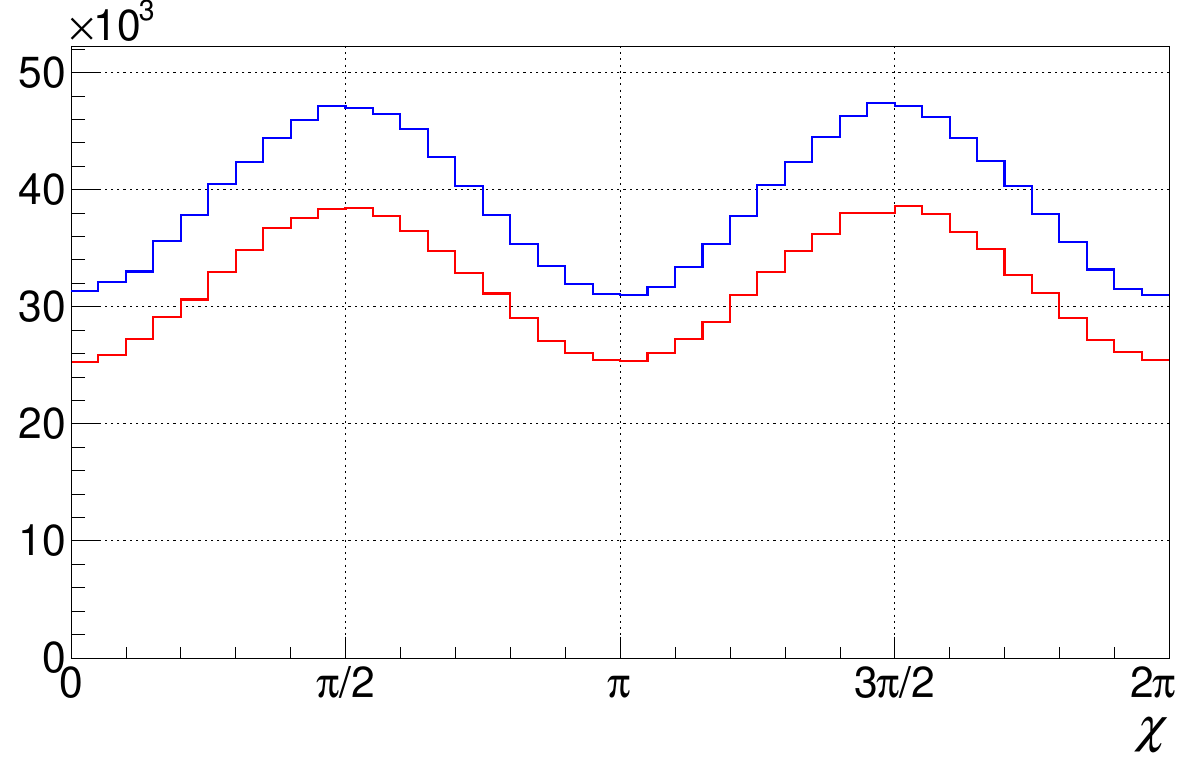}\\
  \includegraphics[width=0.495\columnwidth]{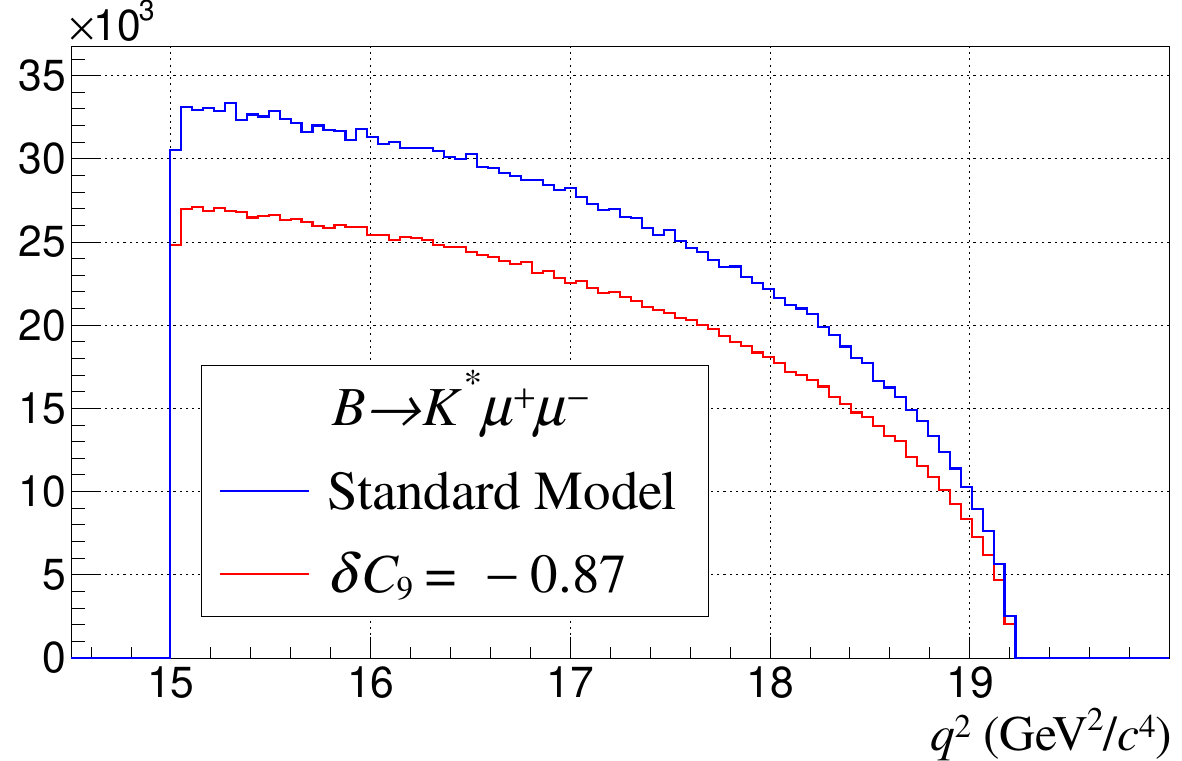}
  \includegraphics[width=0.495\columnwidth]{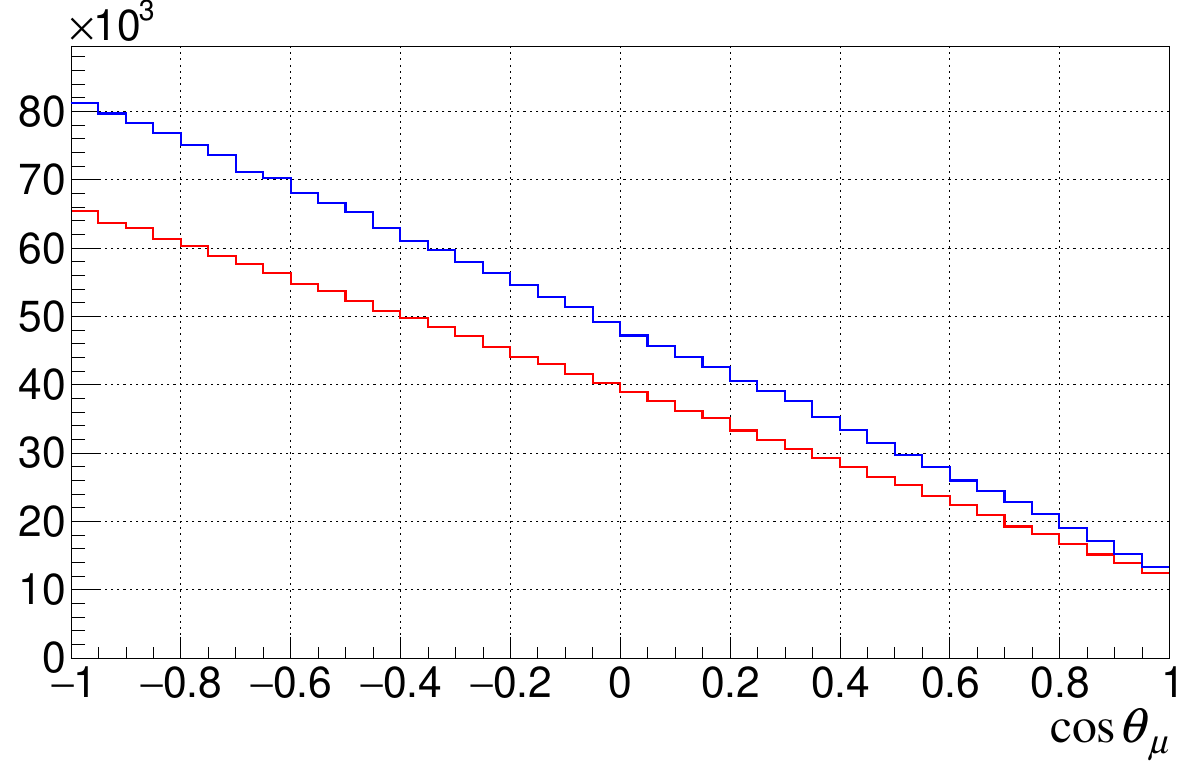}\\ 
  \includegraphics[width=0.495\columnwidth]{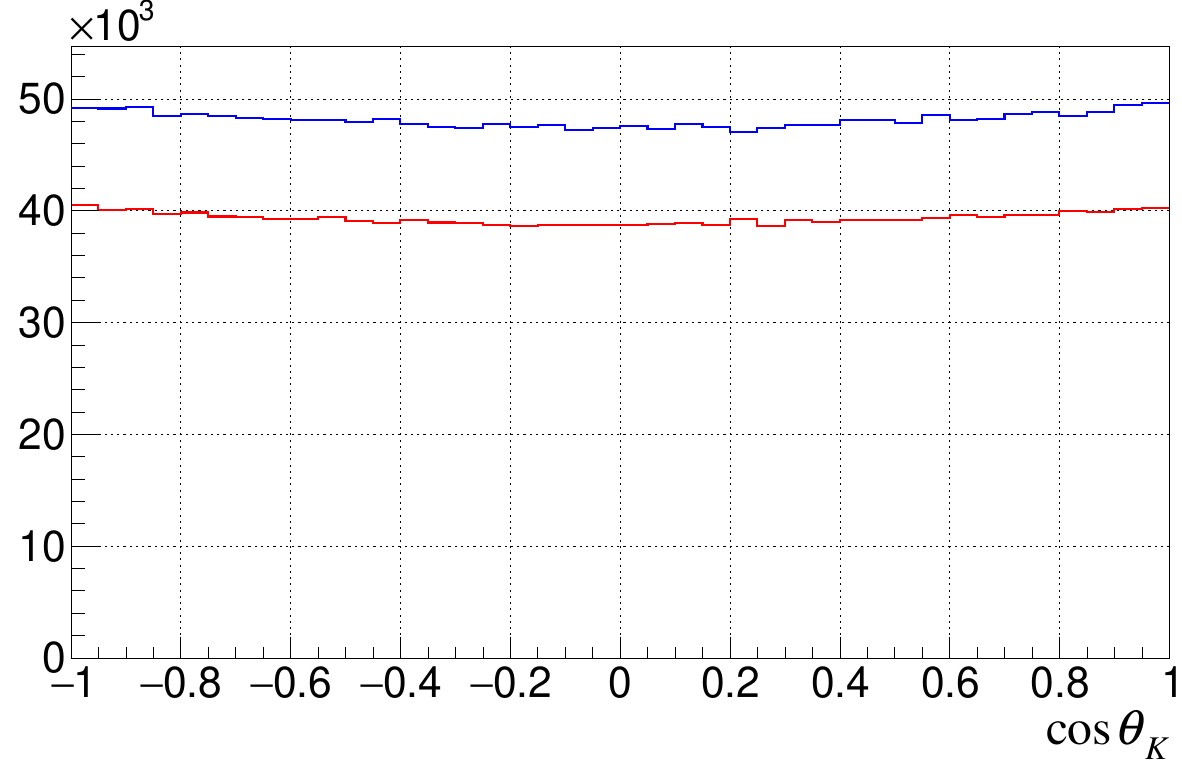}
  \includegraphics[width=0.495\columnwidth]{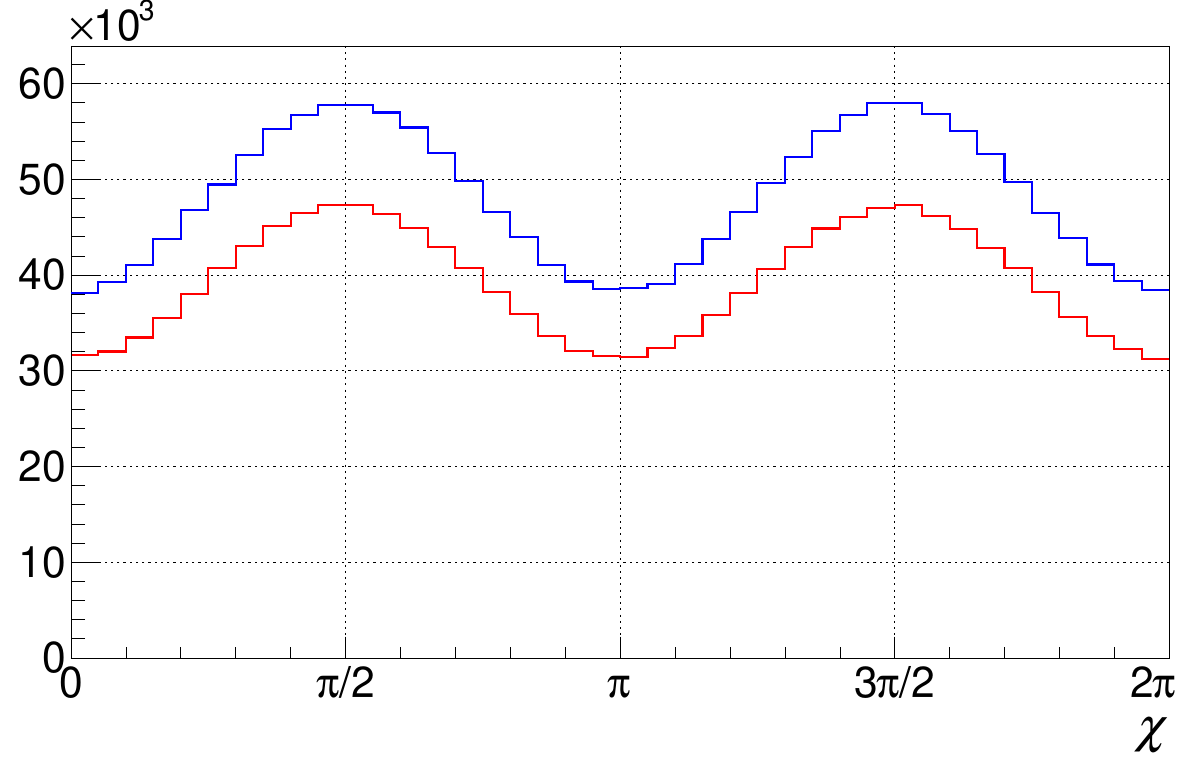}\\
  \caption{\label{fig:ab0ll_q215_20} A comparison of SM and BSM ($\delta
    C_9 = -0.87$) angular distributions in the $q^2>15$ $\GeV^2/c^4$ region for
    the di-electron~(4 upper plots) and di-muon~(4 lower plots) decay modes.}
\end{figure*}

Using a high statistics sample of $B\to K^* e^+ e^-$ and $B\to K^*
\mu^+ \mu^-$ MC simulated events, we can compare the distributions in
the four kinematic variables for the SM and a BSM scenario with
$\delta C_9 =-0.87$ in three selected regions of $q^2$ ($ q^2 <1
~\rm{GeV}^2$, $ 1< q^2 <6 ~\rm{GeV}^2$ and $ q^2>15 ~\rm{GeV}^2$).The
resulting distributions are shown in Figs.~\ref{fig:ab0ll_q200_01},
\ref{fig:ab0ll_q201_06} and \ref{fig:ab0ll_q215_20}. These three
figures show features of BSM physics that were not previously
appreciated. For example, in each $q^2$ region, there are differences
in shape between the SM and BSM scenario in $\cos\theta_{K^*}$ and
$\cos\theta_\ell$. In all three regions, there are changes in the
forward-backward asymmetry of the leptonic system, which are visible
in the $\cos\theta_\ell$ distributions.  At low $q^2$ in $B\to K^* e^+
e^-$, the SM pole contribution arising from ${\cal O}_7$ dominates and
BSM effects are difficult to distinguish. However, in $B\to K^* \mu^+
\mu^-$ the BSM effects are more prominent in the angular
distributions. At higher $q^2$, there are differences in the shape of
the $q^2$ distribution, which are correlated to changes in the angular
distributions. All such effects can be accommodated with
multi-dimensional maximum likelihood fits.

\begin{figure*}[h!]
  \includegraphics[width=0.495\columnwidth]{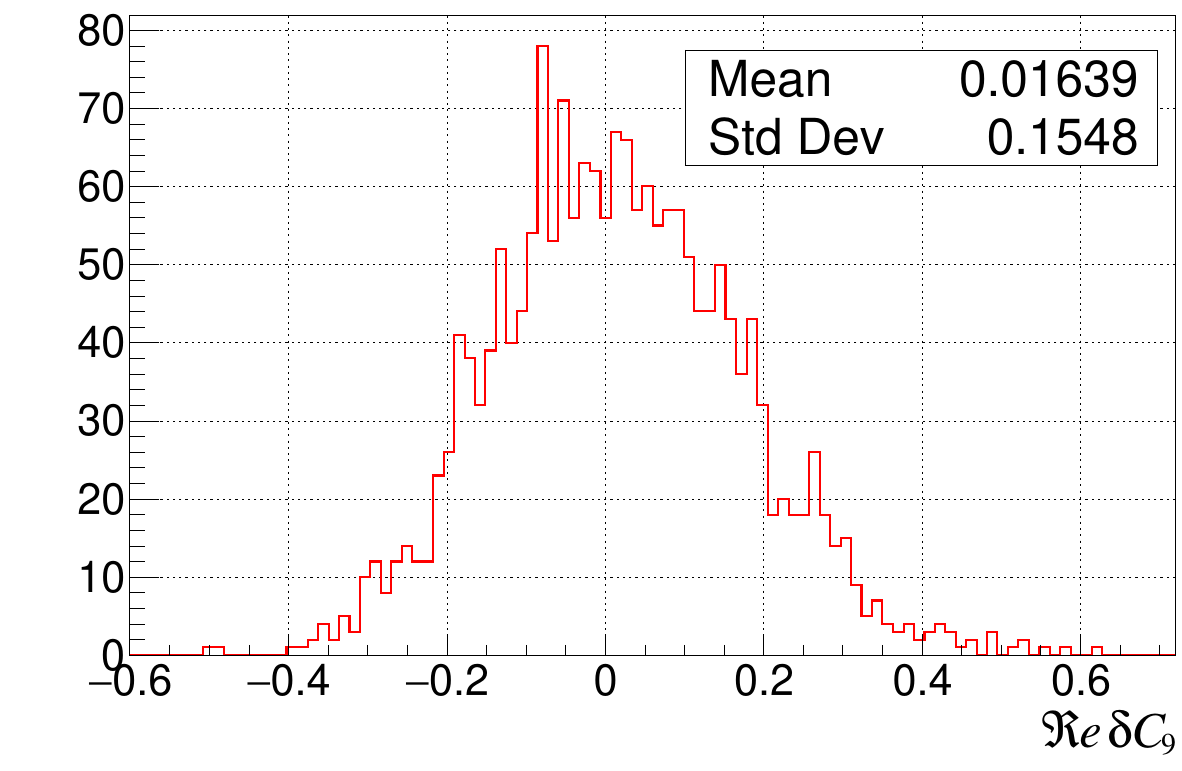} 
  \includegraphics[width=0.495\columnwidth]{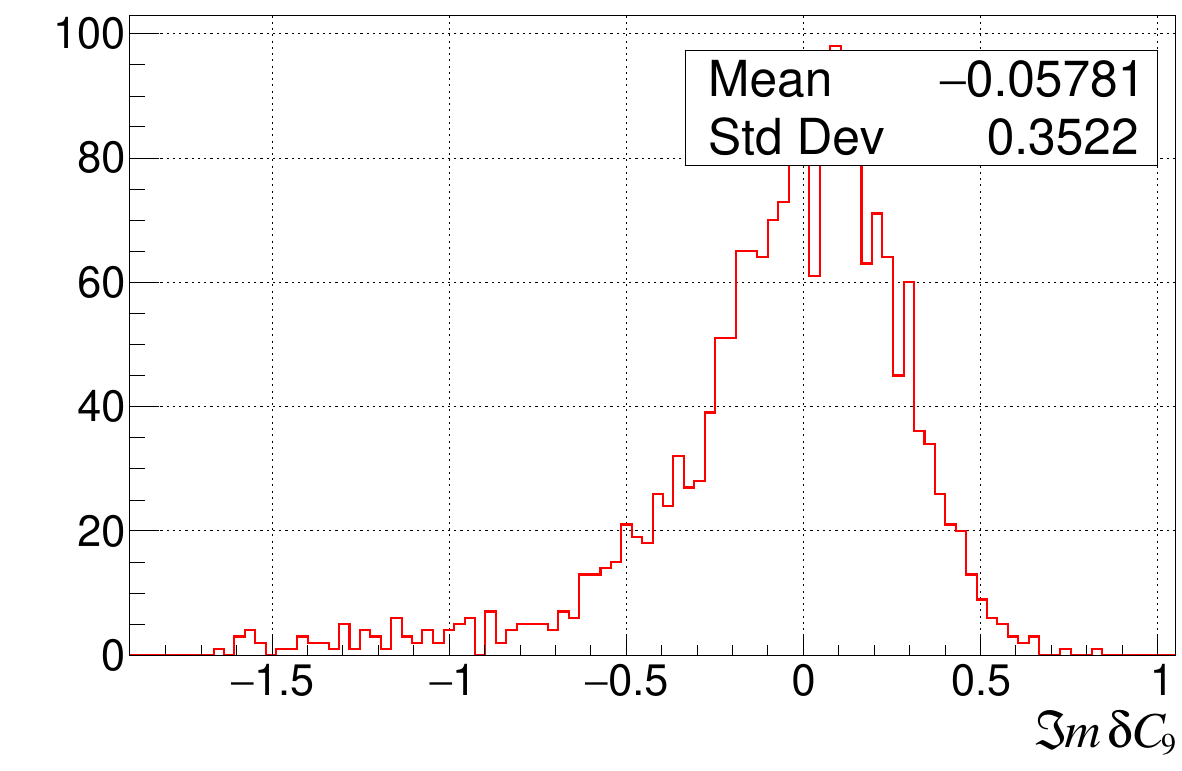} \\
  \includegraphics[width=0.495\columnwidth]{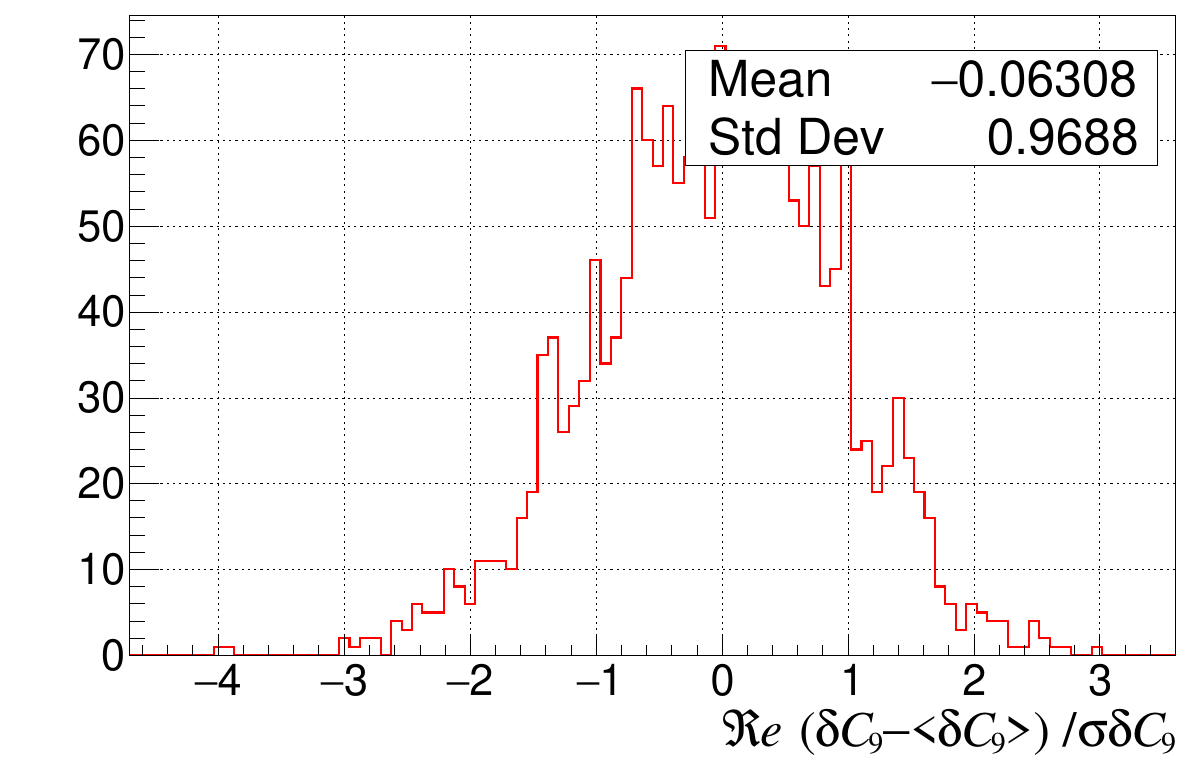} 
  \includegraphics[width=0.495\columnwidth]{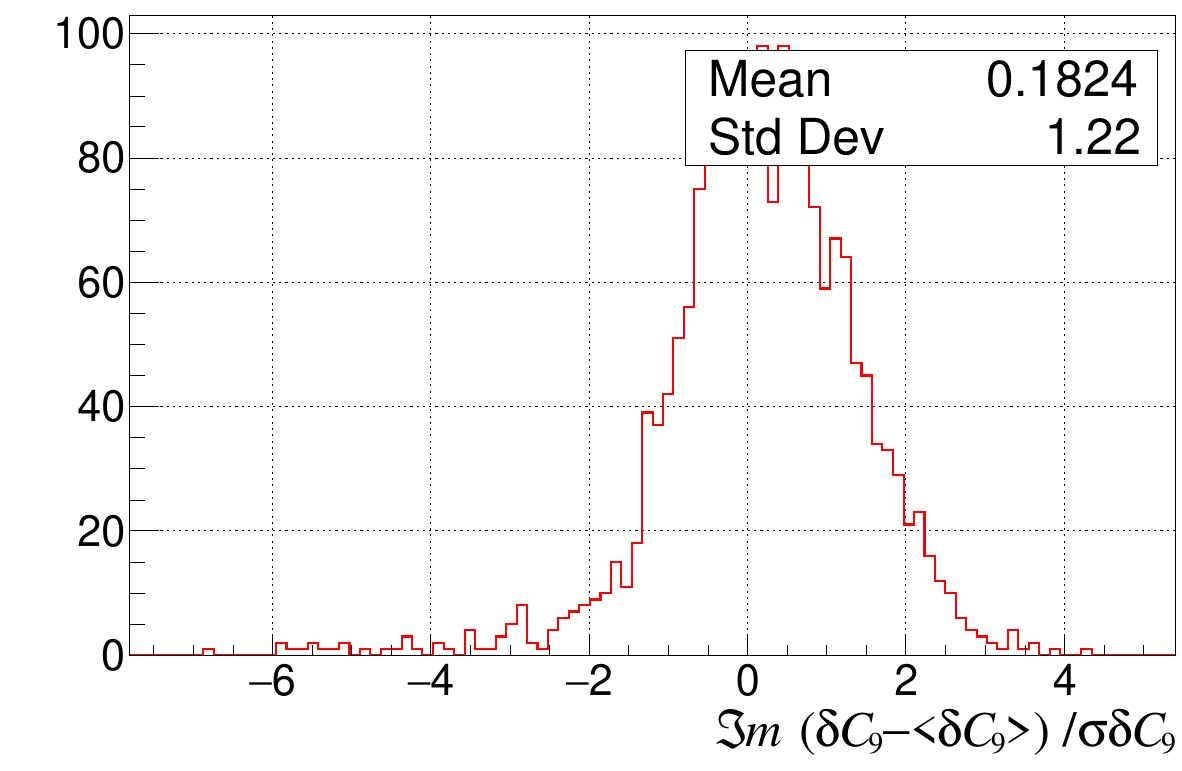} \\
  \caption{\label{fig:dc9mm} Distribution of $\delta C_9$ values
    obtained from unbinned likelihood fits to simulated $B\to K^*
    \mu^+\mu^-$ SM decays in the full $q^2$ range, without resonances
    included. The real and imaginary parts of $\delta C_9$ extracted
    by the fits are shown in the top row. The bottom row shows the
    corresponding pull distributions, which demonstrate that the
    fitter correctly estimates fit uncertainties.  About 2000
    pseudo-experiments are performed, where in each 40 ab$^{-1}$
    luminosity and 25\% efficiency are assumed.}
\end{figure*}

Rather than extracting angular asymmetries in bins of $q^2$, and then
providing these as inputs to a theory-based fitting
package~\cite{Altmannshofer:2021qrr,Alguero:2021anc,Hurth:2021nsi,Ciuchini:2021smi}
that determines $\delta C_i$, we fit directly for $\delta C_i$ using
4D unbinned likelihood fits to $q^2$, $\cos{\theta_\ell}$,
$\cos{\theta_K}$, and $\chi$ with $\delta C_i$ as a free parameter.
For example, Fig.~\ref{fig:dc9mm} shows the distribution of $\delta
C_9$ resulting from about 2000 pseudo-experiments performing such
multidimensional fits repeatedly, assuming for now that $\delta C_9=0$
to check the statistical sensitivity. According to the fit results the
real and imaginary parts of $\delta C_9$ can be constrained with a
standard deviation of 0.15 and 0.35, corresponding to 3 and 7 \% of SM
$|C_9|$, respectively. This is a substantial improvement over the
traditional analysis method. The pull distributions demonstrate that
the obtained fit uncertainties meet expectations. The mitigation of
form factor uncertainties is discussed below.

To further test the fit procedure we generate samples with the BSM
physics contribution $\delta C_9=-0.87$ and repeat the
multidimensional fits.  The fit extracts the real part of the BSM
contribution correctly with 8 and 10 standard deviation significance
for the di-muon and di-electron modes, respectively. The imaginary
part is consistent with zero and its uncertainty roughly matches to
the SM fit results, albeit a small number of samples show relatively
large deviation from zero.  The fitted Wilson coefficient
distributions obtained in the pseudo-experiments are shown in
Appendix~\ref{app:figs}.

\clearpage
\subsection{$\delta C_{10}$ sensitivity}
The statistical sensitivity to $\delta C_{10}$ is estimated from our
multidimensional fit. According to the fit results the real and
imaginary parts of $\delta C_{10}$ can be constrained to be 0.21 and
0.28, corresponding to 5 and 7 \% of $|C_{10}|\approx 4.1$,
respectively. The corresponding distributions are shown in
Appendix~\ref{app:figs}.

\subsection{$C_{9}'$ and $C_{10}'$ sensitivity}
Next, we investigate the statistical sensitivity to BSM physics in the form of
right-handed currents. The right-handed contribution $C_9^{\prime}$
can be constrained at level of 3 \% of $|C_9|$ in the di-muon mode.
Similarly, the right-handed contribution $C_{10}^{\prime}$ can be
constrained at level of 3 \% of $|C_{10}|$ in the di-muon
mode. Details of the fit results are shown in Appendix~\ref{app:figs}.

\subsection{$C_7$, $C_7^{\prime}$ signatures at low $q^2$ and in the $\chi$ angle distribution}
\begin{figure*}[h!]
  \includegraphics[width=0.495\columnwidth]{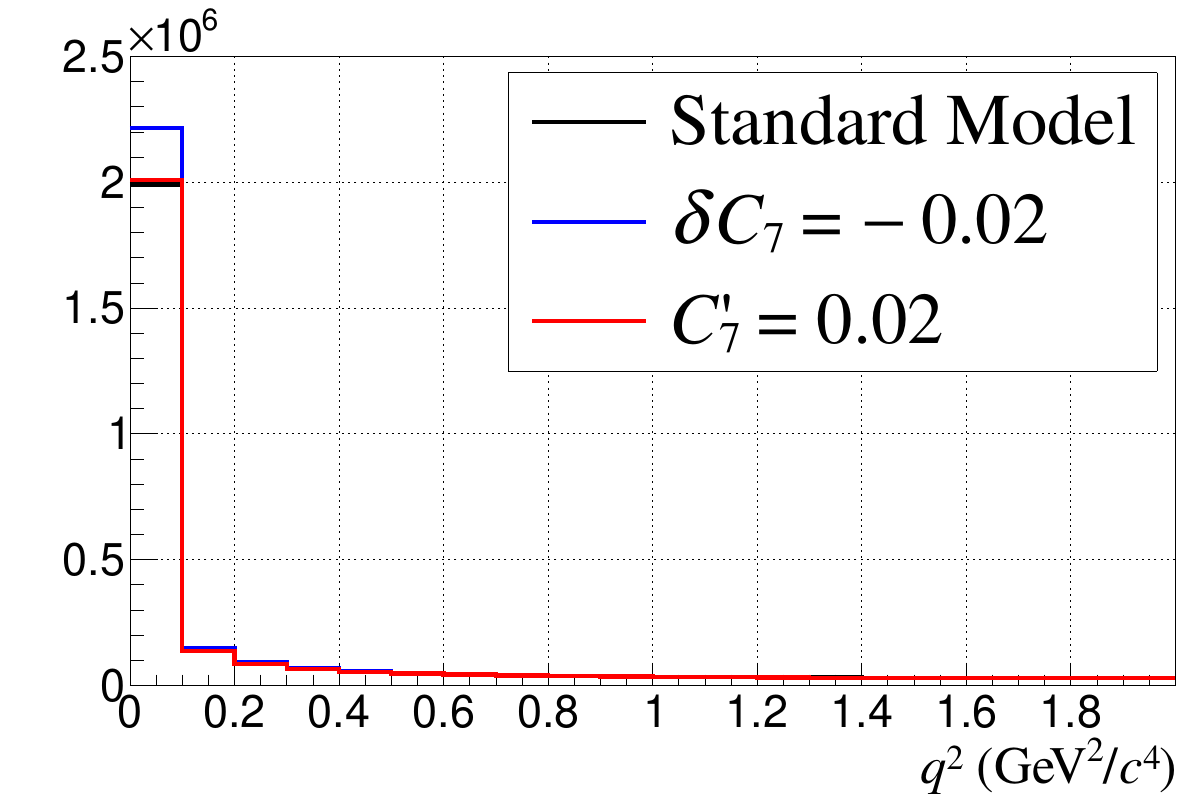}
  \includegraphics[width=0.495\columnwidth]{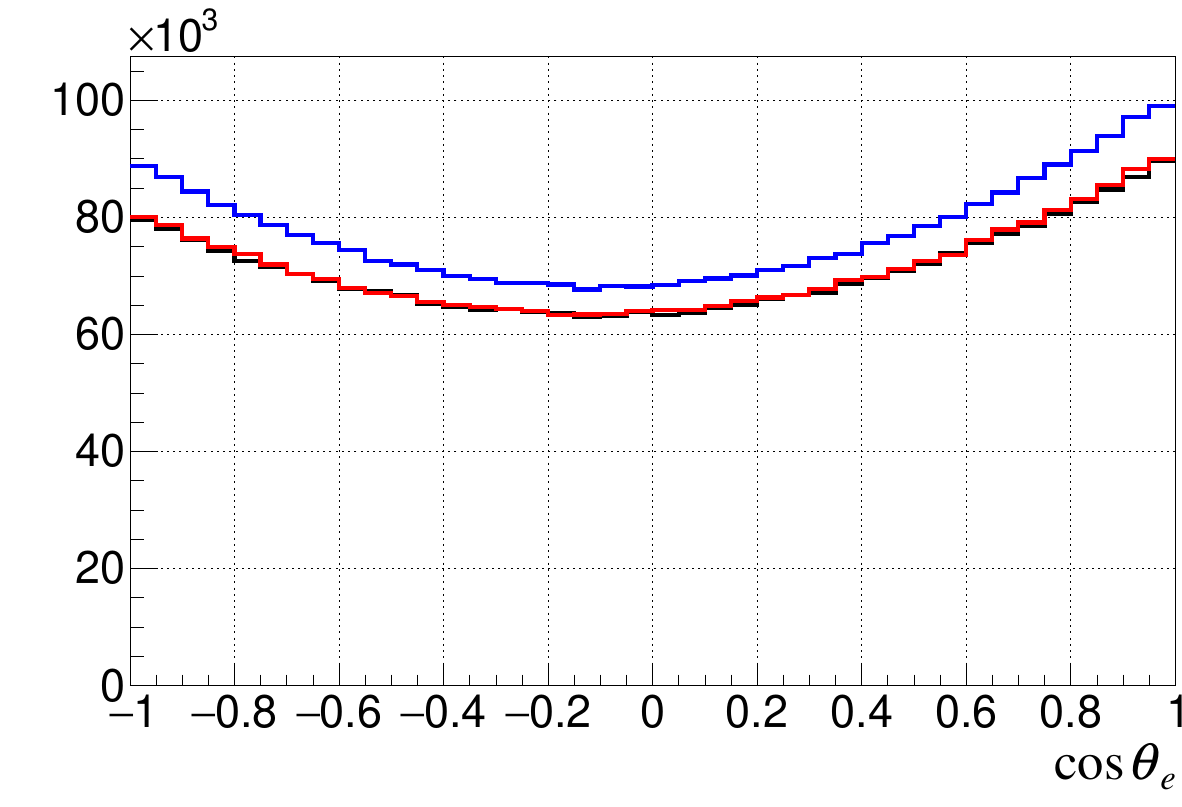}\\
  \includegraphics[width=0.495\columnwidth]{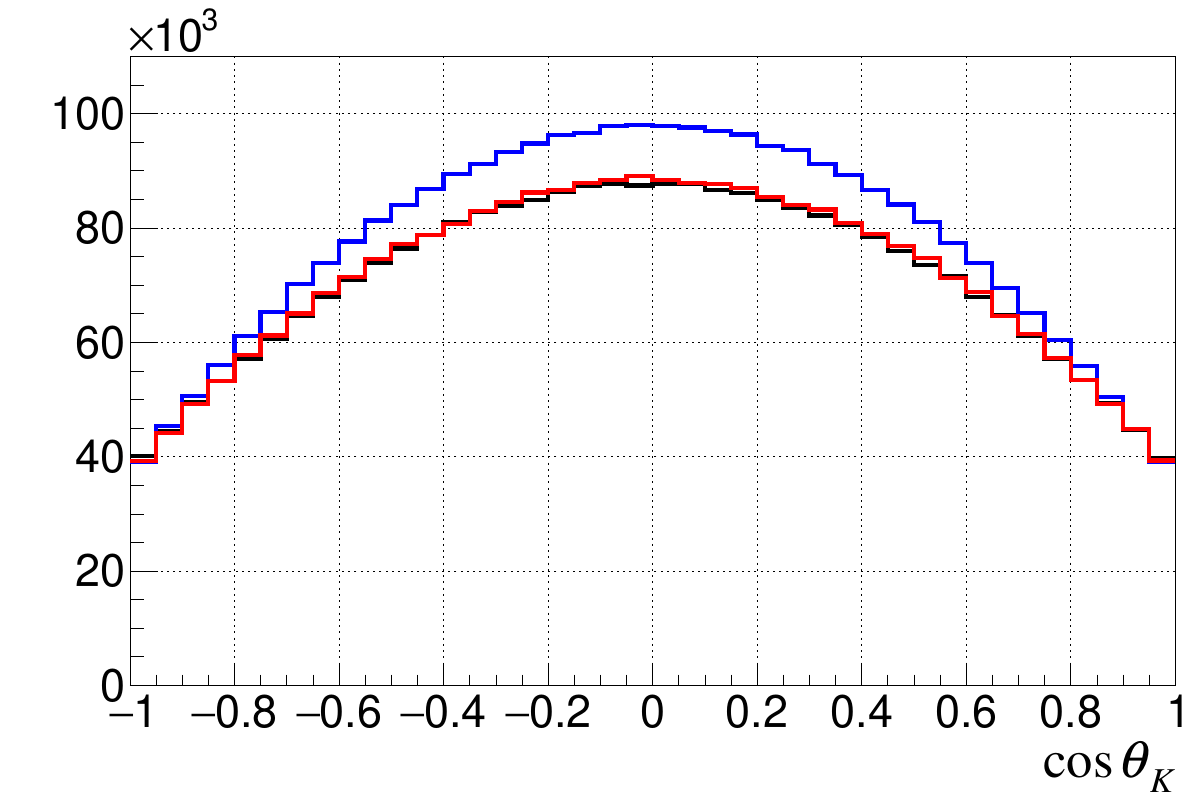}
  \includegraphics[width=0.495\columnwidth]{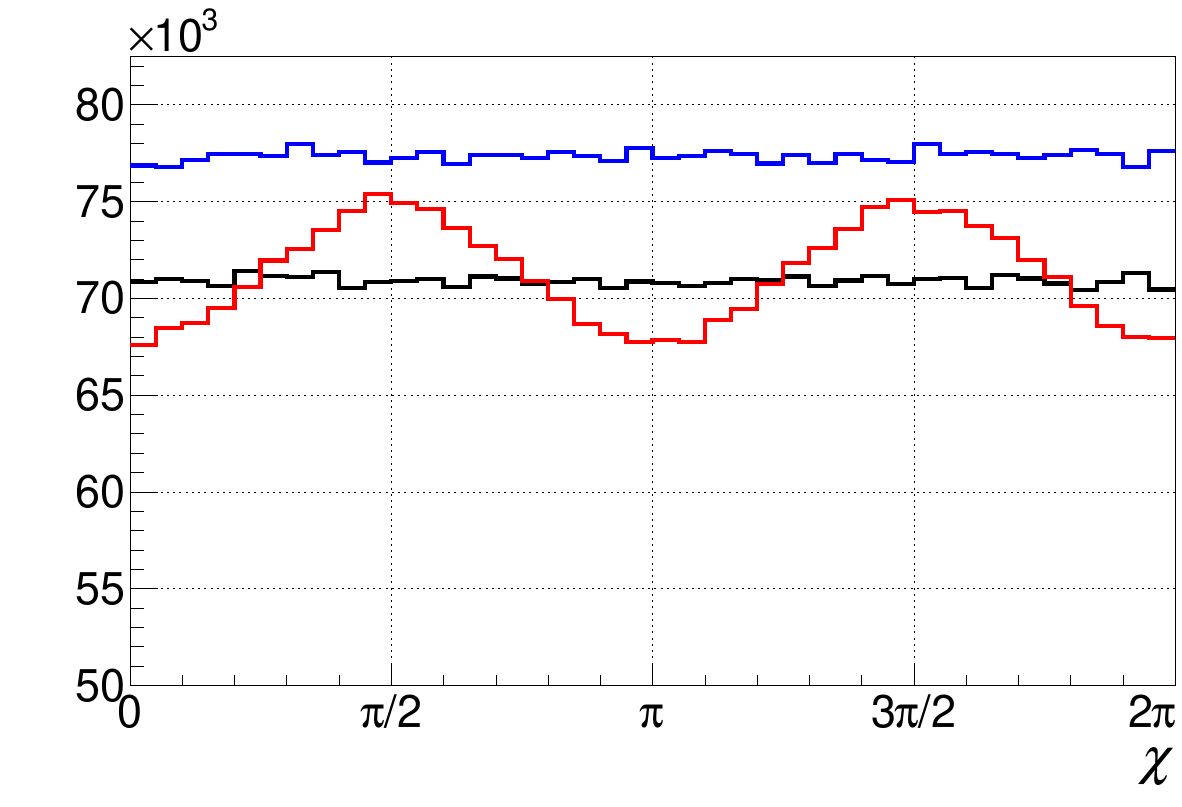}
  \caption{\label{fig:c7_cp7_lowq2_angles} The effect of
    $C_7^{\prime}$ and $\delta C_7$ on angular distributions in the
    di-electron decay mode for $q^2 < 2\,\GeV^2/c^4$. There is a
    striking modulation in the $\chi$ angle distribution due to the
    right-handed BSM physics contribution as well as clear left-handed BSM
    signatures at low $q^2$ and in $\cos\theta_K$.}
\end{figure*}

Sensitivity in the di-electron mode at low $q^2$, $q^2< 2\,\GeV^2/c^4
$, is critical as the new physics terms interfere with the photon
pole. Even moderate non-zero $C_7^{\prime}$ (right-handed BSM physics)
produces a prominent modulation in the $\chi$ distribution, while
left-handed BSM physics ($\delta C_7$) signatures appear at low $q^2$
and in the $\cos\theta_K$ distribution.  These BSM physics signatures
are shown in Fig.~\ref{fig:c7_cp7_lowq2_angles}.

The SM NNLO value of $|C_7|$ is $\approx 0.3$ so we expect, according
to our study, to constrain the real and imaginary parts of $C_7$ to
better than 2.5 and 7.5 \%, respectively from the di-muon mode
alone. The contribution from BSM physics in the form of right-handed
currents, which corresponds to $C_7^{\prime}$, is expected to be
constrained to better than 6 \% of $C_7$ from the di-muon mode
alone. In the di-electron mode where we have better access to the
photon pole we might constrain the real and imaginary parts of $C_7$
to better than 1.5 and 6.5 \%, respectively and $C_7^{\prime}$ will be
constrained to better than 3 \% of $|C_7|$. The distributions for these
fits and the resulting summary table are given in Appendix~\ref{app:figs}.

\clearpage
\section{Disentangling QCD and resonance effects from BSM Physics in $B \to K^* \ell^+ \ell^-$\label{sec:qcd}}
The semileptonic decays $B\to K^* \ell^+ \ell^-$ are expected to have
smaller theoretical uncertainties than purely hadronic $B$ decays.
However, these decay modes still require QCD corrections, which are
not easily calculable.  The experimental final
state $K^* \ell^+ \ell^-$ includes contributions from $c\bar{c}$
resonances such as $B \to J/\psi K^*$, $B \to \psi(2S) K^*$ as well as
other broader $c\bar{c}$ resonances with higher mass, where the
$c\bar{c}$ resonances decay to $\ell^+ \ell^-$. It is well known that
these $b\to c \bar{c} s$ modes have non-factorizable contributions
\cite{Belle:2002otd, BaBar:2007rbr, LHCb:2013vga}. In addition,
complications arise due to non-local contributions from
electromagnetic corrections to purely hadronic
operators~\cite{Beneke:2001at}, which cannot be calculated from first
principles. Only limited success has been achieved in computing the
effects of charm-loop contributions \cite{Khodjamirian:2010vf}. Given
the difficulties in reliably computing all the QCD effects, our choice
of $\Delta$ observables\cite{B2TIPbook} offers the only existing
approach to obtain sensitivity to BSM physics.

The global fits assume that BSM effects depend on lepton flavor and
are present in the muon mode but are negligible in the electron
mode. This is required to explain the deviation in $R_K$ and $R_{K^*}$
from the SM. The SM hadronic uncertainties due to the strong
interaction are the same in the both modes, and thus should cancel to
first order in the difference so that only BSM contributions remain.

We test the assumption that hadronic uncertainties affect the di-muon
and di-electron modes equally, and cancel in the difference.  We fit
to pseudo-experiments with MC statistics roughly matched to the
expectation for future Belle II experimental data. The MC data are
generated using nominal ABSZ form factors~\cite{Bharucha:2015bzk}. In
each fit the hadronic form factors in the unbinned likelihood function
are varied, but taken to be the same for the di-muon and di-electron
modes. Specifically, we randomly pick the parameters of the
$z$-expansion in the ABSZ form factors within the uncertainties
defined by the correct covariance matrices and thus automatically take
into account correlations between parameters~\cite{Bharucha:pc}.

\begin{figure}[h!]
  \includegraphics[width=0.495\columnwidth]{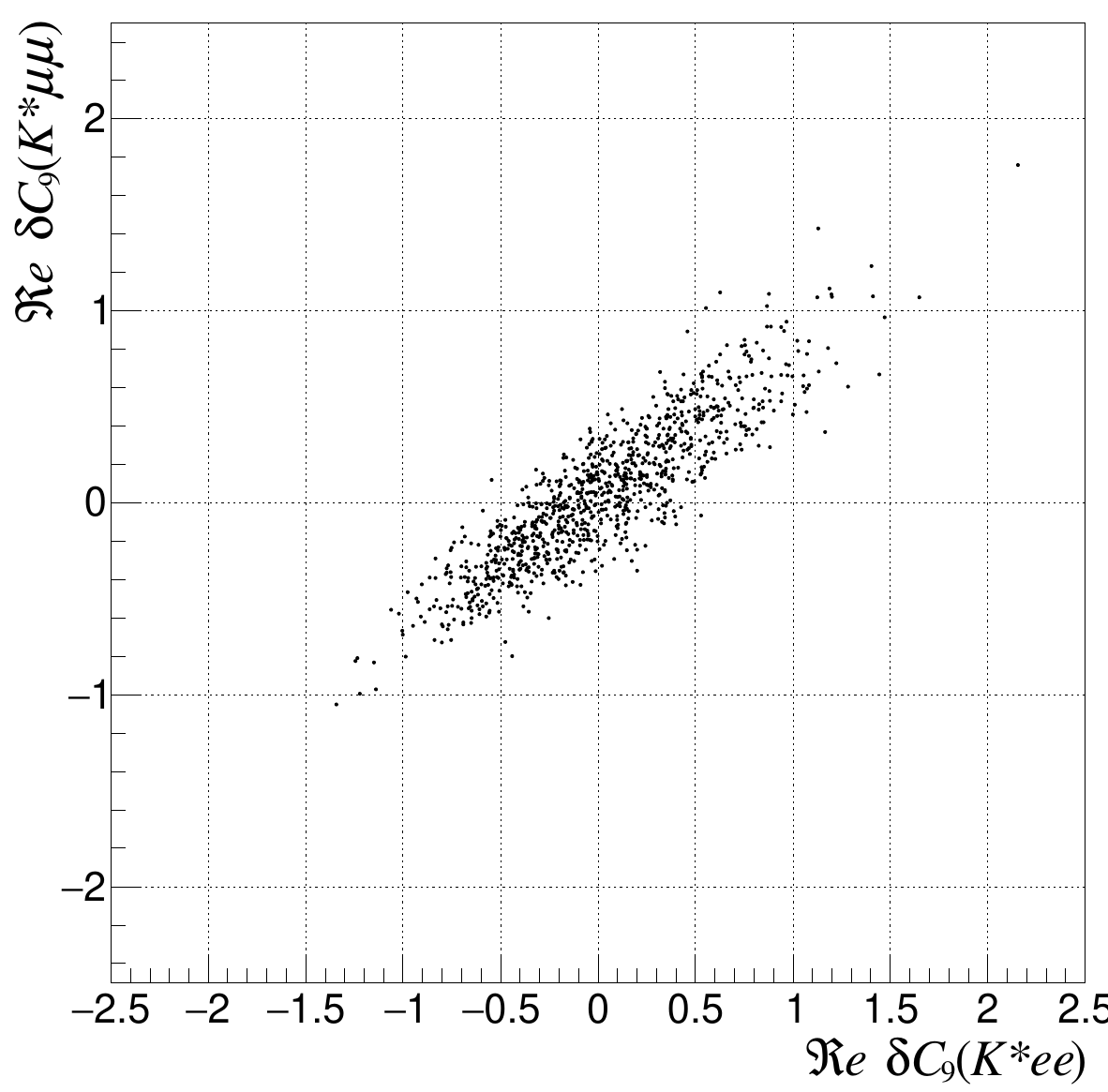} 
  \includegraphics[width=0.495\columnwidth]{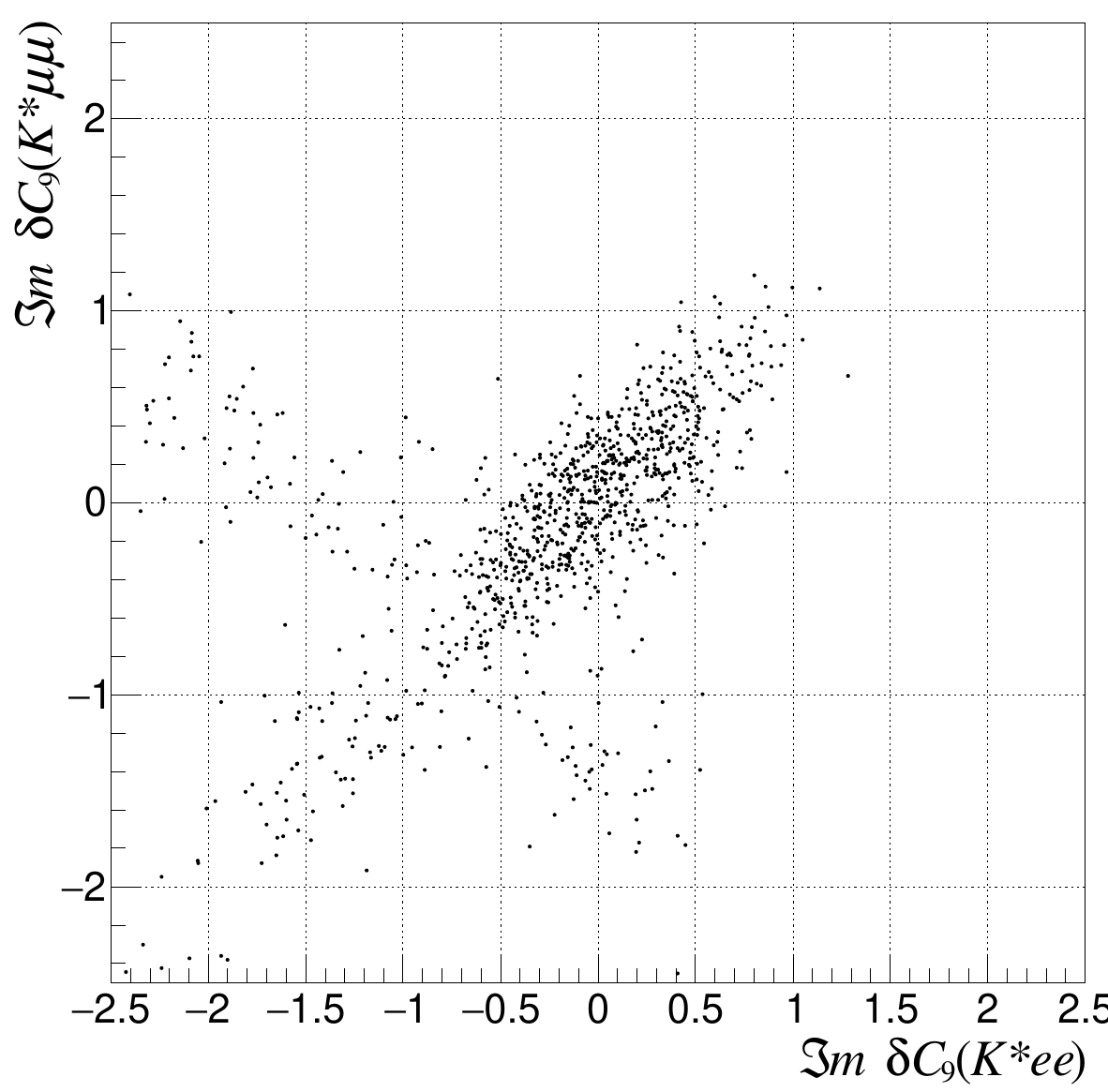}
  \caption{\label{fig:dc9ffrnd} Result of 4-D unbinned likelihood fits
    for the $\delta C_9$ variable using both $B\to K^* e^+e^-$ and
    $B\to K^* \mu^+\mu^-$ MC data. Hadronic form factors are varied
    within their uncertainties.  The fits are performed in the full
    $q^2$ range without including resonances. Note that the bias on the
    fitted $\delta C_9$ values, resulting from form factor
    uncertainties, is strongly correlated for the electron and muon
    modes.}
\end{figure}

\begin{figure}[h!]
  \includegraphics[width=0.495\columnwidth]{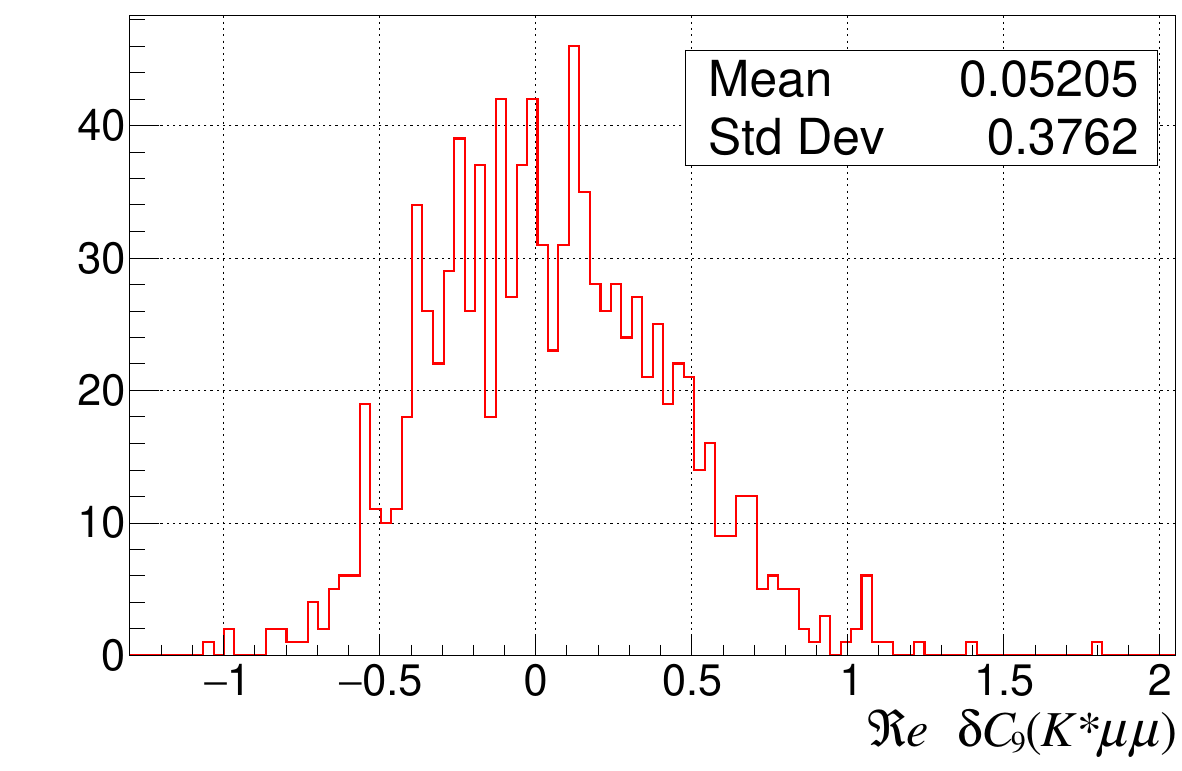}
  \includegraphics[width=0.495\columnwidth]{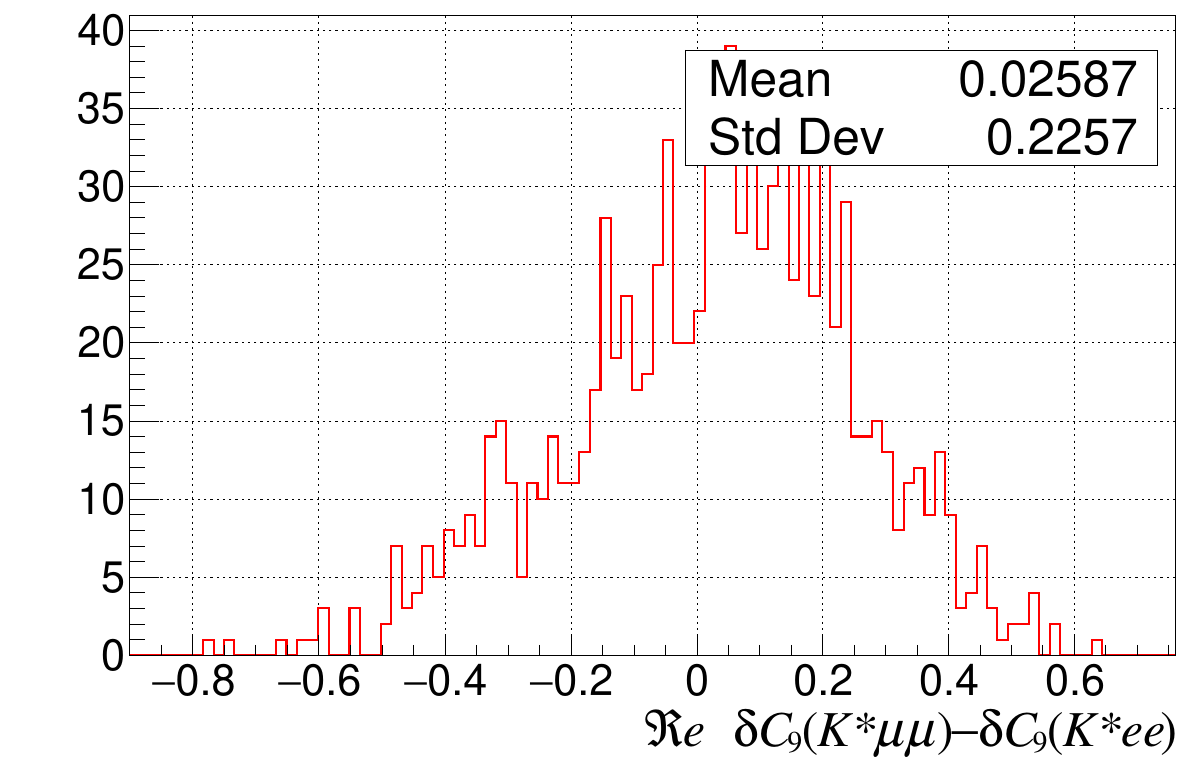}
  \caption{\label{fig:dc9ffrndproj} Distribution of the real part of
    the $\delta C_9$ coefficient obtained from fits to $B\to K^*
    \mu^+\mu^-$ events (left plot) and the differences between those
    values and corresponding results for the $B\to K^* e^+e^-$ mode
    (right plot) for randomly chosen hadronic form factors within the
    uncertainties as shown in Fig.~\ref{fig:dc9ffrnd}. Note that
    uncertainties are reduced by subtracting the fit results for the
    two modes. The remaining uncertainty is due to limited statistics
    in each fit.}
\end{figure}

An example result is shown in Fig.~\ref{fig:dc9ffrnd} for the $\delta
C_9$ variable, where a clear correlation between $\delta C_9$
extracted in the muon and electron modes is seen. The $\delta C_9$
distribution for the di-muon mode as well as the difference with the
di-electron mode is shown in Fig.~\ref{fig:dc9ffrndproj}. The average
value of the difference is close to zero and its standard deviation is
about 40\% less than for the di-muon mode alone, thus demonstrating
the cancellation of hadronic uncertainties.

\begin{figure}[h!]
  \includegraphics[width=0.495\columnwidth]{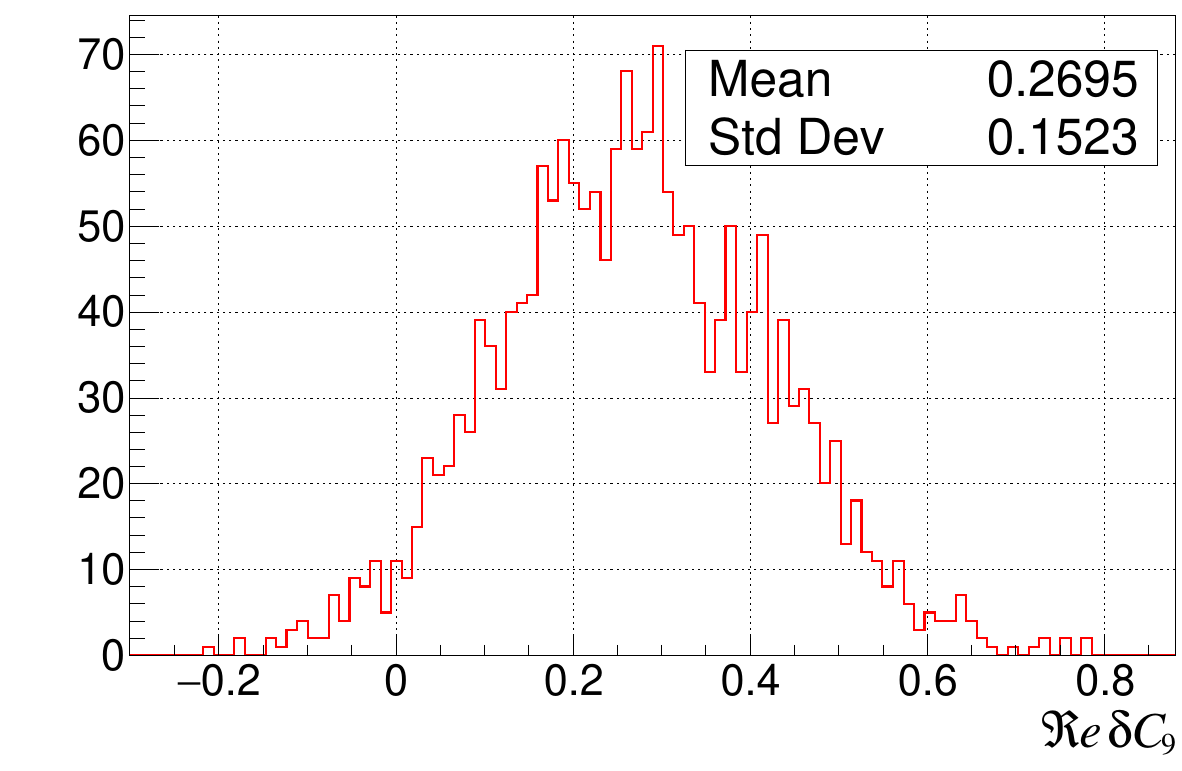} 
  \includegraphics[width=0.495\columnwidth]{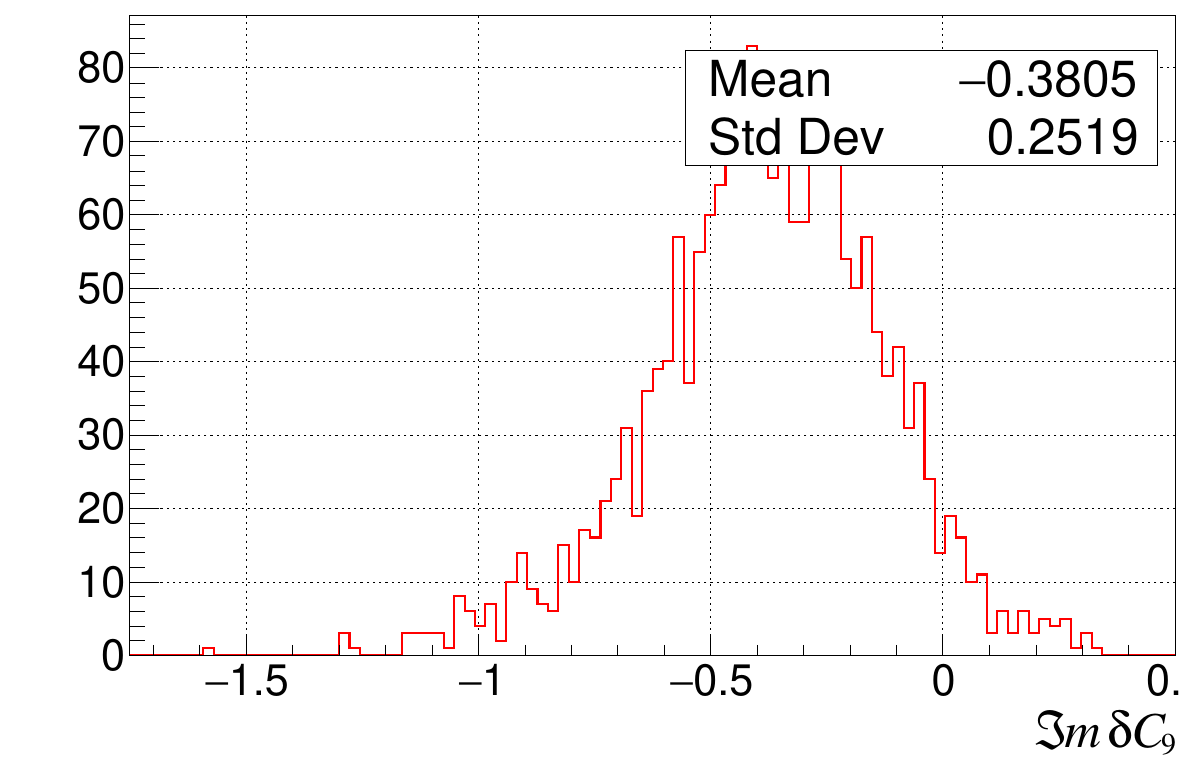}
  \caption{\label{fig:dc9mmffv1} Distribution of $\delta C_9$ obtained
    from unbinned likelihood fits to simulated $B\to K^* \mu^+\mu^-$
    SM decays with a randomly selected but fixed set of 7 hadronic form factors,
    consistent with the SM within theoretical uncertainties. The shift
    of the mean (0.27 in $\Re ~\delta C_9$) from zero cannot be
    distinguished from BSM physics using this $B\to K^*\mu^+\mu^-$ final state
    alone.}
\end{figure}

To illustrate the magnitude of possible bias on fitted Wilson
coefficients from ABSZ hadronic form factor uncertainties, we again
perform pseudo-experiments with form factor parameters in the
likelihood function that differ from the nominal parameters used to
generate the fitted MC data. This time, however, we use the same
parameters for each fit, so that the bias does not average out over
multiple experiments. In Fig.~\ref{fig:dc9mmffv1} the $\delta C_9$
coefficient extraction is shown as an example. We find that the
systematic bias in this coefficient is nearly twice as large as the
statistical uncertainty of the fit. With the expected amount of Belle
II data, SM theoretical uncertainties might mimic BSM physics effects.

\begin{figure}[h!]
  \includegraphics[width=0.95\columnwidth]{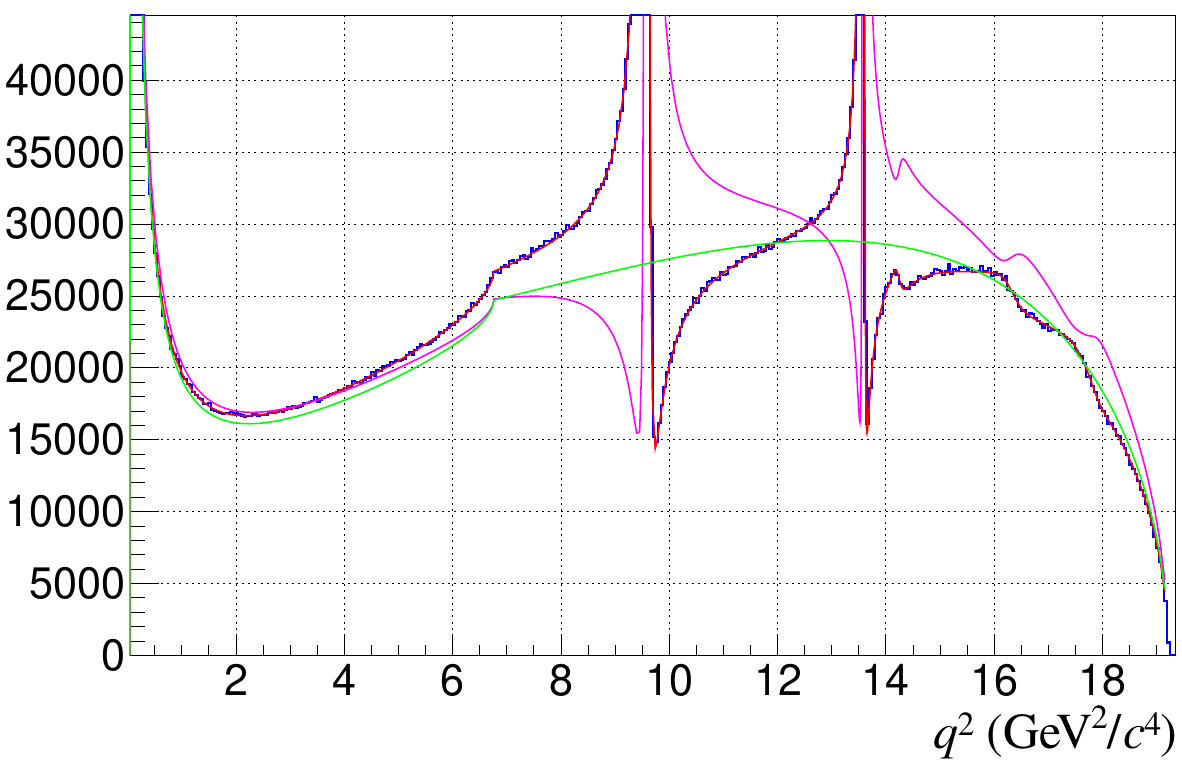} 
  \caption{\label{fig:ab0mmq2reso} The $q^2$ distribution of
    $\bar{B}\to \bar{K}^*\mu^+\mu^-$ decay in the presence of
    $c\bar{c}$ resonances. The histogram is the result from the
    \evtgen generator, the green curve shows the result of the
    likelihood integration without resonances, and the red(magenta)
    curve is the result of the likelihood integration with the strong
    phase 0($\pi$) when resonances are included. The contribution of
    these resonances (and non-factorizable effects) will be a limiting
    uncertainty in the extraction of BSM Wilson coefficients from
    $B\to K^* \mu^+ \mu^-$.}
\end{figure}

Resonance effects are taken into account by modifying the Wilson
coefficient $C_9$ according to Eq.~\ref{eq:hfuncreso} both in the
\evtgen event generator and in the likelihood
function. Figure~\ref{fig:ab0mmq2reso} shows the $q^2$ distribution
for $\bar{B}\to \bar{K}^*\mu^+\mu^-$ decays. We find that resonances
affect $q^2$ regions that extend well beyond the resonance widths, and
thus rather large veto windows are required to mitigate their
presence.

\begin{figure*}[h!]
  \includegraphics[width=0.495\columnwidth]{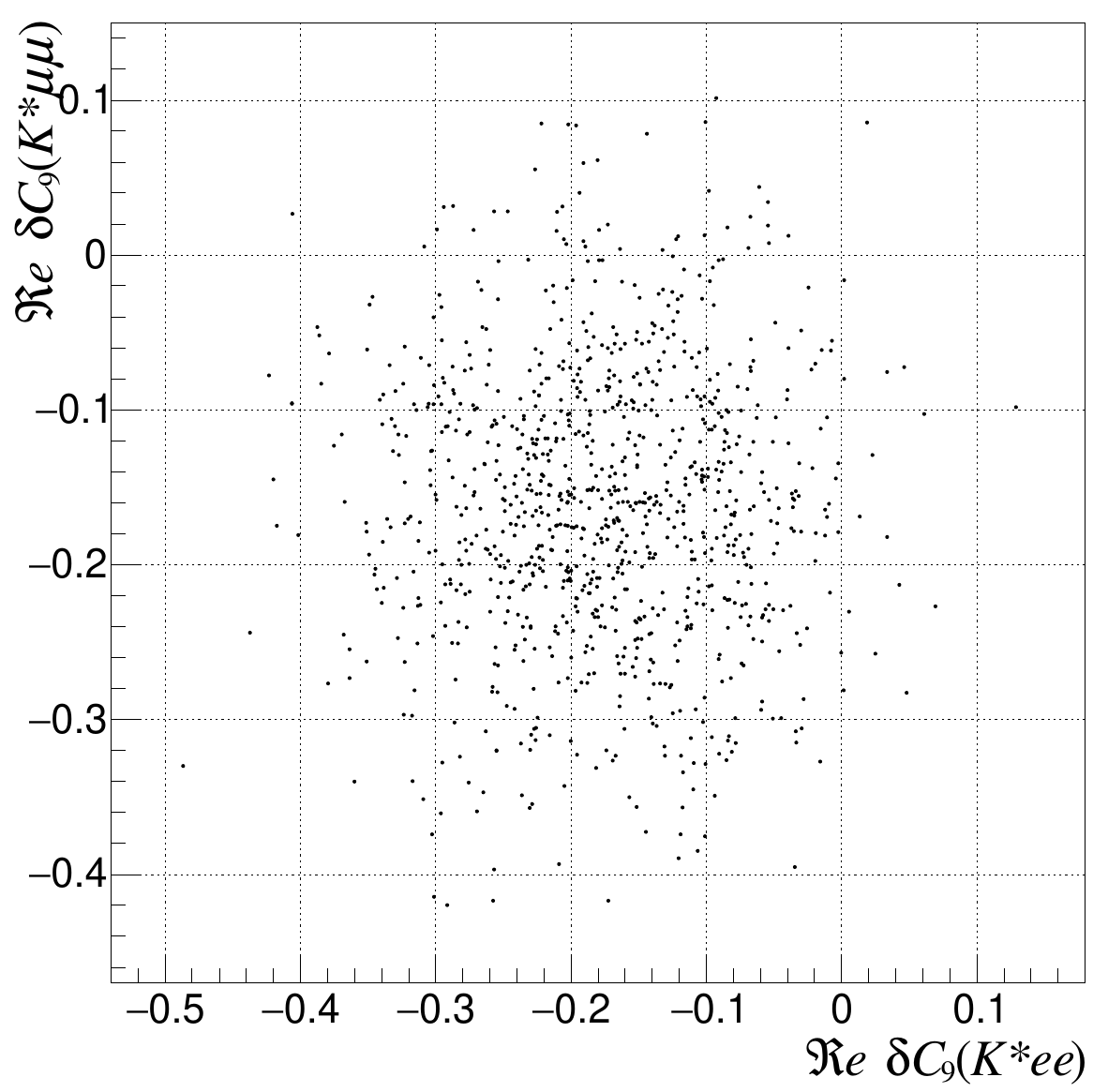} 
  \includegraphics[width=0.495\columnwidth]{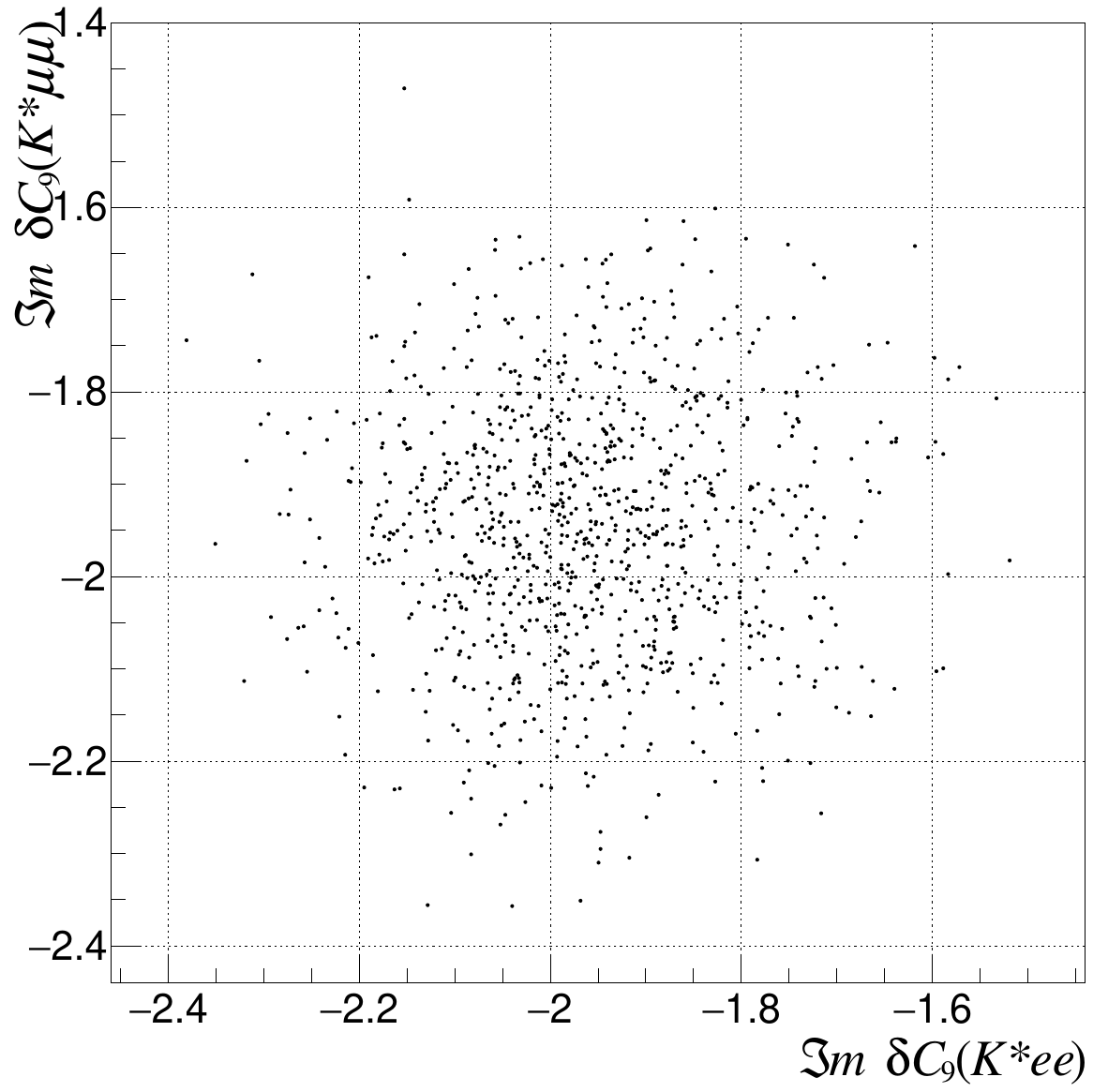} \\
  \includegraphics[width=0.495\columnwidth]{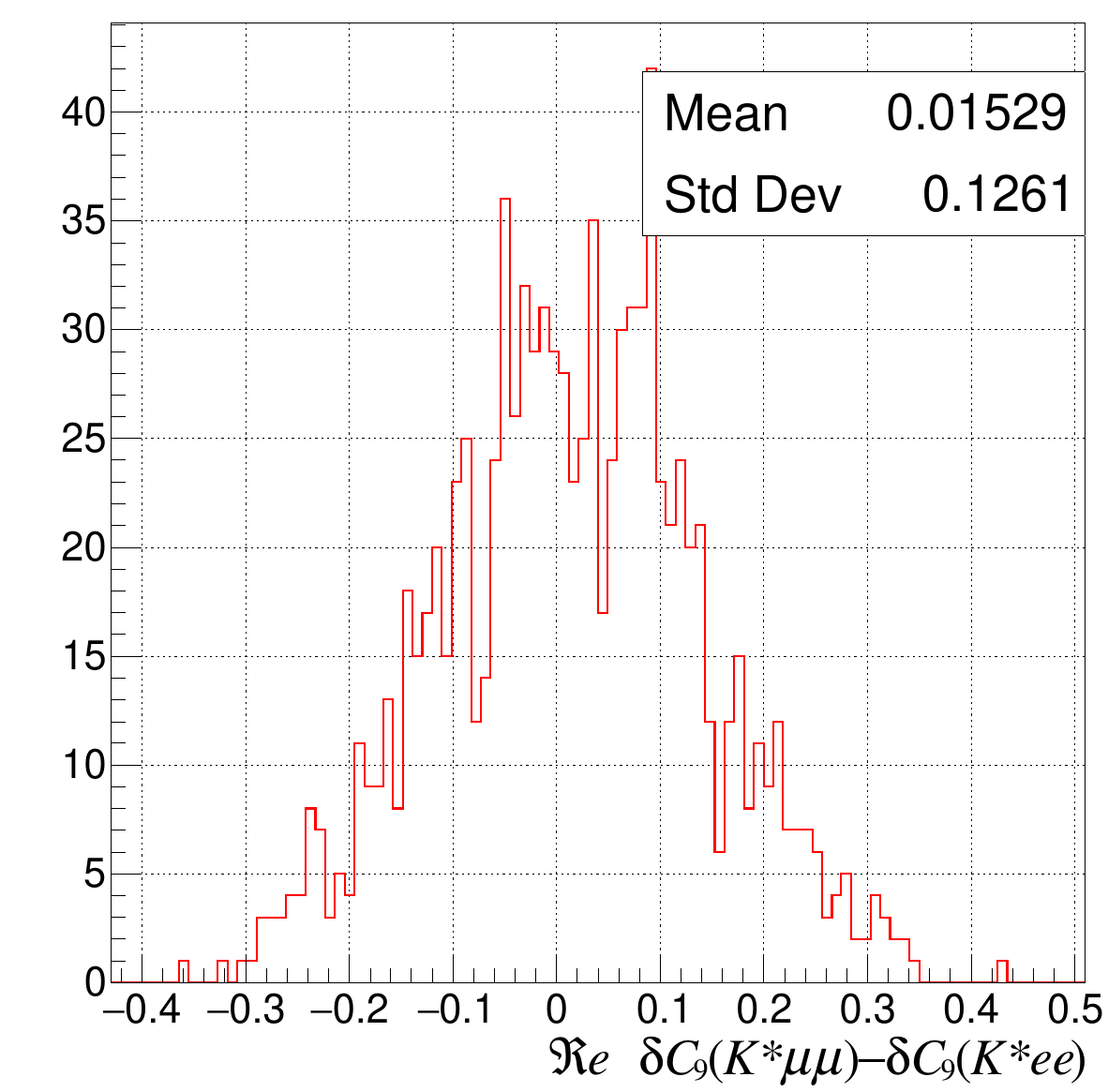} 
  \includegraphics[width=0.495\columnwidth]{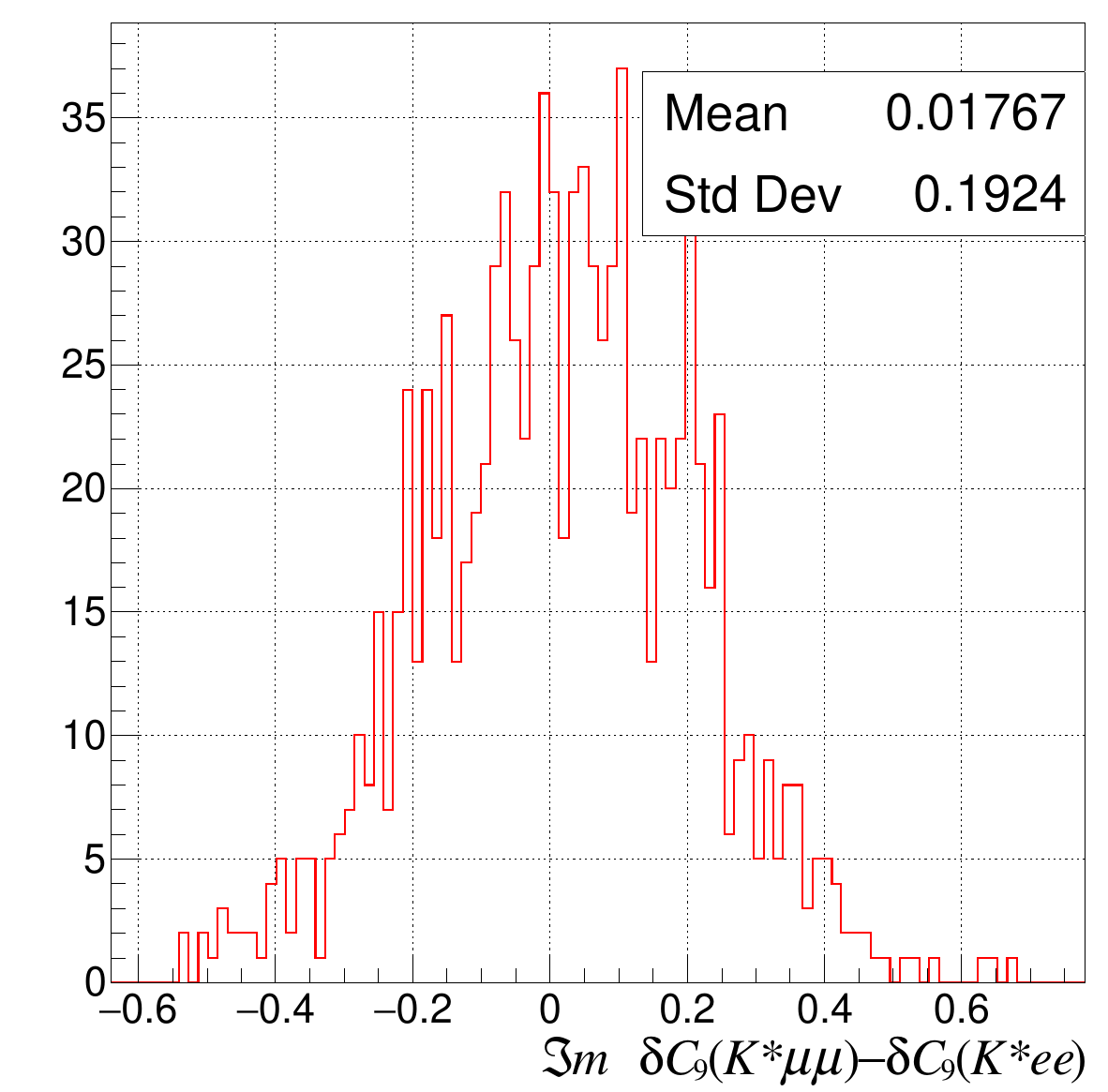} \\
  \caption{\label{fig:diffdc9_2res} A test of the resonance effects
    in the fit for the $\delta C_9$ parameter. The top row shows
    correlation plots for the real and imaginary parts of $\delta
    C_9$. The bottom row shows the difference between modes.}
\end{figure*}

To illustrate how the resonances might affect the $C_9$ extraction
even with large veto windows, we perform likelihood fits for the
region $q^2 \in [1,8]\cup[11,12.5]\cup[15,q^2_\text{max}]$. For
simplicity we use a likelihood function with two resonances only:
$J/\psi$ and $\psi(2S)$.  SM MC data is generated without
resonances. With the above $q^2$ selection applied, the average number
of $K^*\mu^+\mu^-$ and $K^*e^+e^-$ decays in each fit is 4850. The
total branching fraction is constrained with 5~\% precision to the SM
prediction. The results of the fits are shown in
Fig.~\ref{fig:diffdc9_2res} (top row). The effect of the resonances is
significant, biasing the fitted $\delta C_9$ values away from
zero. This effect is particularly large for the imaginary part, where
the bias is substantially larger than the SM value shown in
Fig.~\ref{fig:C9}. The $\Delta$ variables shown
Fig.~\ref{fig:diffdc9_2res} (bottom row), however, behave well and are
centered close to zero with uncertainties much smaller than the original bias
due to the resonances. This again suggests that possible BSM
contributions to the muon mode can be extracted even in the presence
of large theoretical uncertainties by using $\Delta $ observables.

The remaining uncertainty then becomes statistical, rather than being
limited by hadronic uncertainties. This highlights the importance of
the technique utilizing the $\Delta$ variables, and the importance of
large experimental statistics for both muon and electron modes.

\clearpage

\section{Sensitivity prospects with Belle II data}

\begin{figure}[h!]
  \includegraphics[width=0.495\columnwidth]{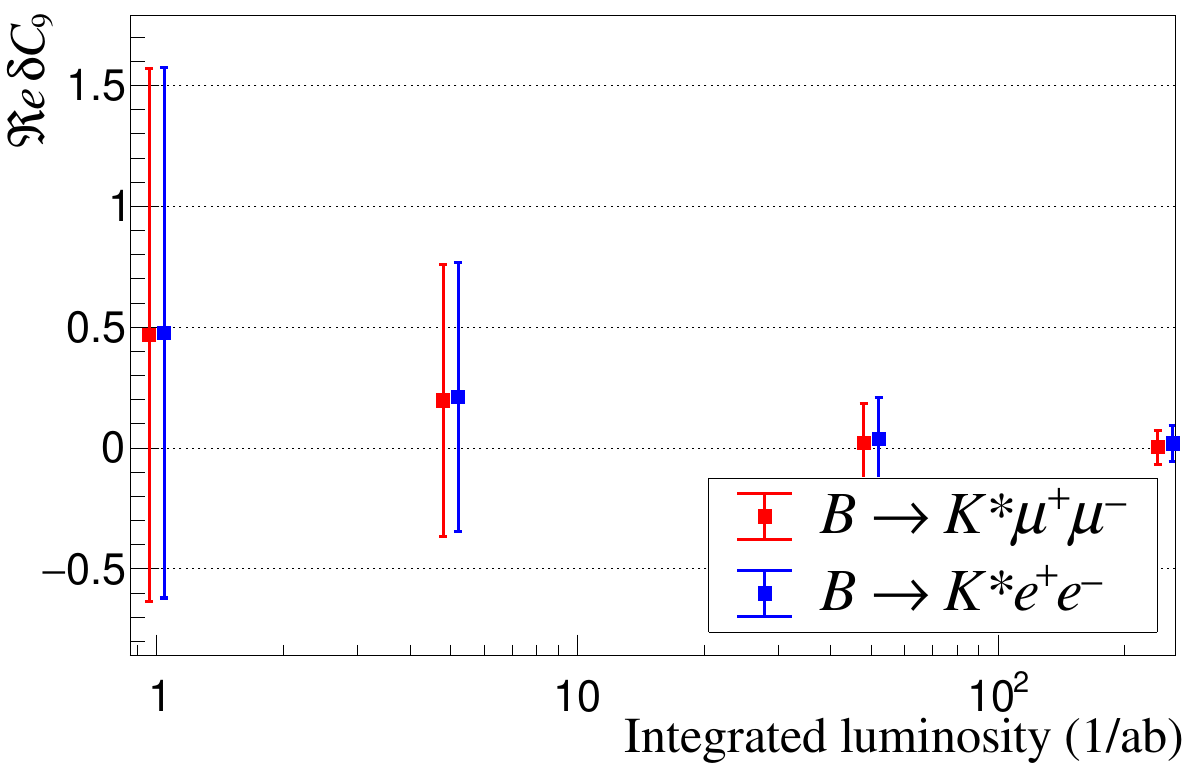}
  \includegraphics[width=0.495\columnwidth]{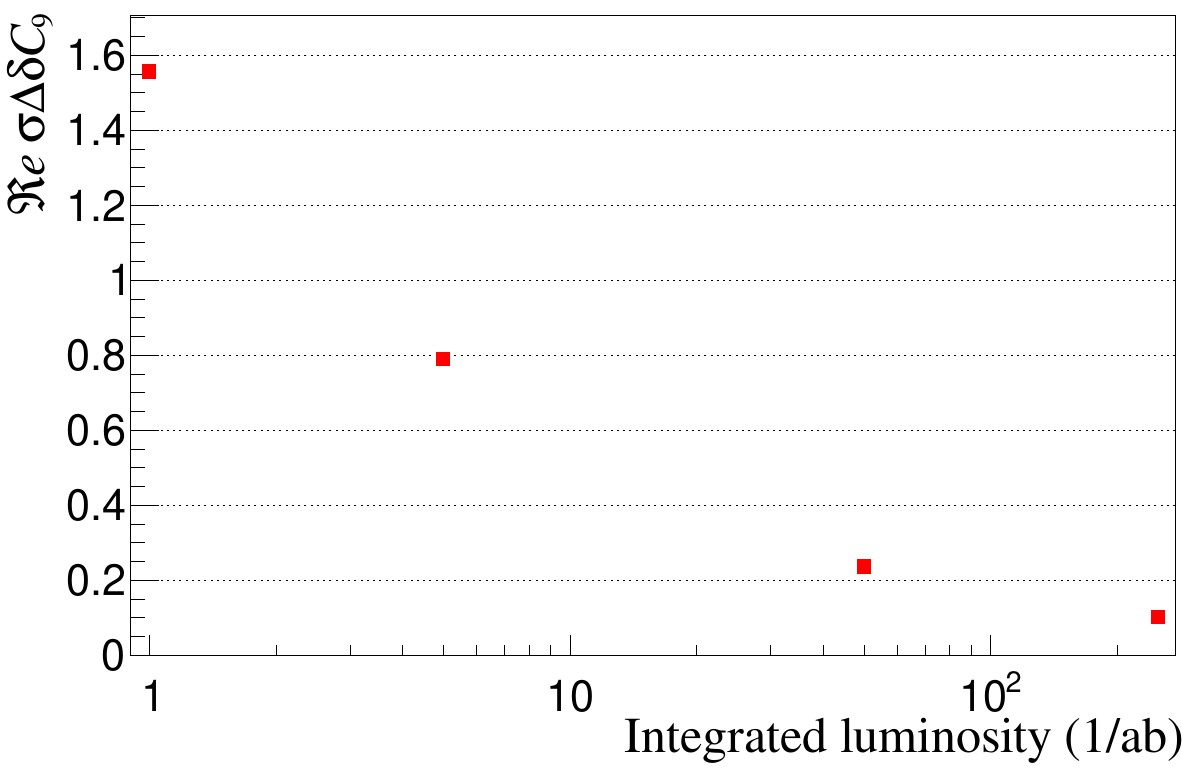}
  \caption{\label{fig:dc9proj} The expected constraint on the $\delta C_9$
    Wilson coefficient based on the unbinned maximum likelihood fit to
    MC data corresponding to various Belle II luminosity milestones. The
    mean value and its uncertainty (error bars) are shown in the left
    plot for the real part of the coefficient. The plot on the right
  shows the total uncertainty in the $\Delta C_9$ variable.}
\end{figure}

\begin{figure}[h!]
  \includegraphics[width=0.495\columnwidth]{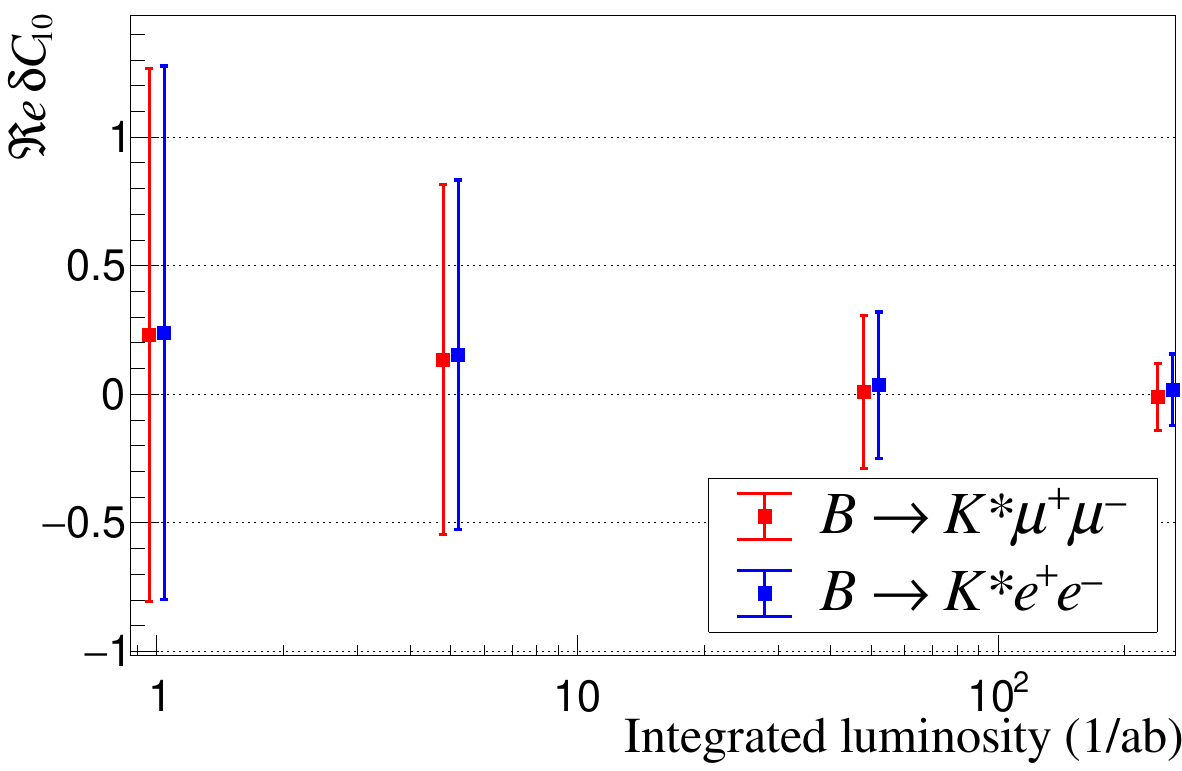}
  \includegraphics[width=0.495\columnwidth]{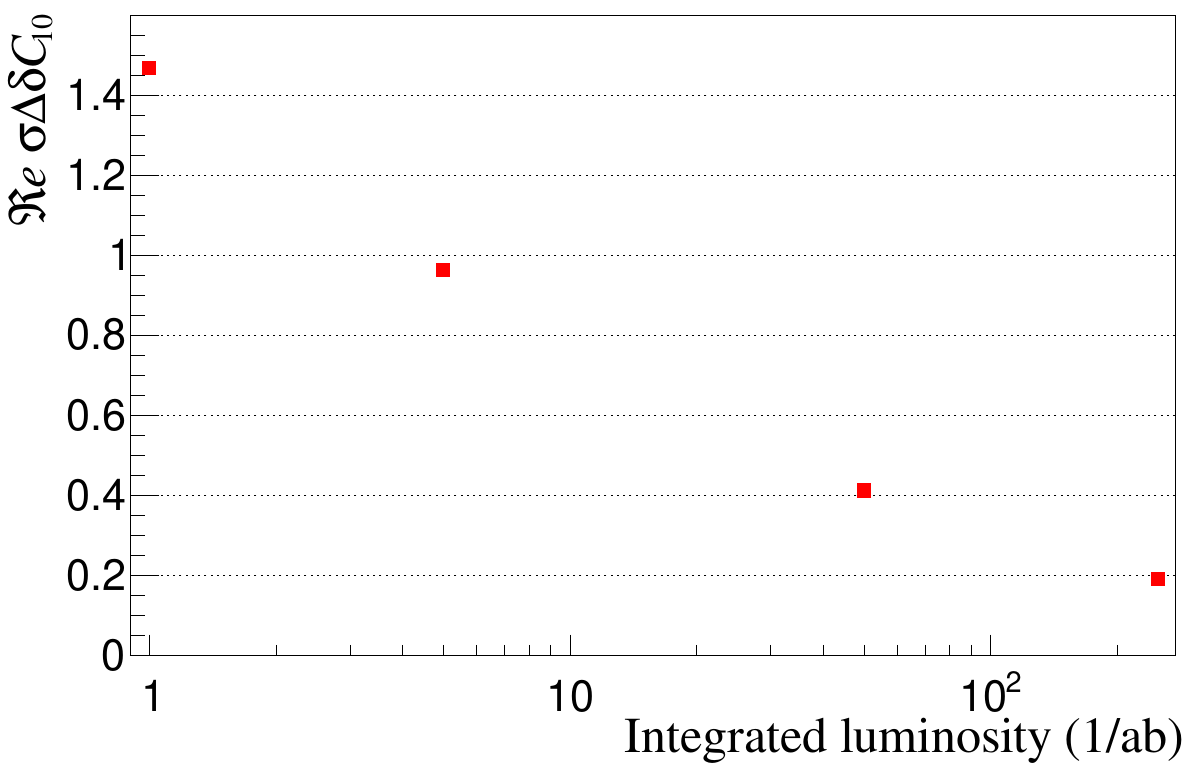}
  \caption{\label{fig:dc10proj} The expected constraint on the $\delta C_{10}$
    Wilson coefficient based on the unbinned maximum likelihood fit to
    MC data corresponding to the Belle II luminosity milestones. The
    mean value and its uncertainty (error bars) are shown in the left
    plot for the real part of the coefficient. The plot on the right
    shows the total uncertainty in the $\Delta C_{10}$ variable.}
\end{figure}

\begin{figure}[h!]
  \includegraphics[width=0.495\columnwidth]{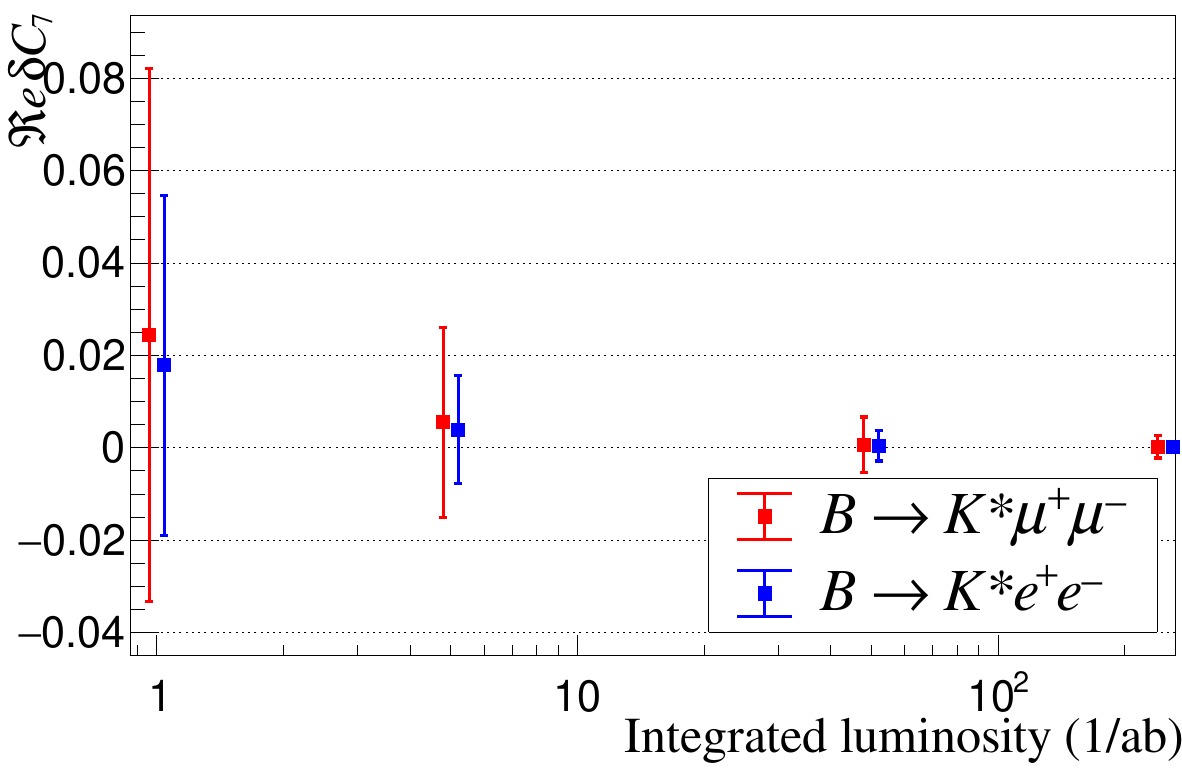}
  \includegraphics[width=0.495\columnwidth]{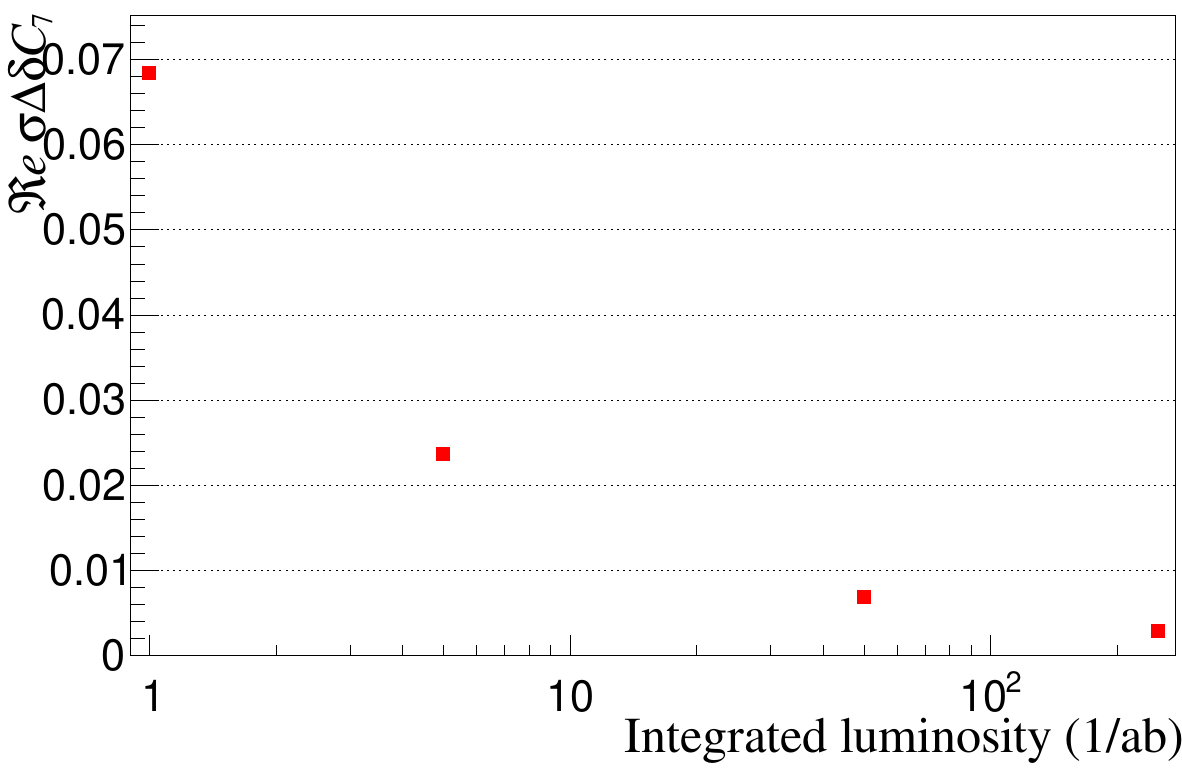}
  \caption{\label{fig:dc7proj} The expected constraint on the $\delta
    C_7$ Wilson coefficient based on unbinned maximum likelihood fits
    to MC data corresponding to various Belle II luminosity
    milestones. The mean value and its uncertainty (error bars) are
    shown in the left plot for the real part of the Wilson
    coefficient. The plot on the right shows the total uncertainty in
    the $\Delta C_7$ variable.}
\end{figure}

\begin{figure}[h!]
  \includegraphics[width=0.495\columnwidth]{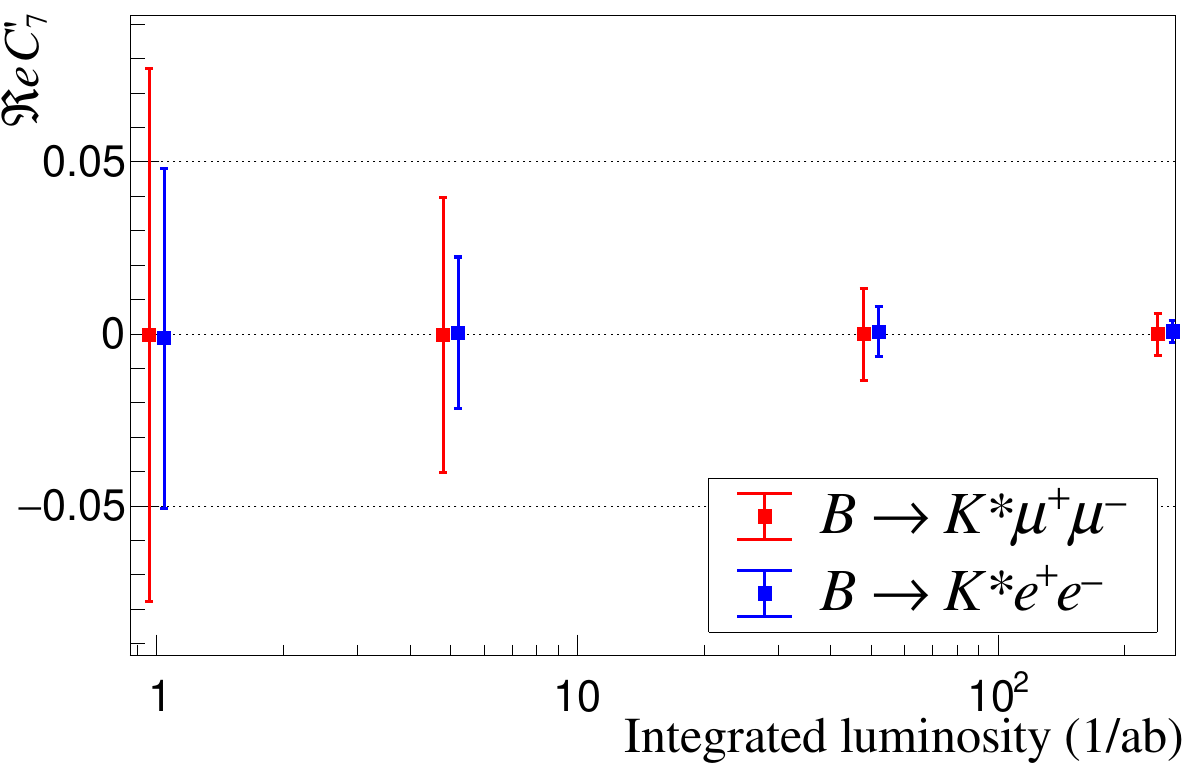}
  \includegraphics[width=0.495\columnwidth]{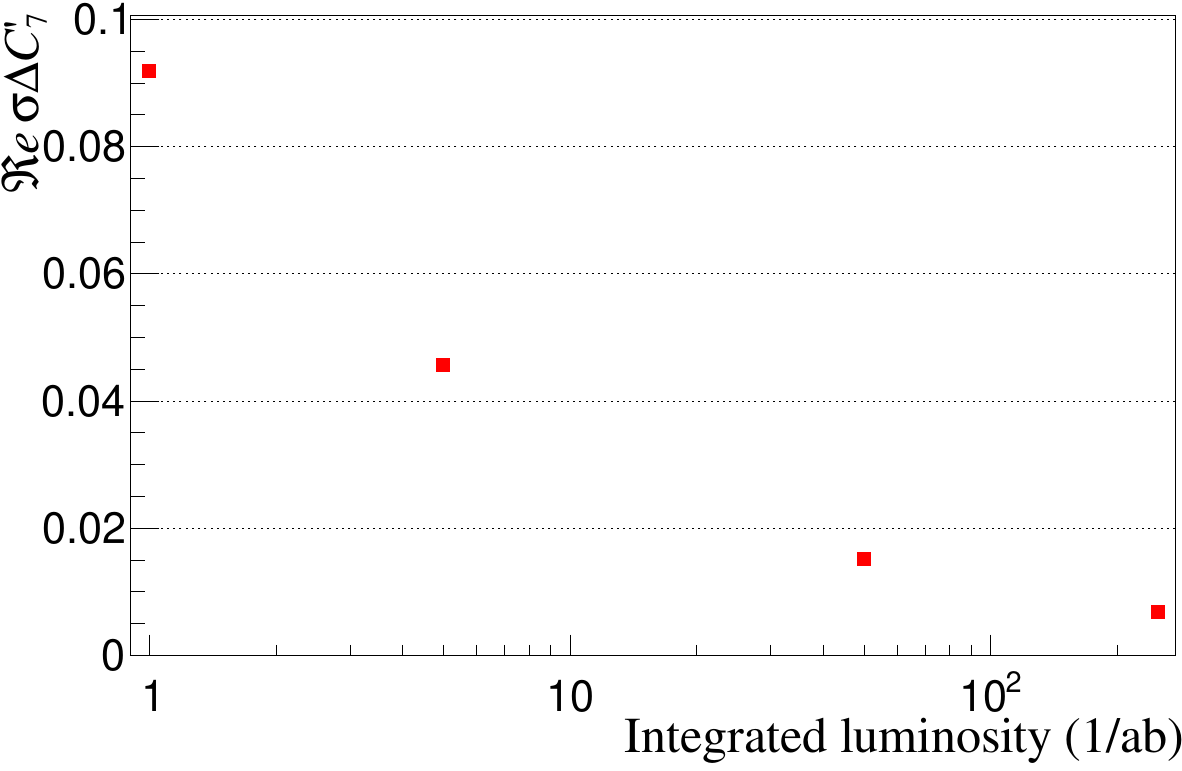}
  \caption{\label{fig:cp7proj} The expected constraint in the $C_7'$
    Wilson coefficient based on unbinned maximum likelihood fits to MC
    data corresponding to various Belle II luminosity milestones. The
    mean value and its uncertainty (error bars) are shown in the left
    plot for the real part of the coefficient. The plot on the right
    shows the total uncertainty in the $\Delta C_7'$ variable.}
\end{figure}

To estimate possible statistical sensitivity to BSM physics, we perform fits to
generated data corresponding to several Belle II luminosity
milestones.  The generated data include contributions of $c\bar{c}$
resonances and thus we exclude the region $\pm 0.25\ \text{GeV}^2/c^4$
around the $J/\psi$ and $\Psi(2S)$ resonances from the fit data. In
addition, for $C_9$, $C_9'$, $C_{10}$, and $C_{10}'$ coefficients the
region $q^2 < 1\, \text{GeV}^2/c^4$ is excluded to remove the photon
pole effects whereas the fits to extract the $\delta C_7$ and $C_7'$
coefficients still use the photon pole data. The reconstruction
efficiency is assumed to be about 25\% based on the Belle results. The
result of the fits are graphically shown in Fig.~\ref{fig:dc9proj},
\ref{fig:dc10proj}, \ref{fig:dc7proj}, and \ref{fig:cp7proj}.

We find that some fits are biased with low statistics albeit the bias
is still significantly smaller than the statistical uncertainty and
rapidly decreases with the integrated luminosity. As expected, the
bias is the same both in the di-electron and di-muon modes and thus
cancels in the $\Delta$ observable.

\clearpage
\section{Conclusions}
We have upgraded the widely used MC event generator \evtgen to model
$B\to K^* \ell^+ \ell^-$ with improved SM decay amplitudes and
amplitudes for possible BSM physics contributions, implemented in the
operator product expansion in terms of Wilson coefficients.  This
newly developed event generator model, the `Sibidanov Physics
Generator', implemented in the \evtgen framework, has been used to
investigate the potential experimental signatures of
lepton-flavor-violating BSM physics signals in $B\to K^*\ell^+
\ell^-$.  We describe the advantages and illustrate the potential of
the new physics generator model, which implements the most general BSM
contributions. This new generator can improve sensitivity of BSM
experimental searches, and clarify BSM physics signatures. Advantages
for experimental work include properly simulating BSM scenarios,
interference between SM and BSM amplitudes, and correlations between
various BSM signatures as well as acceptance bias. To address SM
uncertainties, due to the strong interaction, which might mimic BSM
effects, we have implemented $b\to c \bar{c} s$ resonant contributions
and up-to-date SM hadronic form factors~\cite{Bharucha:2015bzk}.

In addition to these modifications to \evtgen, we also reviewed and
optimized the computational efficiency of \evtgen for $B\to
K^*\ell^-\ell^+$.  We achieved a cumulative performance gain of two
orders of magnitude. 

We have shown that directly using the correlation between angular
observables improves experimental statistical sensitivity.  For
example, in the case of BSM $C_7$, $C_9$, or $C_{10}$ contributions,
there must be correlated signatures between $A_{FB}(q^2)$ and
$S_5(q^2)$ angular asymmetries. The correlation in
Eq.~\eqref{eq:ratio} depends on the Wilson coefficients. Therefore,
the correlation improves the direct determination of Wilson
coefficients.  The use of a four-dimensional unbinned maximum
likelihood fit to data naturally takes this correlation into account
and improves the experimental sensitivity to BSM physics compared to
independent binned fits to angular asymmetries. Note that theoretical
fits to binned data reported in literature also take into account this
correlation.
  
QCD uncertainties due to form factors and resonances currently limit
the sensitivity of fits for BSM physics in $B\to K^* \ell^+\ell^-$. We
have demonstrated the experimental feasibility of an approach using
$\Delta$-observables, which allow one to unambiguously distinguish
between hadronic effects and BSM physics in the case of LFV BSM
physics.  The $\Delta$-observables appear ideally suited for
experimental analysis in Belle II with the large, 50 ab$^{-1}$, data
sets expected in the next decade.  For instance, $\Delta C_9$ will be
constrained to better than 3\%.  We find that Belle II also should
have excellent statistical sensitivity to BSM physics in $C_7$ and
$C_7'$, which appear at low $q^2$ in the di-electron channel. For
example, in the di-electron mode $C_7^{\prime}$ should be constrained
to 3\% of $C_7$.  A new MC generator for direct simulation of Wilson
coefficients in $B\to K^* \ell^+\ell^-$ will be a useful tool for
experimenters measuring these decays and determining the acceptance
corrections for BSM physics. We demonstrated an application for
lepton-flavor violating BSM physics. Further work will use this MC
generator for experimental feasibility studies of ML (Machine
Learning) based approaches to $C_i$ determination as well as to more
difficult BSM scenarios with LFU (Lepton Flavor Universal) couplings
in which the effects of interference with $B\to J/\psi K^*$ must be
simulated and constrained.

Note Added: This paper is a significantly modified and improved
version of Ref.~\cite{Sibidanov:2022gvb}, which was submitted to the
US Community Summer Study on the Future of Particle Physics (Snowmass
2021).  Ref.~\cite{Sibidanov:2022gvb} will not, however, appear in the
Snowmass proceedings.  Improvements in this paper include, but are not
limited to, calculations of correlations between several experimental
observables and discussions of prospects for BSM-sensitive observables
with several benchmark values of Belle II integrated luminosity.

\section{Acknowledgments}
We thank Emi Kou (IJCLab) for useful discussions concerning
correlations between angular observables.  T.E.B., S.D., S.K., A.S.,
and S.E.V. thank the DOE Office of High Energy Physics for support
through DOE grant DE-SC0010504. R.M. acknowledges the support of SERB
Grant SPG/2022/001238. S.S. acknowledges the support of SERB Grant
SRG/2022/001608.

\appendix
\section{\label{App:theory}Description of the theoretical framework}

The matrix element introduced in Eq.~\ref{eq:matrixelement} can
further be extended via the inclusion of four dimension six operators
namely
$$\mathcal{O}_S^{(\prime)}= (\bar{s}P_{R(L)}b) (\bar{\mu}\mu)~~{\rm
  and}~~\mathcal{O}_P^{(\prime)}= (\bar{s}P_{R(L)}b)
(\bar{\mu}\gamma_5\mu) $$ with the corresponding Wilson coefficients
$C_S^{(\prime)}$ and $C_P^{(\prime)}$ which can arise in presence of
BSM physics.  The SM prediction for $C_i$ is known to NNLL accuracy and in the
amplitude for $B\to K^*(\to K\pi)\ell^+\ell^-$ certain combinations of
these Wilson coefficients appear, which are given
by~\cite{Altmannshofer:2008dz}
\begin{eqnarray}
  C_7^{\rm eff} & = & 
  C_7 -\frac{1}{3}\, C_3 -\frac{4}{9}\, C_4 - \frac{20}{3}\, C_5\, -\frac{80}{9}\,C_6\,,
  \nonumber\\
  C_9^{\rm eff} & = & C_9 + Y(q^2)\,,{\rm with}\nonumber\\
  Y(q^2) & = & h(q^2,m_c) \left( \frac{4}{3}\, C_1 + C_2 + 6 C_3 + 60 C_5\right)
  \nonumber\\
  & & {}-\frac{1}{2}\,h(q^2,m_b) \left( 7 C_3 + \frac{4}{3}\,C_4 + 76 C_5
  + \frac{64}{3}\, C_6\right)
  \nonumber\\
  & & {}-\frac{1}{2}\,h(q^2,0) \left( C_3 + \frac{4}{3}\,C_4 + 16 C_5
  + \frac{64}{3}\, C_6\right)
  \nonumber\\
  & & {} + \frac{4}{3}\, C_3 + \frac{64}{9}\, C_5 + \frac{64}{27}\,
  C_6\,. \label{eq:Yfun}
\end{eqnarray} 
The function
\begin{eqnarray}
  \label{eq:hfunc}
  \hspace*{-0.4cm}&&h(q^2,m_q)  = -\frac{4}{9}\, \left( 
  \ln\,\frac{m_q^2}{\mu^2} - \frac{2}{3} 
  - z \right)\qquad{  }\nonumber\\ 
  &&\quad{}- \frac{4}{9}\,(2+z) \sqrt{|z-1|} \times 
  \left\{
  \begin{array}{l@{\quad}l}
    \displaystyle\arctan\, \frac{1}{\sqrt{z-1}} & z>1\\[10pt]
    \displaystyle\ln\,\frac{1+\sqrt{1-z}}{\sqrt{z}} - \frac{i\pi}{2} & z \leq 1
  \end{array}
  \right.
\end{eqnarray}
with $z=4 m_q^2/q^2$, is related to the basic fermion loop. The
limiting function $h(q^2,0)$ is given by
\begin{equation}
  \label{eq:hfunclim}
  h(q^2,0) = \frac{8}{27} + \frac{4}{9} \left( \ln\frac{\mu^2}{q^2} +i\pi \right).
\end{equation}

\begin{figure*}[tbh]
  \includegraphics[width=0.495\columnwidth]{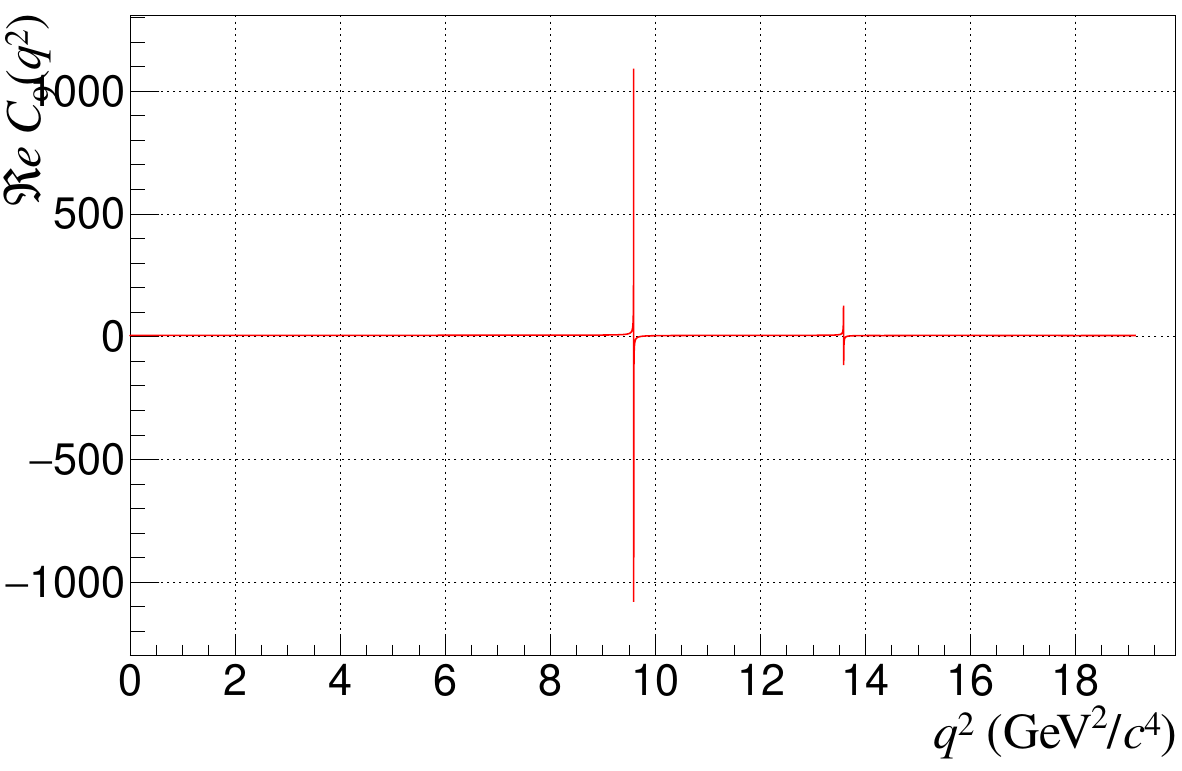}
  \includegraphics[width=0.495\columnwidth]{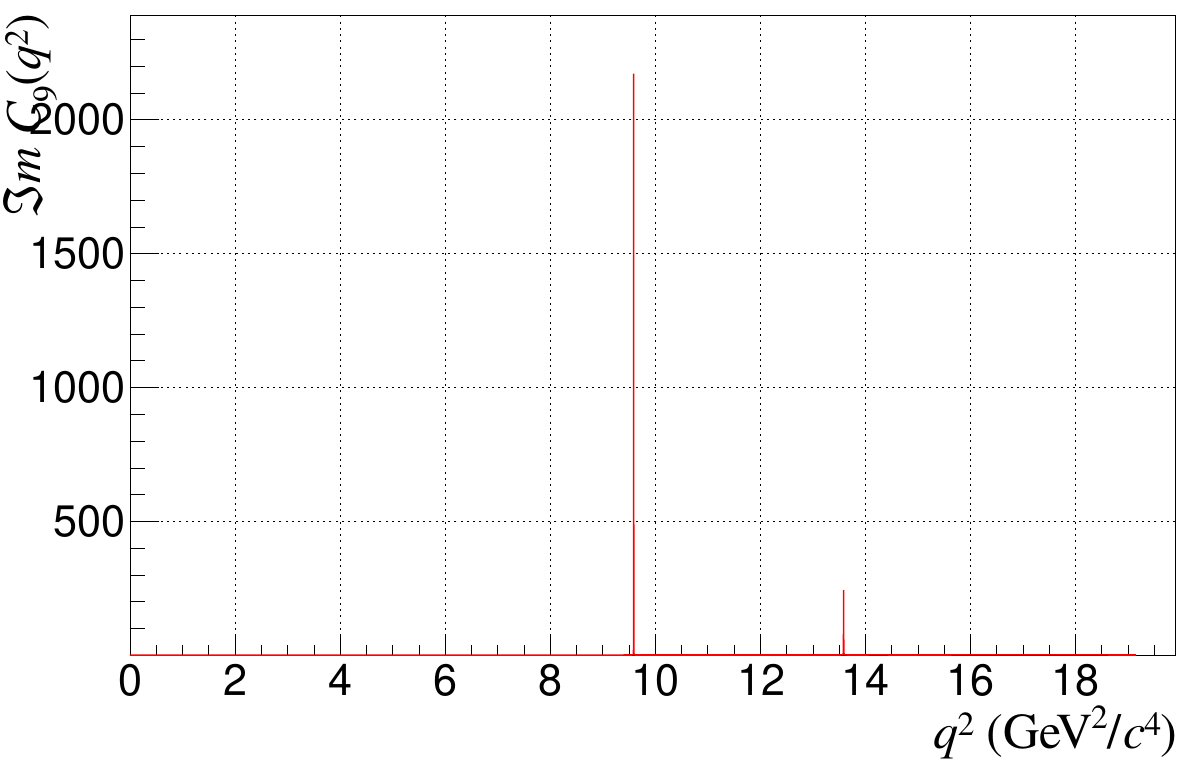}\\
  \includegraphics[width=0.495\columnwidth]{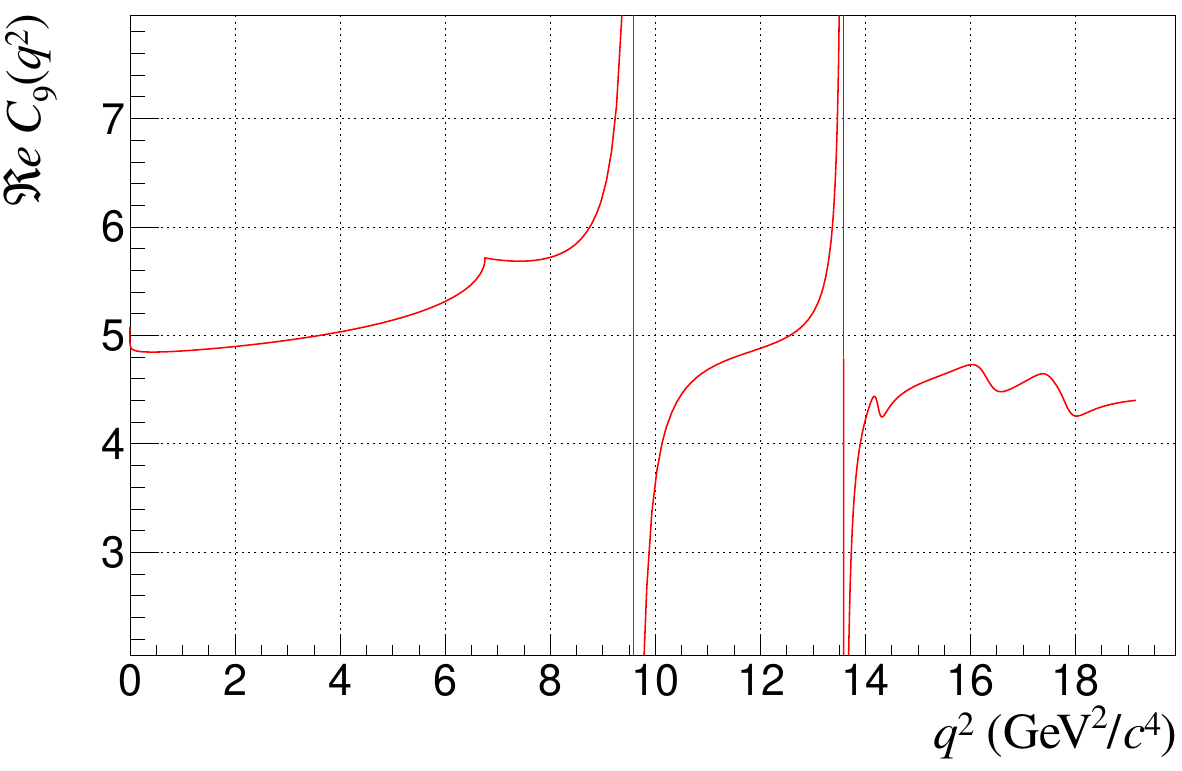}
  \includegraphics[width=0.495\columnwidth]{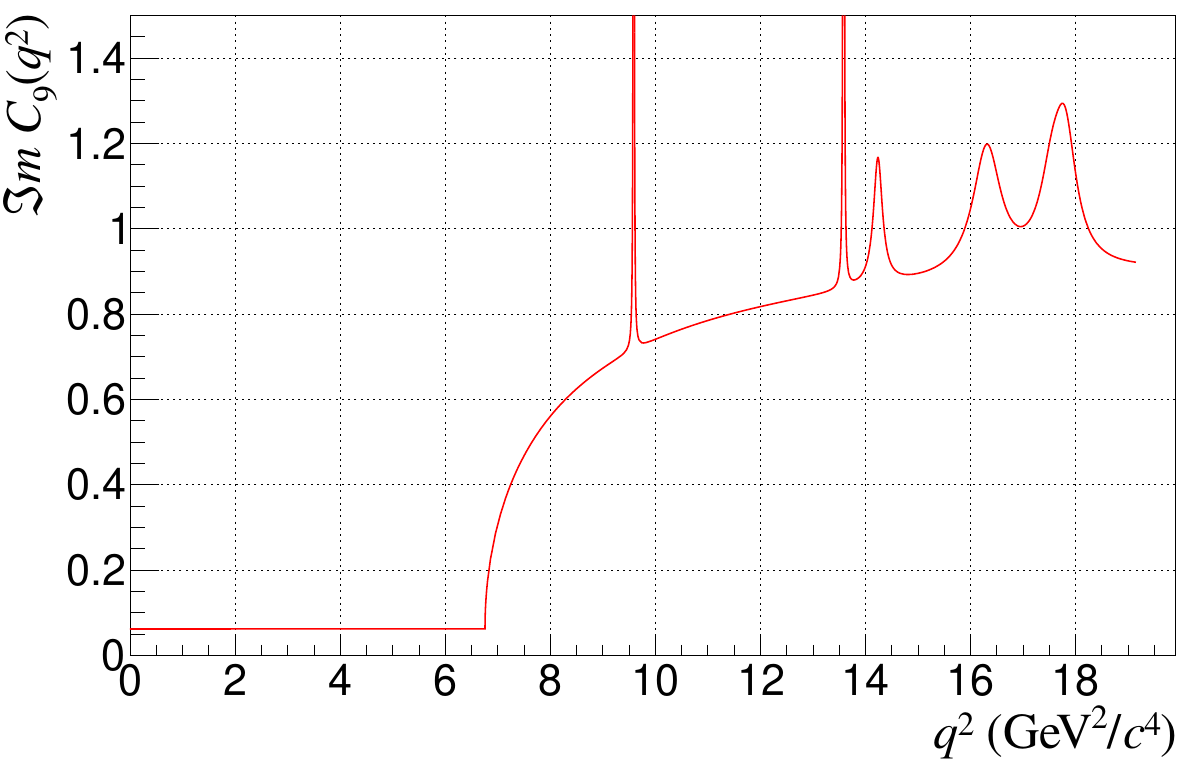}
  \caption{\label{fig:C9reso} The SM Wilson coefficient $C_9$ versus
    $q^2$ (see Eq.~\ref{eq:WC}) with $c\bar{c}$ resonances
    included. The top row shows the full range and the bottom row
    shows a zoomed-in version. The contribution of these resonances
    will be a limiting uncertainty in the extraction of BSM Wilson
    coefficients from $B\to K^* \mu^+ \mu^-$.}
\end{figure*}

One possible but simplistic approach to include the resonance effects
is to use a Breit-Wigner ansatz where the $h$ function in
Eq.~\ref{eq:Yfun} for the $c\bar{c}$ contributions should be replaced
by~\cite{Kruger:1996cv}
\begin{eqnarray}
\hspace*{-0.4cm} &&h(m_c,q^2)  \rightarrow  h(m_c,q^2)\nonumber\\
  & & \qquad{}-\frac{3\pi}{\alpha^2} \sum_{V=J/\psi,\psi^\prime,..} 
  \hspace*{-0.4cm}  \frac{m_V \text{Br}(V\rightarrow 
  \ell^+\ell^-)\Gamma^V_{\text{total}}}{q^2-m_V^2+i m_V 
  \Gamma^{V}_{\text{total}}}\,,
  \label{eq:hfuncreso}
\end{eqnarray}
where $m_V$, $\Gamma^V_\text{total}$ and $\Gamma^V_\text{had}$ are the
mass, total decay width and hadronic decay width of the vector boson
$V$, respectively. Equation \ref{eq:hfuncreso} assumes no
  non-factorizable contributions, and therefore no relative strong
phase between the two terms. Using this approach $C_9$ is modified by
the $c\bar{c}$ contributions as shown in Fig.~\ref{fig:C9reso}.

The differential decay rate dependence on $q^2$, $\cos\theta_\ell$, 
$\cos\theta_K$, and $\chi$ assuming zero-width $K^*$ can be written as
\begin{eqnarray}
  &  \label{eq:helicity} 
  \displaystyle\frac{\dd{}^4\Gamma(\bar{B}^0\to \bar{K}^{0*}\ell^+\ell^-)}{\dd{q^2}\, \dd{\cos\thl}\,
    \dd{\cos\thK}\, \dd\chi} =  I(q^2,\thl,\thK,\chi) = \\
  & \frac{9}{32\pi}
  \Big[ I_1^s \sin^2\thK + I_1^c\cos^2\thK + (I_2^s \sin^2\thK + I_2^c \cos^2\thK) \cos 2\thl \nn \\
    & + I_3 \sin^2\thK \sin^2\thl \cos 2\chi + I_4 \sin
    2\thK \sin 2\thl \cos\chi + I_5 \sin 2\thK \sin\thl \cos\chi  \nn \\
    & + I_6^s \sin^2\thK \cos\thl\nn + I_6^c \cos^2\thK \cos\thl+ I_7 \sin 2\thK \sin\thl \sin\chi \nn \\
    & + I_8 \sin 2\thK \sin 2\thl \sin\chi + I_9 \sin^2\thK \sin^2\thl
    \sin 2\chi
    \Big]\,. \nn
\end{eqnarray}

The angular coefficients can be expressed in terms of transversity amplitudes as
\begin{align}
  \label{eq:I1s}
  I_1^s = &  \frac{(2+\beta^2)}{4} \Big[|\apeL|^2 + |\apaL|^2 +
    (L\to R) \Big]+  \frac{4m^2}{q^2}\Re(\apeL^{}\apeR^{*}+\apaL^{}\apaR^{*}), \\ 
  \label{eq:I1c}
  I_1^c = & |\azeL|^2 \!+\!|\azeR|^2\!+\!\frac{4m^2}{q^2}\big[\abs{{\cal A}_t}^2
    \!+\!2\Re(\azeL^{}\azeR^{*})\big] + \beta^2 {|\cal{A}_S|}^2,\\
  I_2^s = &\frac{\beta^2}{4}\Big[ |\apeL|^2+ |\apaL|^2 + (L\to
    R)\Big],\\ 
  I_2^c = & -\beta^2 \Big[|\azeL|^2 + (L\to R)\Big],\\
  I_3 = & \frac{\beta^2}{2}\Big[ |\apeL|^2 - |\apaL|^2  + (L\to
    R)\Big],\\  
  I_4 = & \frac{\beta^2}{\sqrt{2}}\Big[\Re (\azeL^{}\apaL^*) +
    (L\to R)\Big],\\
  I_5 = & \sqrt{2}\beta\Big[\Re(\azeL^{}\apeL^*) - (L\to R)
    - \frac{m}{\sqrt{q^2}}\, \Re(\apaL {A_S^*}+\apaR {A_S^*}) \Big], \\
  I_6^s = & 2\beta\Big[\Re (\apaL^{}\apeL^*) - (L\to R) \Big],\\
  I_6^c = &
  4 \beta  \frac{m}{\sqrt{q^2}}\, \Re \left[ \azeL {A_S^*} + (L\to R) \right],\\
  I_7 = & \sqrt{2}\beta\Big[\Im (\azeL^{}\apaL^*) - (L\to R) + \frac{m}{\sqrt{q^2}}\, {\Im}(\apeL {A_S^*}+\apeR {A_S^*})
    \Big],\\
  I_8 = & \frac{1}{\sqrt{2}}\beta^2\Big[\Im(\azeL^{}\apeL^*) +
    (L\to R)\Big],\\
  I_9 = & \beta^2\Big[\Im (\apaL^{*}\apeL) + (L\to R)\Big],
\end{align}
where $$\beta=\sqrt{1-\frac{4\,m^2}{q^2}}$$ and we have dropped the
explicit $q^2$ dependence of the transversity amplitudes ${\cal
  A}_{\perp,\parallel,0}^{L,R}$ and ${\cal A}_t$ for notational
simplicity. Note that throughout the paper, we retain lepton masses and 
include non-zero
lepton masses.

\begin{align}
  A_{\perp L,R}  &=  N \sqrt{2} \lambda^{1/2} \bigg[ 
    \left[ (C_9^\eff + C_9^{\eff\prime}) \mp (C_{10}^\eff + C_{10}^{\eff\prime}) \right] \frac{ V(q^2) }{ m_B + m_\kstar} \nn\\
    &+ \frac{2m_b}{q^2} (C_7^\eff + C_7^{\eff\prime}) T_1(q^2)
    \bigg], \\
  A_{\parallel L,R}  & = - N \sqrt{2}(m_B^2 - m_\kstar^2) \bigg[ \left[ (C_9^\eff - C_9^{\eff\prime}) \mp (C_{10}^\eff - C_{10}^{\eff\prime}) \right] 
    \frac{A_1(q^2)}{m_B-m_\kstar}\nn\\
    &+\frac{2 m_b}{q^2} (C_7^\eff - C_7^{\eff\prime}) T_2(q^2)
    \bigg],\\
  A_{0L,R}  &=  - \frac{N}{2 m_\kstar \sqrt{q^2}}  \bigg\{ 
  \left[ (C_9^\eff - C_9^{\eff\prime}) \mp (C_{10}^\eff - C_{10}^{\eff\prime}) \right]
  \nonumber\\
  & \qquad \times 
  \bigg[ (m_B^2 - m_\kstar^2 - q^2) ( m_B + m_\kstar) A_1(q^2) 
    -\lambda \frac{A_2(q^2)}{m_B + m_\kstar}
    \bigg] 
  \nonumber\\
  & \qquad + {2 m_b}(C_7^\eff - C_7^{\eff\prime}) \bigg[
    (m_B^2 + 3 m_\kstar^2 - q^2) T_2(q^2)
    -\frac{\lambda}{m_B^2 - m_\kstar^2} T_3(q^2) \bigg]
  \bigg\},\label{3.48} \\
  A_t  &= \frac{N}{\sqrt{q^2}}\lambda^{1/2} \left[ 2 (C_{10}^\eff - C_{10}^{\eff\prime}) + \frac{q^2}{m_\mu} (C_{P} - C_{P}^\prime)  \right] A_0(q^2) ,\\
  \label{3.50}
  A_S  &= - 2N \lambda^{1/2} (C_{S} - C_{S}^\prime)  A_0(q^2) ,
\end{align}
where
\begin{equation}
  N= V_{tb}^{\vphantom{*}}V_{ts}^* \left[\frac{G_F^2 \alpha^2}{3\cdot 2^{10}\pi^5 m_B^3}
    q^2 \lambda^{1/2}
    \beta \right]^{1/2},
\end{equation}
with
\begin{equation}
  \lambda = \lambda(q^2) = m_B^4  + m_{K^*}^4 + q^4 - 2 (m_B^2 m_{K^*}^2+ m_{K^*}^2 q^2  + m_B^2 q^2).
  \label{eq:kallen}
\end{equation}

The differential distribution for the $CP$-conjugate mode $B^0 \to
K^{0*} \ell^+\ell^- $ can be written as
\begin{align}
  \label{eq:helicitybar}
  \frac{\dd{}^4\Gamma(B^0\to K^{0*}\ell^+\ell^-)}{\dd{q^2}\, \dd{\cos\thl}\,
    \dd{\cos\thK}\, \dd\chi}
  & =
  \bar{I}(q^2,\thl,\thK,\chi) \,,
\end{align}
where $\bar{I}(q^2,\thl,\thK,\chi)$ is obtained with the following
replacements in Eq.~\ref{eq:helicity}: $I_{1,2,3,4,7} \to
\bar{I}_{1,2,3,4,7}$ and $I_{5,6,8,9} \to
-\bar{I}_{5,6,8,9}$~\cite{Kruger:1999xa}. Here $I_i$ is equal to
$\bar{I}_i$ with all weak $CP$ phases conjugated.

\section{\label{app:figs} Detailed fit result distributions}
In Section~\ref{sec:bsmsignals} we studied sensitivity to various
signatures of BSM physics.  Here we provide the detailed results for
the extracted Wilson coefficients with the maximum likelihood fit to
MC data sets produced by the generator in SM and various BSM
scenarios. A numerical summary of the fit results is shown in Table \ref{table:fitsummary}.

\begin{table}
\centering\begin{tabular}{ccccc}
\hline
  Mode & \multicolumn{2}{c}{Electron} & \multicolumn{2}{c}{Muon} \\ \hline
  Wilson coefficient & Mean & RMS  & Mean & RMS \\ \hline
  $\Re e ~\delta C_9$ & $-0.8642$ & $0.0872$& $-0.8623$ & $0.1088$ \\
  $\Im m ~\delta C_9$ & $-0.0389$ & $0.3000$& $-0.0446$ & $0.3184$ \\ \hline
  $\Re e ~\delta C_{10}$ & $-$ & $-$ & $0.0032$ & $0.2147$ \\
  $\Im m ~\delta C_{10}$ & $-$ & $-$ & $-0.0556$ & $0.2786$ \\
  $\Re e~ \delta C_{9}'$ &  $-$ & $-$ & $0.0045$ & $0.1039$ \\
  $\Im m~ \delta C_{9}'$ & $-$ & $-$ & $-0.0003$ & $0.1565$ \\
  $\Re e~\delta C_{10}'$ & $-$ & $-$ & $0.0033$ & $0.0947$ \\
  $\Im m~\delta C_{10}'$ & $-$ & $-$ & $-0.0037$ & $0.1394$ \\
  $\Re e~\delta C_7$ & $0.0006$ & $0.0040$& $0.0010$ & $0.0074$ \\
  $\Im m~\delta C_7$ & $0.0001$ & $0.0020$& $0.0014$ & $0.0230$ \\
  $\Re e~\delta C_7'$ & $0.0006$ & $0.0088$& $0.0000$ & $0.0168$ \\
  $\Im m~\delta C_7'$ & $0.0000$ & $0.0091$& $0.0003$ & $0.0186$ \\ \hline
\end{tabular}
\caption{\label{table:fitsummary}Summary of the fit results for 1000
  samples of 50 ab$^{-1}$ assuming 25\% experimental reconstruction
  efficiency. For extraction of the $C_9$ coefficient the samples are
  simulated with the NP case $\delta C_9 = -0.87$. Other values are
  extracted assuming the SM. }
\end{table}

\begin{figure*}[h!]
  \includegraphics[width=0.495\columnwidth]{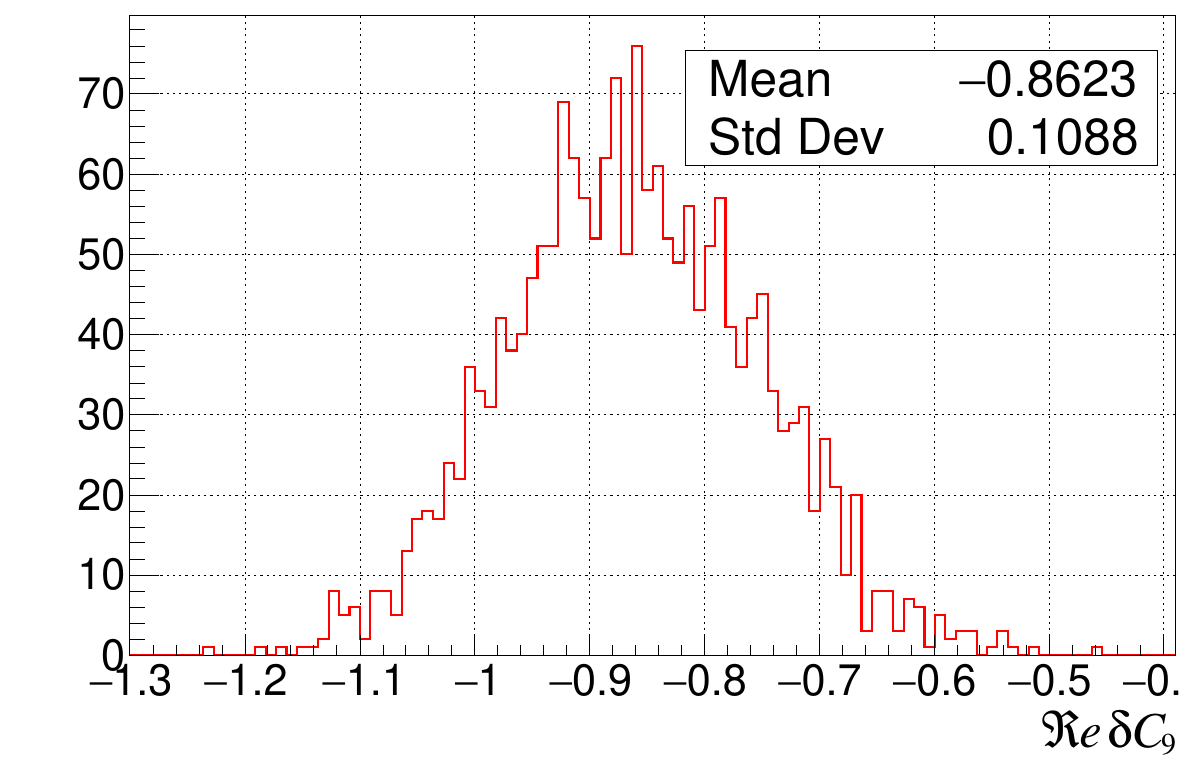} 
  \includegraphics[width=0.495\columnwidth]{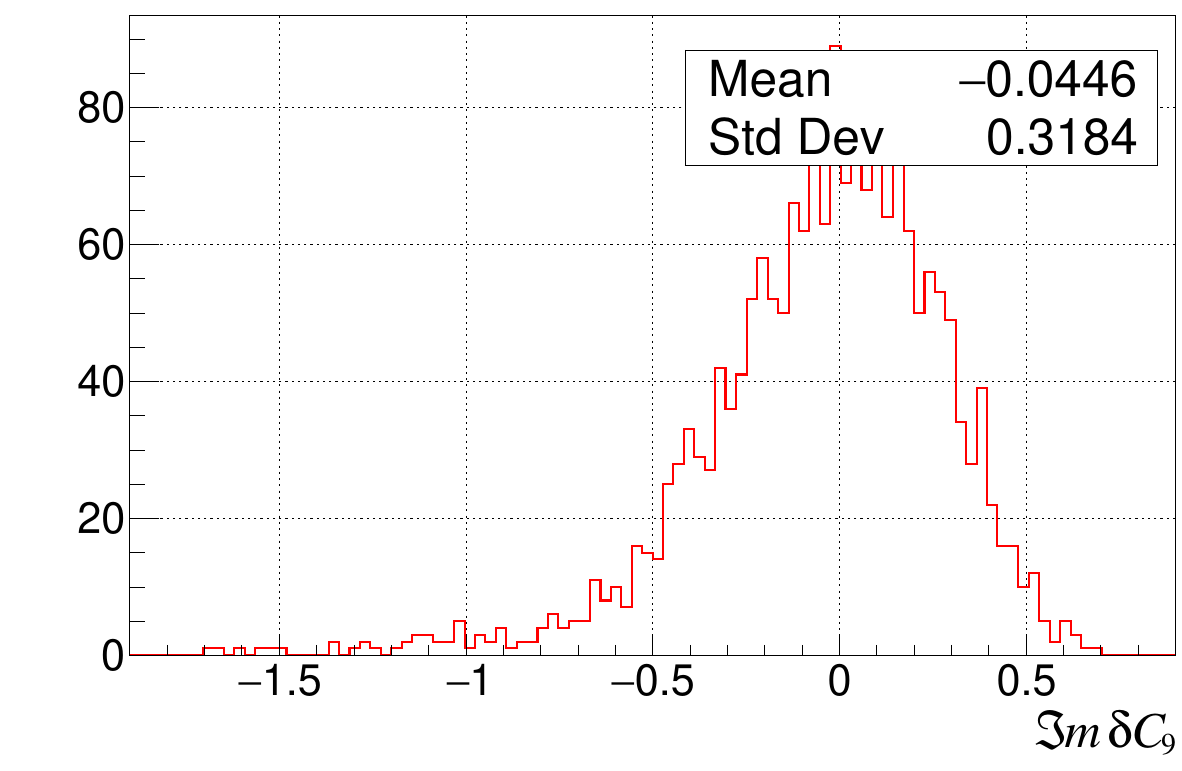}
  \caption{\label{fig:dc9mm_m0_87} Distribution of $\delta C_9$ values
    obtained from unbinned likelihood fits to $B\to K^*\mu^+\mu^-$
    decays generated with the BSM contribution $\delta C_9 = -0.87$ in
    the full $q^2$ range, without resonances included.}
\end{figure*}

\begin{figure*}[h!]
  \includegraphics[width=0.495\columnwidth]{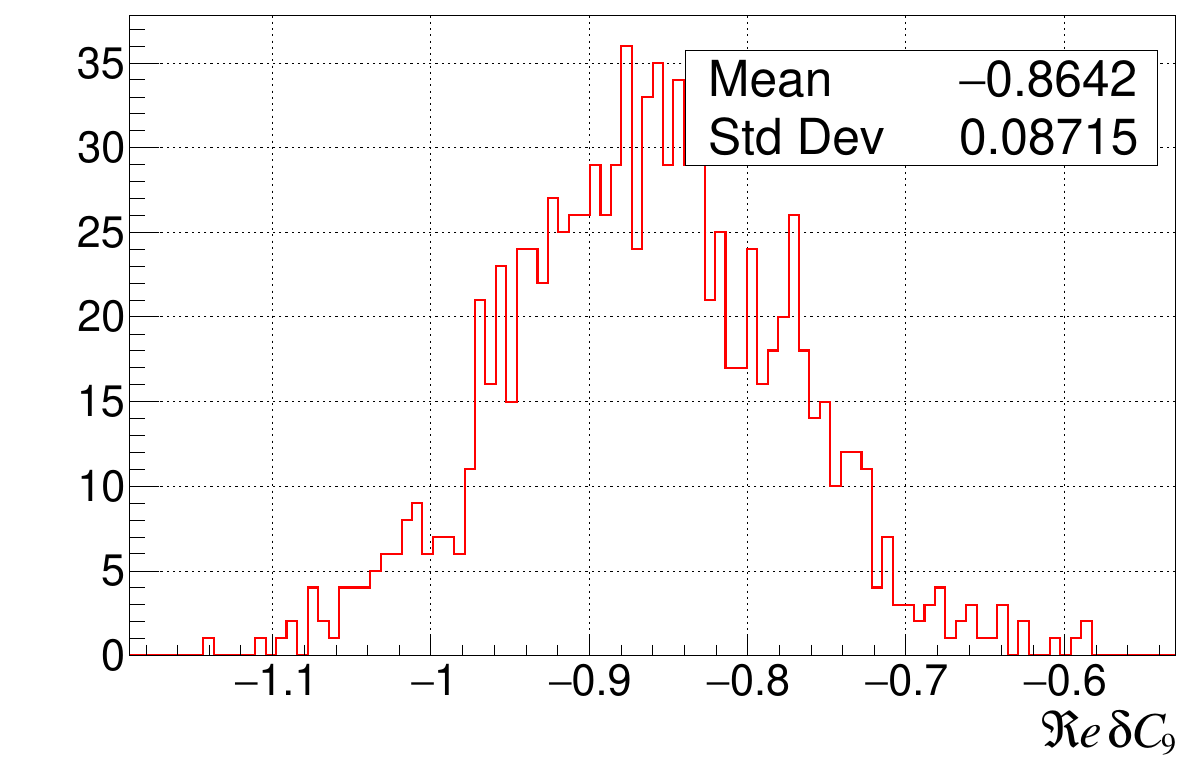} 
  \includegraphics[width=0.495\columnwidth]{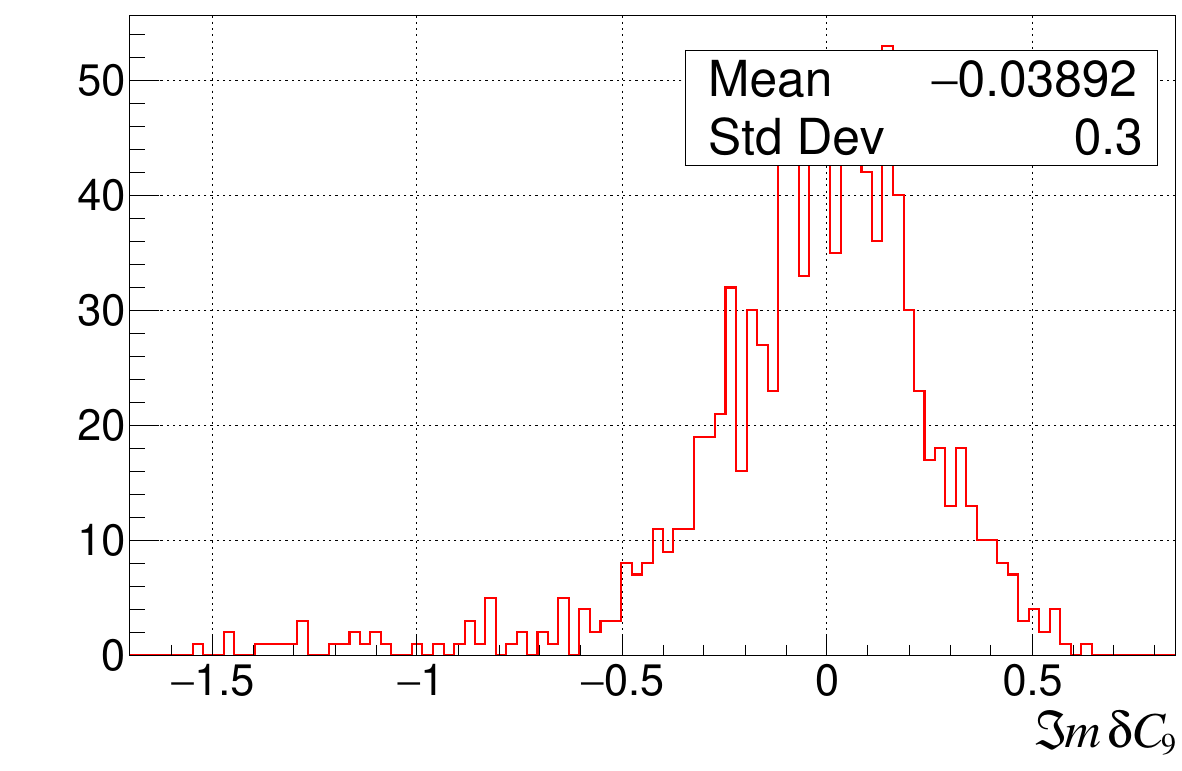}
  \caption{\label{fig:dc9ee_m0_87} Distribution of $\delta C_9$ values
    obtained from unbinned likelihood fits to $B\to K^* e^+e^-$ decays
    generated with the BSM contribution $\delta C_9 = -0.87$ in the
    full $q^2$ range, without resonances included.}
\end{figure*}

\begin{figure*}[h!]
  \includegraphics[width=0.495\columnwidth]{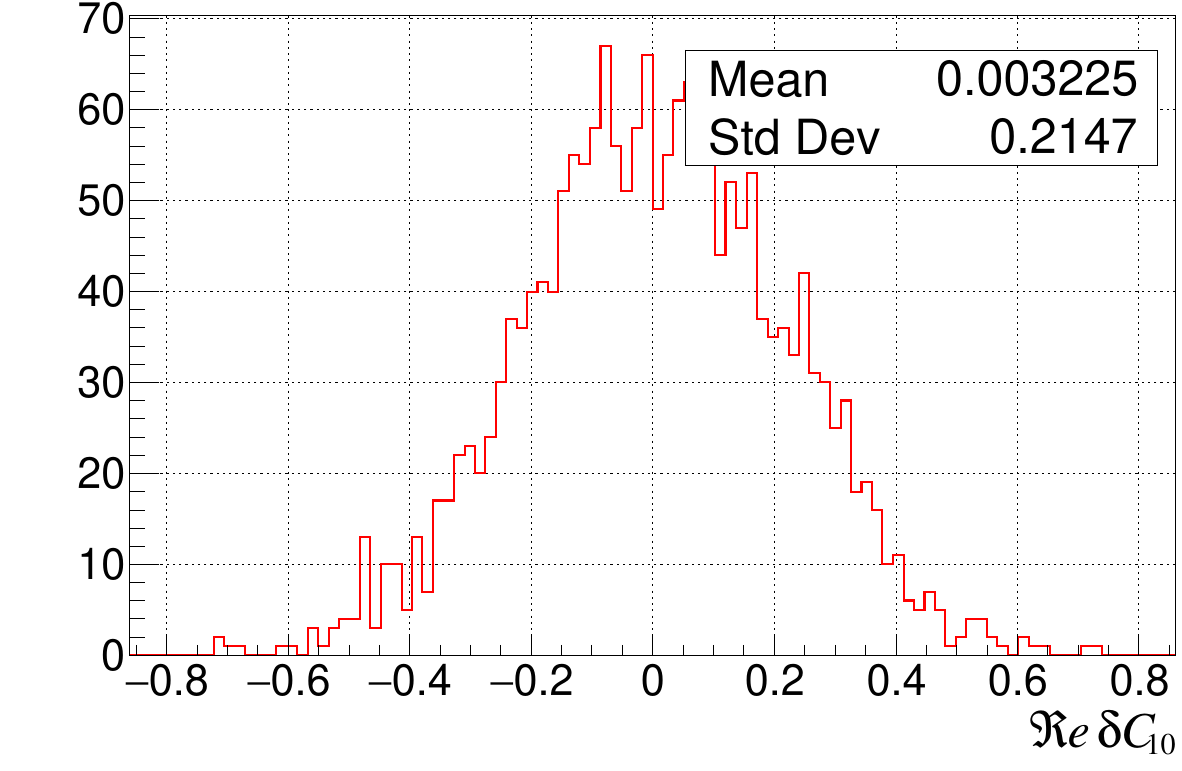} 
  \includegraphics[width=0.495\columnwidth]{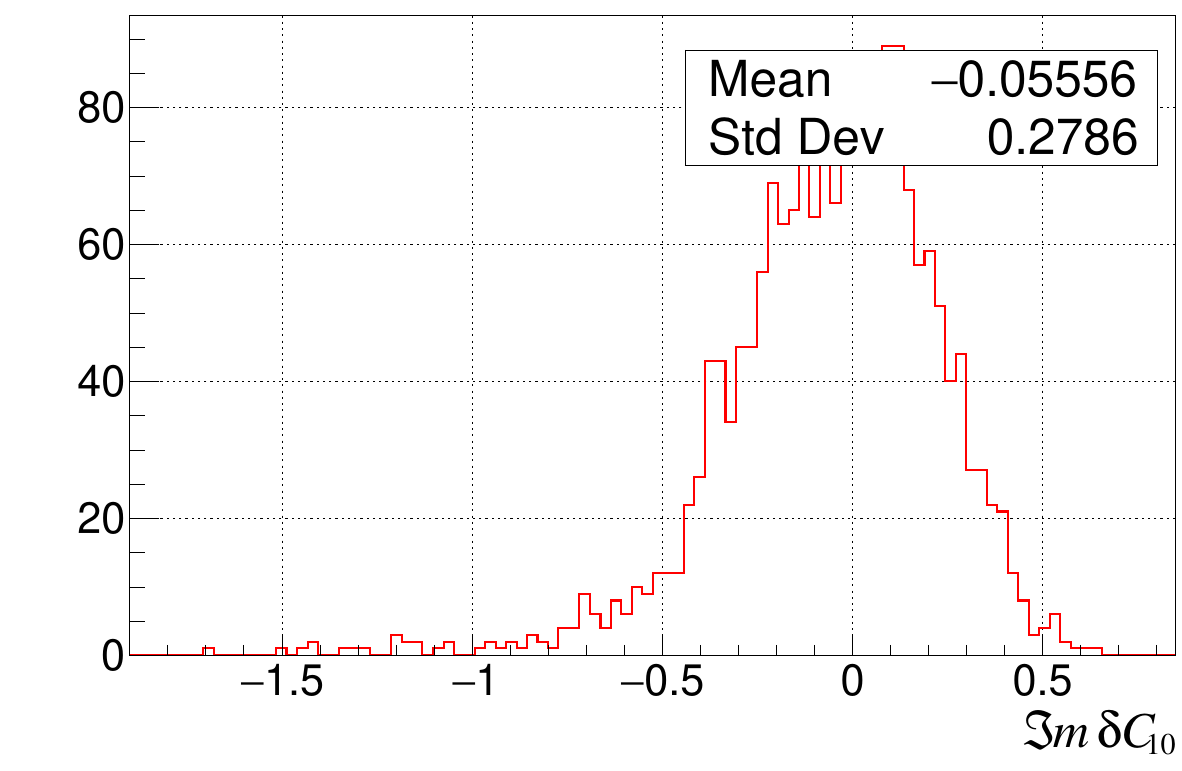}
  \caption{\label{fig:dc10mm} Distribution of $\delta C_{10}$ values
    obtained from unbinned likelihood fits to simulated $B\to K^*
    \mu^+\mu^-$ SM decays in the full $q^2$ range, without including resonances. }
\end{figure*}

\begin{figure*}[h!]
  \includegraphics[width=0.495\columnwidth]{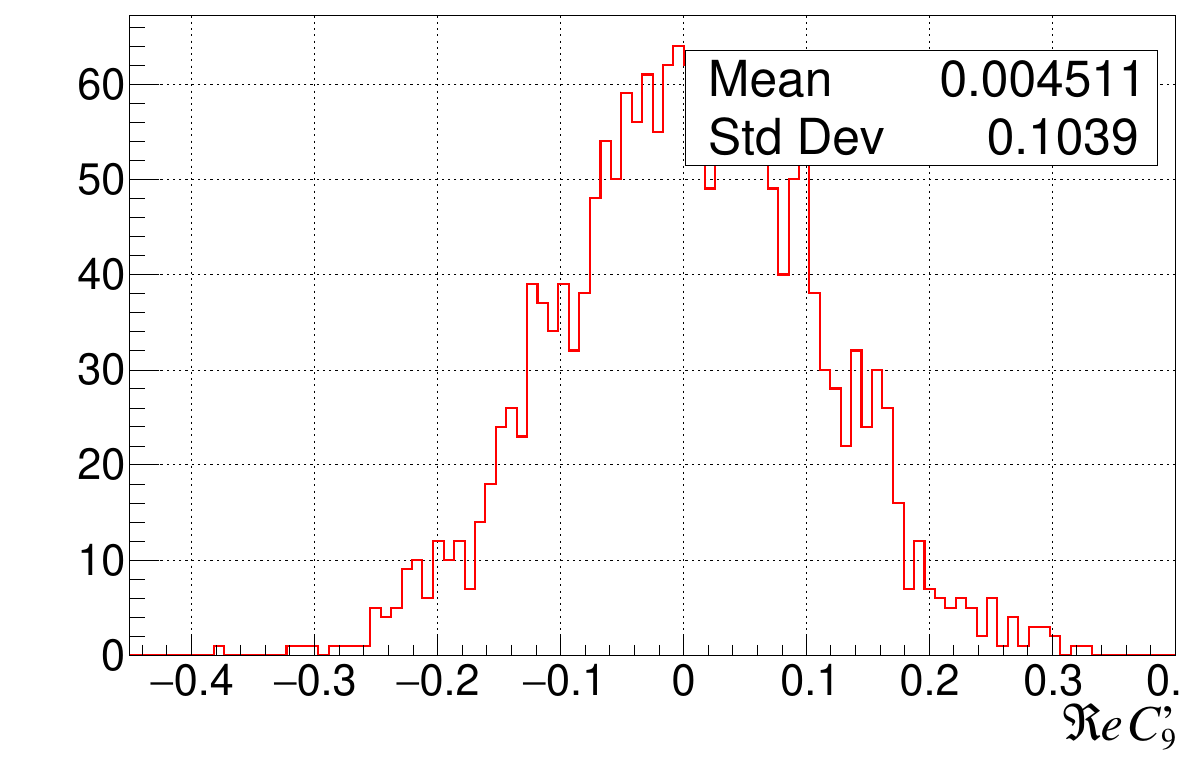} 
  \includegraphics[width=0.495\columnwidth]{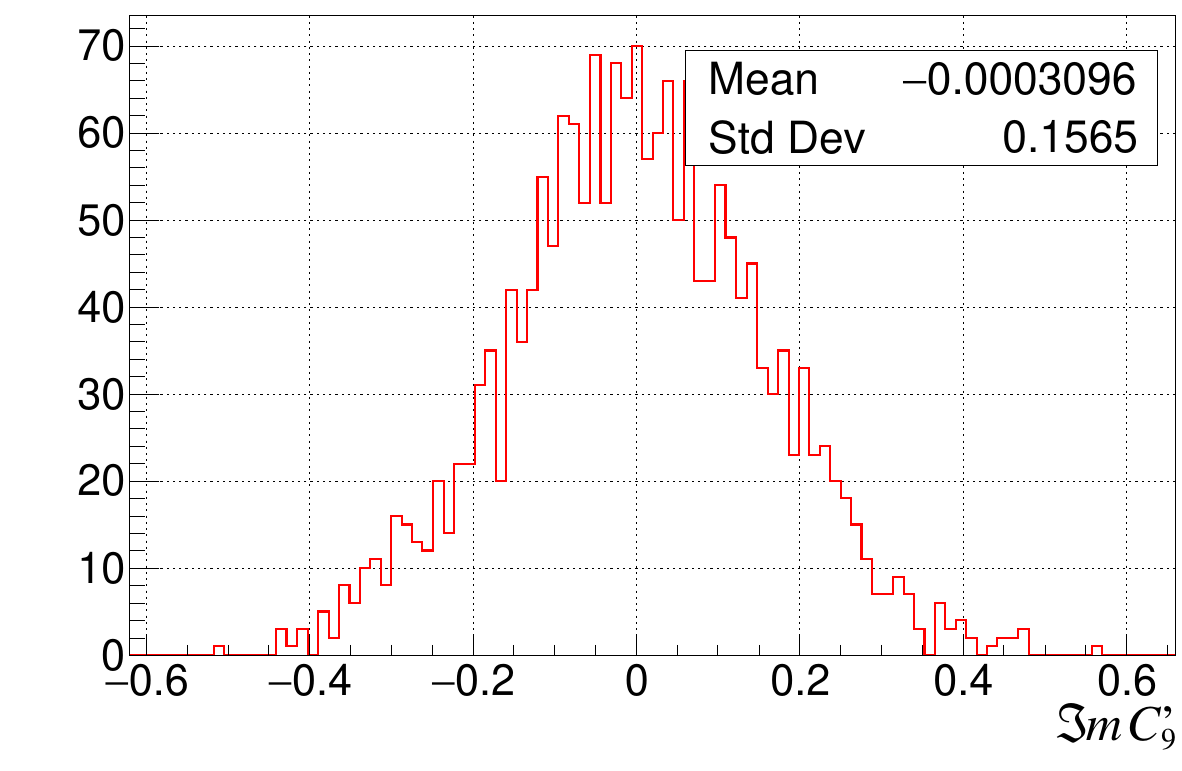}
  \caption{\label{fig:cp9mm} Distribution of $C_9^{\prime}$ values
    obtained from unbinned likelihood fits to simulated samples of $B\to K^*
    \mu^+\mu^-$ SM decays in the full $q^2$ range, without resonances
    included.}
\end{figure*}

\begin{figure*}[h!]
  \includegraphics[width=0.495\columnwidth]{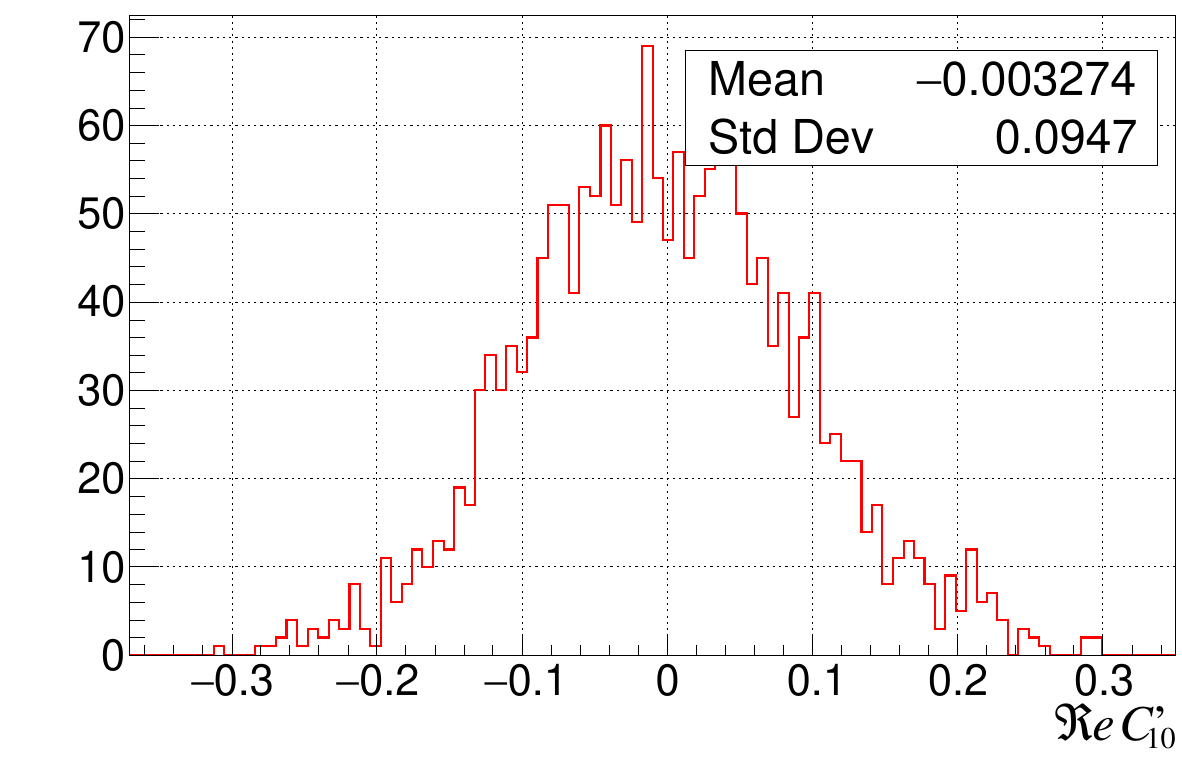} 
  \includegraphics[width=0.495\columnwidth]{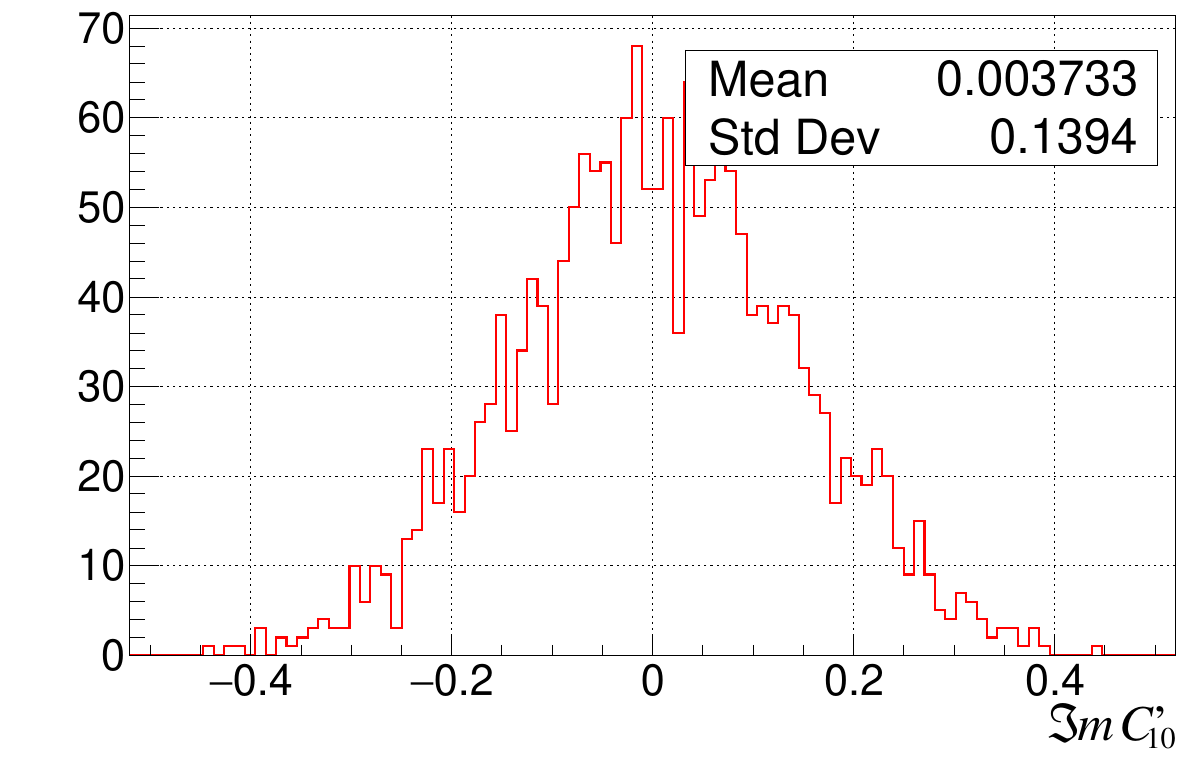}
  \caption{\label{fig:cp10mm} Distribution of $C_{10}^{\prime}$ values
    obtained from unbinned likelihood fits to simulated $B\to K^*
    \mu^+\mu^-$ SM decays in the full $q^2$ range, without resonances
    included.}
\end{figure*}

\begin{figure*}[h!]
  \includegraphics[width=0.495\columnwidth]{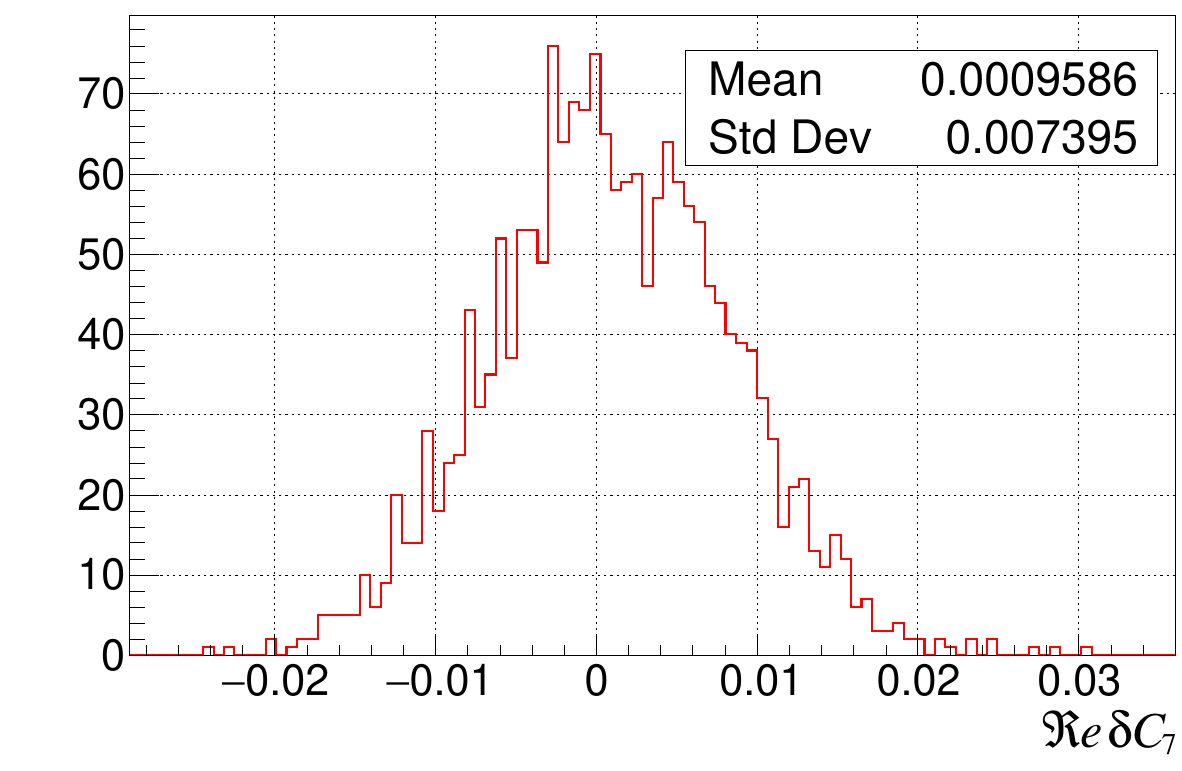} 
  \includegraphics[width=0.495\columnwidth]{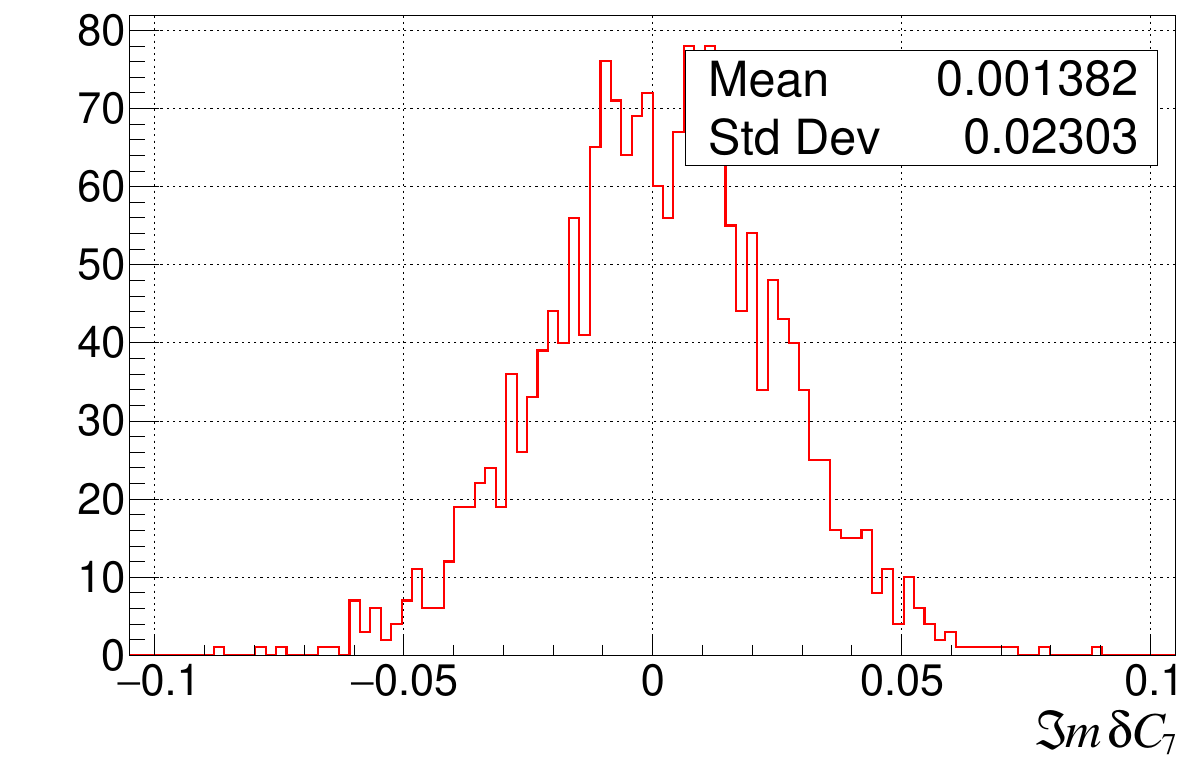}
  \caption{\label{fig:dc7mm} Distribution of $\delta C_7$ values
    obtained from unbinned likelihood fits to simulated $B\to K^*
    \mu^+\mu^-$ SM decays in the full $q^2$ range, without resonances
    included.}
\end{figure*}

\begin{figure*}[h!]
  \includegraphics[width=0.495\columnwidth]{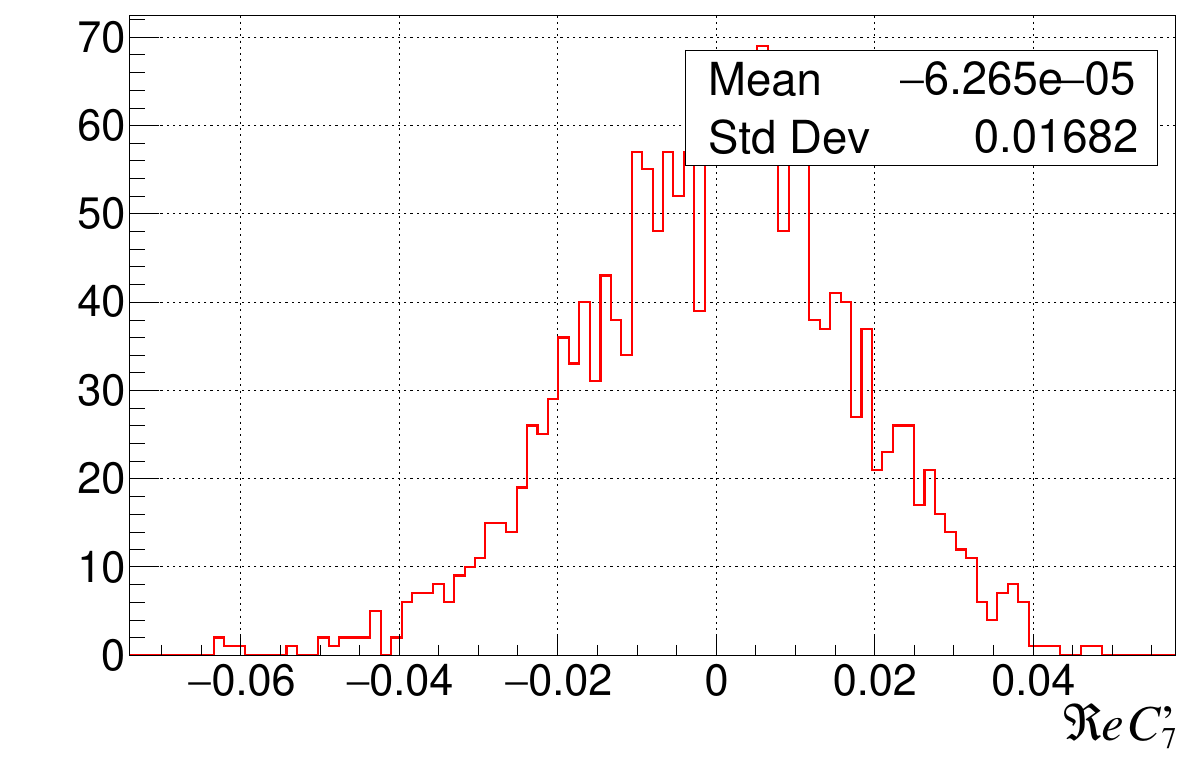} 
  \includegraphics[width=0.495\columnwidth]{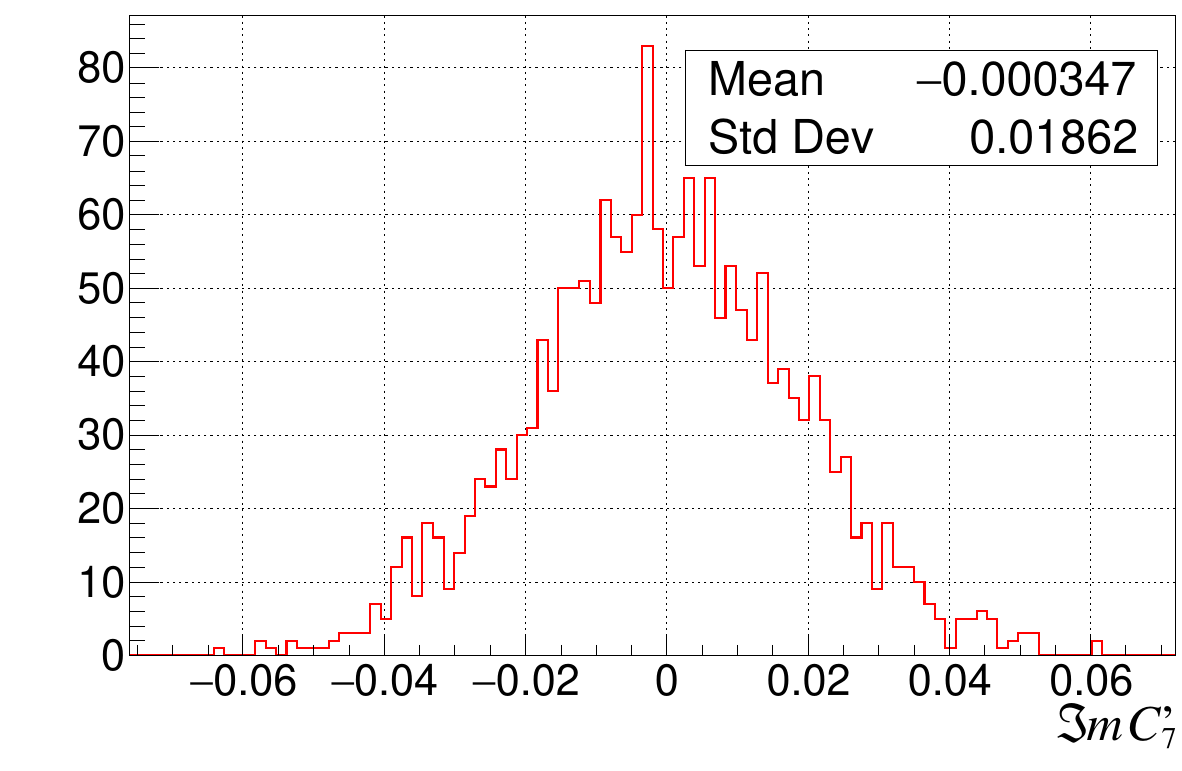}
  \caption{\label{fig:cp7mm} Distribution of $C_7'$ values obtained
    from unbinned likelihood fits to simulated $B\to K^* \mu^+\mu^-$
    SM decays in the full $q^2$ range, without including resonances.}
\end{figure*}

\begin{figure*}[h!]
  \includegraphics[width=0.495\columnwidth]{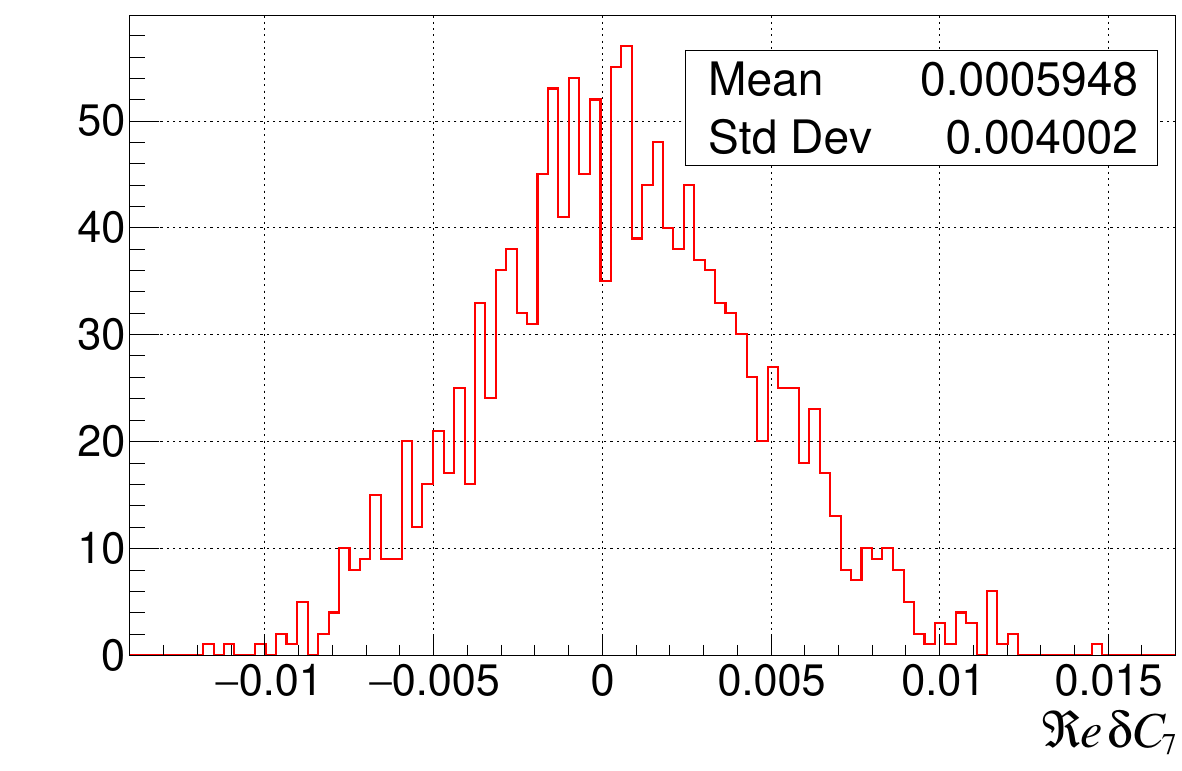} 
  \includegraphics[width=0.495\columnwidth]{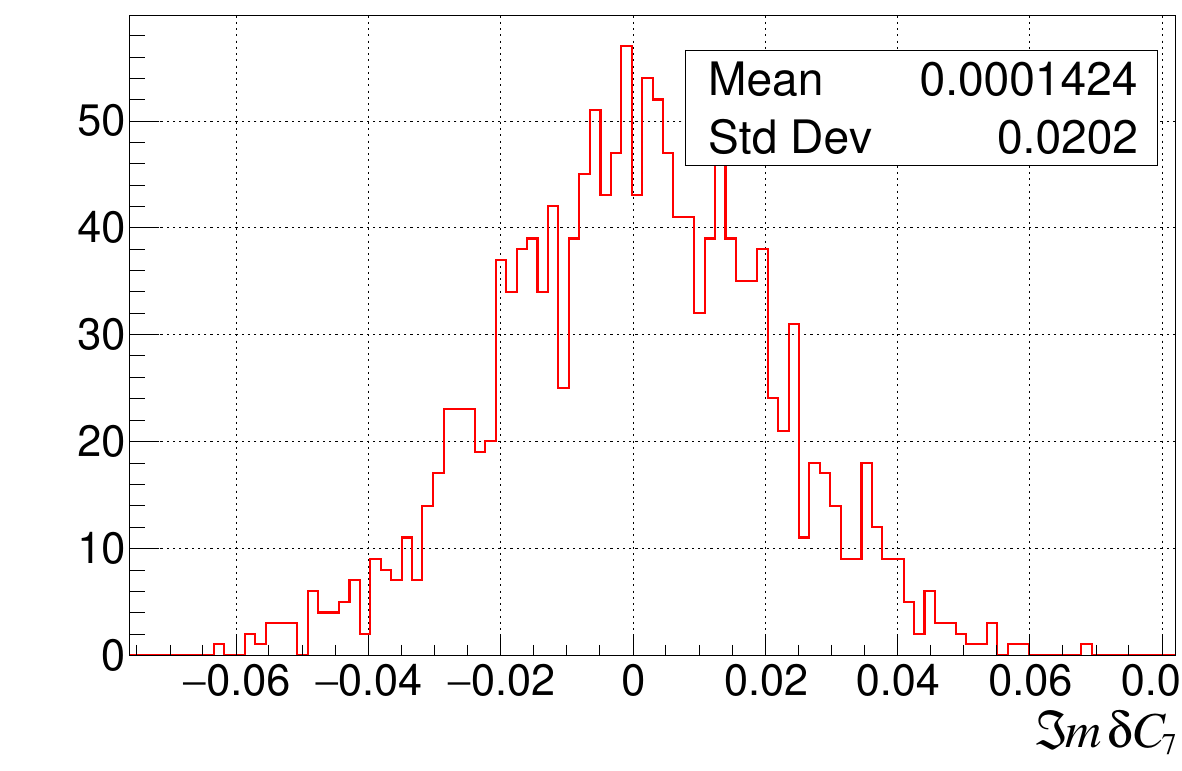}
  \caption{\label{fig:dc7ee} Distribution of $\delta C_7$ values
    obtained from unbinned likelihood fits to simulated $B\to K^*
    e^+e^-$ SM decays in the full $q^2$ range, without including resonances.}
\end{figure*}

\begin{figure*}[h!]
  \includegraphics[width=0.495\columnwidth]{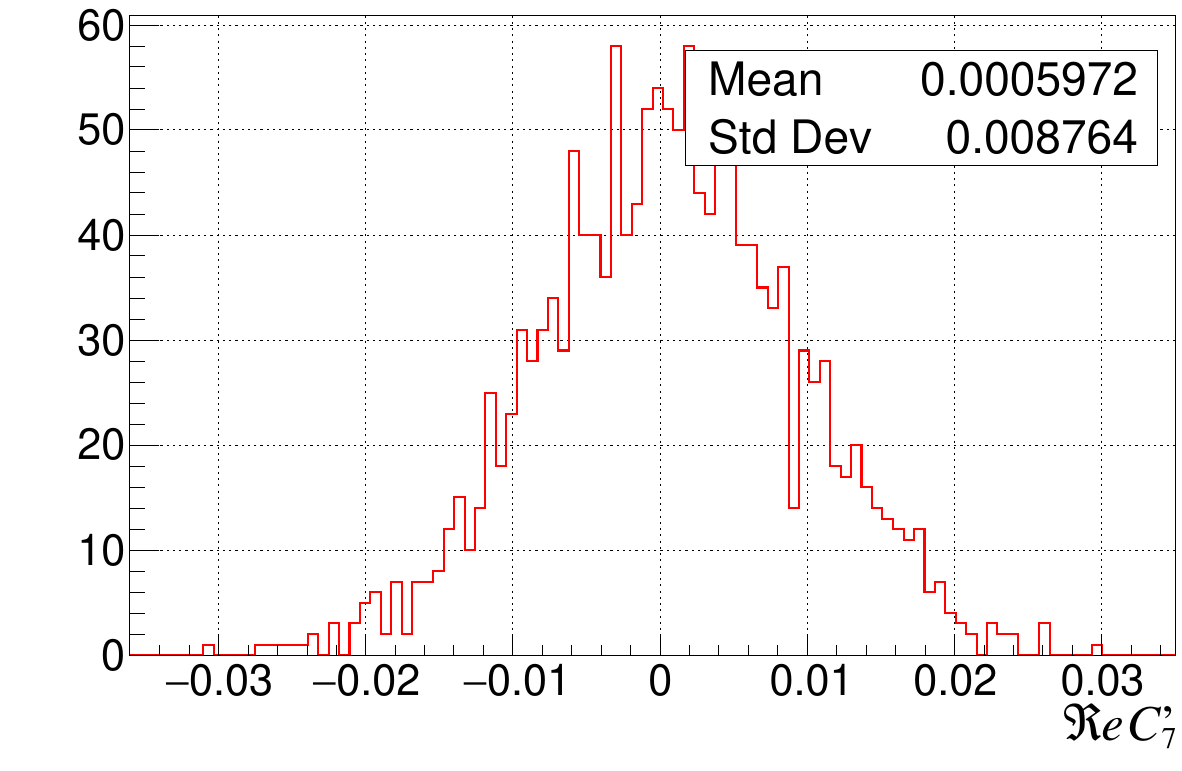} 
  \includegraphics[width=0.495\columnwidth]{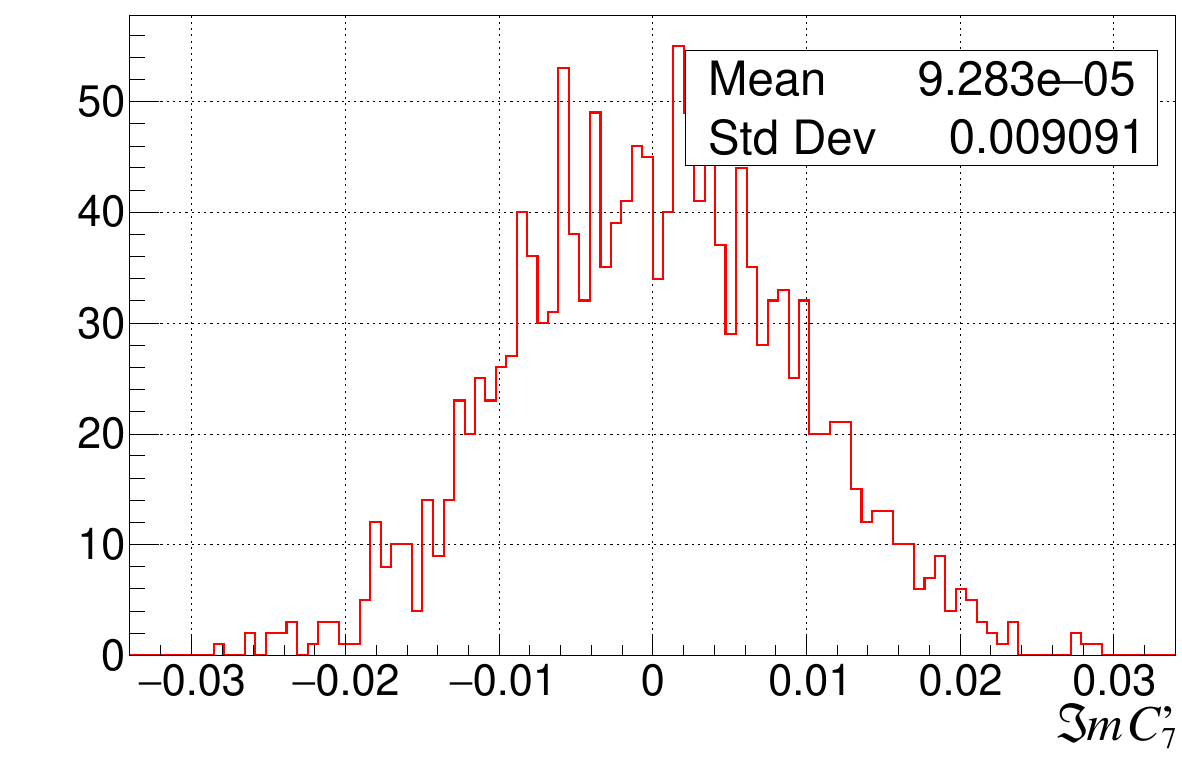}
  \caption{\label{fig:cp7ee} Distribution of $C_7^{\prime}$ values
    obtained from unbinned likelihood fits to simulated $B\to K^*
    e^+e^-$ SM decays in the full $q^2$ range, without including resonances.}
\end{figure*}

\bibliographystyle{JHEP}
\bibliography{myreferences}  

\end{document}